\documentclass[ascmo, manuscript]{copernicus}
\nolinenumbers
\usepackage{graphicx}
\usepackage{csquotes}
\usepackage{placeins}
\usepackage{changes}
\usepackage{xcolor}
\usepackage{amsmath, bm, amssymb, amsthm}
\usepackage{booktabs}
\usepackage{xr-hyper}
\usepackage{tikz}
\usetikzlibrary{positioning, arrows.meta, calc, shapes.geometric, decorations.pathreplacing}
\usepackage{makecell}
\usepackage{scalerel}

\setaddedmarkup{\textcolor{blue}{#1}}
\setdeletedmarkup{\textcolor{red}{\sout{#1}}}
\setcommentmarkup{\textcolor{purple}{\textbf{[Comment: #1]}}}

\begin{document}
\title{Stochastic weather generators for high-frequency wind vector time series}

\Author[1]{Mingshi}{Cui}
\Author[1]{Kevin}{Eng}
\Author[1]{Justin T.}{Greene}
\Author[1]{Zern}{Ke}
\Author[1]{Abolfazl}{Sodagartojgi}
\Author[1]{Zhiqiu}{Xia}
\Author[1]{Gemma E.}{Moran}
\Author[1][ms2870@stat.rutgers.edu]{Michael L.}{Stein}

\affil[1]{Department of Statistics, Rutgers University, Piscataway, NJ, USA}

\received{}
\pubdiscuss{}
\revised{}
\accepted{}
\published{}

\runningtitle{Stochastic Wind Generators}
\runningauthor{M. Cui et al.}

\firstpage{1}

\maketitle

\begin{abstract}
Surface winds can vary substantially from one minute to the next, so there is scope for studying its variation on this fine time scale. 
Restricting to the month of June to minimize seasonality,
this work develops a range of machine learning models for generating realistic time series of surface wind vectors at a site in Lamont, Oklahoma based on more than 30 years of high quality measurements at the minute time scale.
Such a generator could be used as an input into models from a range of disciplines, notably for wind energy, but also wildfire spread and aviation, among others.
The data show complex diurnal structures in both wind speed and direction that would be challenging to capture with standard time series models, so we consider a number of machine learning approaches to producing a stochastic wind generator based on time vector-quantized variational autoencoders.
We consider generating a day's worth of data at a time and generating a day of wind vectors conditional on the previous day's winds.
We also study methods for incorporating a discrete weather state variable in the generator.
We evaluate the generators using a wide range of formal and informal methods.
The best of these generators can capture many but not all of the complex features present in the observational data.
In particular, the best of our approaches accurately mimic diurnal changes in wind volatility but struggle to match the observed distribution of extreme wind speeds.

\end{abstract}

\introduction
%\deleted{Recent years have seen a growing interest in the use of statistical and machine learning approaches to the analysis of wind data, to a large extent driven by a need to forecast wind speeds for predicting energy production from wind turbines \mbox{\citep{gneiting2006, hering2010, sloughter2010, zhu2012,  bessac2016, AZIMI2016208, sharma2018, bali2019, DEMOLLI2019111823, liu2020, lenzi2020, DING202158, peng2021ealstm, wang2024, zhang2024}}.These works use a wide range of statistical and machine learning methods, with many addressing simultaneous prediction at multiple locations }

%(\cite{gneiting2006, gneiting2008, hering2010, sloughter2010, zhu2012, jiang2013, hering2015, bessac2016, AZIMI2016208, Pessanha2017, sharma2018, bali2019, DEMOLLI2019111823, liu2020, lenzi2020, DING202158, Ding2022, peng2021ealstm, wang2024, WindSpeedDistributions2021, zhang2024, price2025probabilistic}).

%\comment{There are too many citations for the wind examples. I have cut it down to four (two highly cited foundational + 2 modern).}

Recent years have seen a growing interest in the use of statistical and 
machine learning approaches to the analysis of wind data. Much of this has been driven by the wind energy sector, which relies on accurate wind speed forecasts as key inputs into wind turbine power generation models
\citep{gneiting2006, hering2010, zhang2024, wang2025}. However,
a second use of these models, and the focus of this paper, is as stochastic
wind generators. These generators are intended to produce realistic synthetic time series
of weather variables which can be used as inputs into a wide range of models, including those for soil erosion \citep{skidmore1990}, wildfire spread \citep{masoudian2021}, and aviation \citep{rhudy2024}.

In many models where wind is an important input, knowing the high-frequency behavior of wind is desirable. For instance, \citet{zhu2012} and \citet{liu2020} note that maintaining a constant balance between electricity supply and demand in the power grid makes variations in wind speed on time scales of 5 to 10 minutes highly relevant, and \citet{jiang2013} demonstrate the value of wind speed predictions at the 10-second scale. Beyond wind speed, high-frequency variability in wind direction substantially impacts wind farm energy yields due to wake effects \citep{porte2013}, and wind turbines may struggle to maintain optimal orientation against rapidly shifting wind directions
\citep{dallas2024}.

Historically, stochastic wind generators were designed for applications where coarse, low-resolution data sufficed. Even as the value of high-frequency simulation has grown, capturing minute-by-minute temporal dependence remains a formidable methodological challenge. As a result, the existing literature largely reflects varying trade-offs between the fidelity of temporal dynamics and the granularity of temporal resolution. For instance, there is 
extensive literature on estimating the marginal probability distribution
of wind vectors \citep{carta2009, WindSpeedDistributions2021, shah2025}.
These methods can easily handle high-frequency wind data; however, because they 
inherently ignore temporal dynamics, they cannot be used to simulate time series data.
Conversely, models that do incorporate temporal dynamics typically operate
at resolutions of one hour or coarser \citep{brown1984, bessac2016, nikolaev2019}. 
Existing methods that attempt to model at higher frequencies, such as 10-minute intervals, \citep{hering2015, yunus2016, wang2024}, generally rely on restrictive parametric frameworks, such as linear autoregressive processes, to generate synthetic data.
This reliance on rigid mathematical structures is especially problematic when modeling the diurnal cycle. For instance, the widely used WINDGEN model forces the entire diurnal cycle into a simple cosine function \citep{skidmore1990}. As we will see in Sect.~\ref{sec:Data}, such rigid structures would inherently fail to capture the extreme volatility and complex intraday variations present in our wind data. In general, because this chaotic volatility becomes highly pronounced at the minute and sub-minute scales, traditional parametric generators are ill-suited for high-frequency simulation.
While recent statistical work has begun to directly analyze observational data at these sub-minute scales \citep{yin2026}, developing stochastic models capable of generating realistic minute-by-minute data remains an open challenge.

To address this gap, we propose a deep generative model for simulating horizontal wind vectors at the minute time scale. The training data consists of roughly 30 years of high-quality meteorological measurements taken at a site in Lamont, Oklahoma as part of the US Department of Energy's
ARM (Atmospheric Radiation Measurement) program \citep{kyrouac2025}. Because capturing minute-by-minute dynamics is a formidable challenge even at a single location, we focus our analysis entirely on the Lamont facility. Furthermore, to isolate the dynamics of the diurnal cycle from broader seasonal shifts, we restrict our modeling to the month of June. To effectively capture high-frequency patterns, we leverage time vector-quantized variational autoencoders \citep[Time VQ-VAE,][]{lee2023vector}. Under this framework, we first develop a model capable of generating highly realistic, standalone days of high-frequency wind vectors. We then extend this architecture to simulate continuous, multi-day weather sequences by conditioning each newly generated day on the final hour of the preceding day. This allows us to effectively embed extreme intraday volatility within a coherent multi-day simulation.
To assess the quality of these generated sequences, we use a wide range of quantitative and graphical methods, most notably a discriminative evaluation framework utilizing a bidirectional Long Short-Term Memory (LSTM) classifier  \citep{yoon2019time} to distinguish between synthetic and real wind vector time series.

The remainder of this paper is organized as follows. Sect.~\ref{sec:Data} describes the 
wind data. Sect.~\ref{sec:Exploratory} provides an exploratory analysis demonstrating the 
complex diurnal patterns in wind vectors and their first differences. Sect.~\ref{sec:Methodology} details the deep generative model used to construct synthetic wind data. 

Sections~\ref{sec:discriminative}--\ref{sec:Stochastic-volatility} each provide distinct ways of evaluating the fidelity of our wind generators.
Sect.~\ref{sec:discriminative} evaluates the quality of the simulated data using a discriminative evaluation framework.
Sect.~\ref{sec:energy_scores} uses energy scores \citep{gneiting2008}, a scoring rule for multivariate outcomes, to further evaluate the realism of our simulated data.
Sect.~\ref{sec:graph-simulated} compares the ability of two of our wind generators to reproduce the patterns found in the graphs shown in Sect.~\ref{sec:Exploratory}.
Sect.~\ref{sec:Stochastic-volatility} uses quantile regression on absolute values of changes in wind speed to evaluate how well our wind generators mimic some aspects of the stochastic volatility in wind speed.

\section{Data}
\label{sec:Data}
The data described in this paper comes from the ARM facility in Lamont, Oklahoma \citep{kyrouac2025}. Specifically, we examine
data generated by the surface meteorological system (MET) \citep{met2021}, which is a cluster of sensors that record barometric pressure, temperature, relative humidity, wind speed and wind direction at a rate of 1 Hz. Although the Lamont facility has many specialized sensors and recordings of other weather features, in this paper, when referring to "Lamont data'', we always mean data from the MET system.
%This facility is run by the US Department of Energy as part of the ARM (Advanced Radiation Measurement) program.
Wind speed and direction are measured at 10 meters above the surface. 
Only minute-by-minute summaries of these 1 Hz measurements are recorded.
For winds, both the average of the 60 measured wind speeds and the norm of the average of the 60 measured horizontal wind vectors are provided.
By Jensen's inequality, the average wind speed is always greater than or equal to the norm of the average wind vector, but the differences are generally small for one-minute periods and we chose to use the norm of the average wind vector in this work.
%The difference between the first two variables is important: the average wind speed refers to the sample average of the magnitude of the measured wind vector at each second while the average wind vector is the sample average of measured wind vectors. 
%The latter is a bivariate measurement consisting of a magnitude and direction.
Surface meteorology has been measured at a network of stations in the region around Lamont since 1994.
We selected the Lamont station because of its continuous operation since 1994 with a relatively small fraction of missing observations.
The relatively high frequency of the recorded observations together with the long period of data collection at this site makes these data an invaluable resource for careful study of the statistical characteristics of time series of surface meteorology.
The high frequency of measurements is particularly relevant for wind, which can vary substantially on short time scales.

%A major modeling challenge was accounting for the volatility of the wind speed. 
A number of past studies of winds have found it helpful to define states or regimes to capture patterns in winds \citep{hering2015, bessac2016}.
To assess the utility of including weather states in a stochastic wind generator at the minute time scale, we define weather states based on the precipitation and barometric recordings coming from the MET system at the Lamont station.
These weather states are used as an indicator for adverse weather events. 
Depending on how the weather states are incorporated into the model, we indeed find that they can lead to more accurate generators.

The MET based weather states were constructed in a $2^4$ factorial fashion with four factors: 1) direction of change in atmospheric pressure; 2) large change in atmospheric pressure; 3) presence of a period of heavy rain; 4) presence of rain. For example, the code $(-, +, + , +)$ corresponds to the event of a large decrease in atmospheric pressure accompanied by heavy rain. While these definitions are
coarse, they allow for comprehensive weather state coverage. Indeed, it would be technically possible to construct very fine, informative weather states as Lamont has a treasure trove of over twenty different high-fidelity weather instruments (e.g. ceilometer, radar wind profiler, LIDAR, disdrometer). However, many of these instruments do not have recordings that are comprehensive in both time and coverage. For instance, some instruments only have recent recordings, and sensitive instruments may undergo frequent and lengthy maintenance.

For determining what constitutes a large change in pressure, we used .02 hPa, which corresponds to approximately the third quartile of barometric changes over 10-minute intervals. Conditions dealing with rain are calculated on a minute-by-minute basis. Heavy rain is defined as the presence of rainfall exceeding 0.25 mm min$^{-1}$, which is the third quartile among all minutes where rain was recorded. The frequencies of the discrete weather states are shown in Table~\ref{tab:state_counts}.  
These weather states were generated using the barometric pressure and tipping bucket (a device that measures the rate of rainfall) recordings. States were calculated for each 10-minute block, and thus have a lower resolution than the recorded wind data which is minute-by-minute. This was done because the barometric pressure recordings are rounded. This rounding makes it impossible to reliably calculate the difference in pressure between two time points by summing over the minute-by-minute differences. %For example, the true sequence of atmospheric pressure readings might be $(3.0, 3.2, 3.5)$, but rounding to the nearest half results in $(3.0, 3.0, 3.5)$.

The weather states were designed specifically to capture adverse weather events rather than to maximize discriminative power between any two time blocks. The cutoffs defining these states were based on established scientific principles. For instance, fast-moving weather fronts cause turbulence and precipitation, which correspond with rapid changes in atmospheric pressure. Therefore, we expect high-volatility events like thunderstorms to co-occur with rapid pressure fluctuations. We hypothesized this state design would improve our stochastic generator,
which we indeed found to be the case.
%; early experiments indicated the model struggled to learn the high-volatility wind regimes accompanying these short-duration events, likely treating them as noise.

\begin{table}[!htbp]
    \caption{Weather state counts over the entire dataset. The 5 omitted states have counts of 0; 4 of these are impossible (presence of heavy rain but no presence of rain).}
    \label{tab:state_counts}
    \centering
    \captionsetup{justification=centering}
    \begin{tabular}{cc}
        \toprule
        \textbf{Weather State} & \textbf{Count} \\
        \midrule
        $(-,-,-,-)$ & 313110\\
        $(+,-,-,-)$ & 211175\\
        $(-,+,-,-)$ & 116966\\
        $(+,+,-,-)$ & 45508\\
        $(-,-,-,+)$ & 2808\\
        $(-,+,+,+)$ & 1974\\
        $(+,-,-,+)$ & 1604\\
        $(+,+,+,+)$ & 1430\\
        $(-,-,+,+)$ & 542\\
        $(+,-,+,+)$ & 401\\
        $(+,+,-,+)$ & 2\\
        \bottomrule
    \end{tabular}
    \vspace{8pt}
    
\end{table}

%
%\textcolor{red}{consider paragraph below for removal. WMO codes where used in the 
%first pass. Codes were switched out later to be only based on lamont data.}
%\textcolor{red}{MS: Do we actually ever use these WMO codes?}
%

%diurnal pattern was clear, this was often broken by weather events such as thunder storms or rain.
%To account for changes in wind patterns due to weather conditions, 
%our most complex simulations used auxiliary data in the form of hourly
%World Meteorological Organization (WMO) codes from the Open Meteo API \citep{Zippenfenig_Open-Meteo}.
%We do not consider this data a primary source since the WMO codes are derived from a reanalysis using
%nearby weather stations. The reanalysis uses the ERA5 model \citep{Hersbach_ERA5}, which has a 
%spatial resolution ranging between 9 to 25km and a temporal resolution of one hour. Even though 
%the Lamont site contains instruments necessary to
%provide a WMO code down to the minute, data is often missing from one or more instrument. Thus a major benefit
%of using the reanalysis is that it provides complete records. \textcolor{blue}{Justin: I thought Kevin defined the weather states we ultimately used without relying on the WMO codes or reanalysis.}

As part of our analysis, we carefully examined the quality of the data using a combination of documentation
and heuristics. The data set spans a period of 32 years, so inconsistencies due to sensor upgrades, malfunctions, maintenance and general changes to data protocols are to be expected. 
See Sect.\ \ref{supsec:zeroes} for details on wind speed values we decided were suspect and removed from the dataset.

\begin{figure}[!htbp]
    \centering
    \includegraphics[width=.88\linewidth]{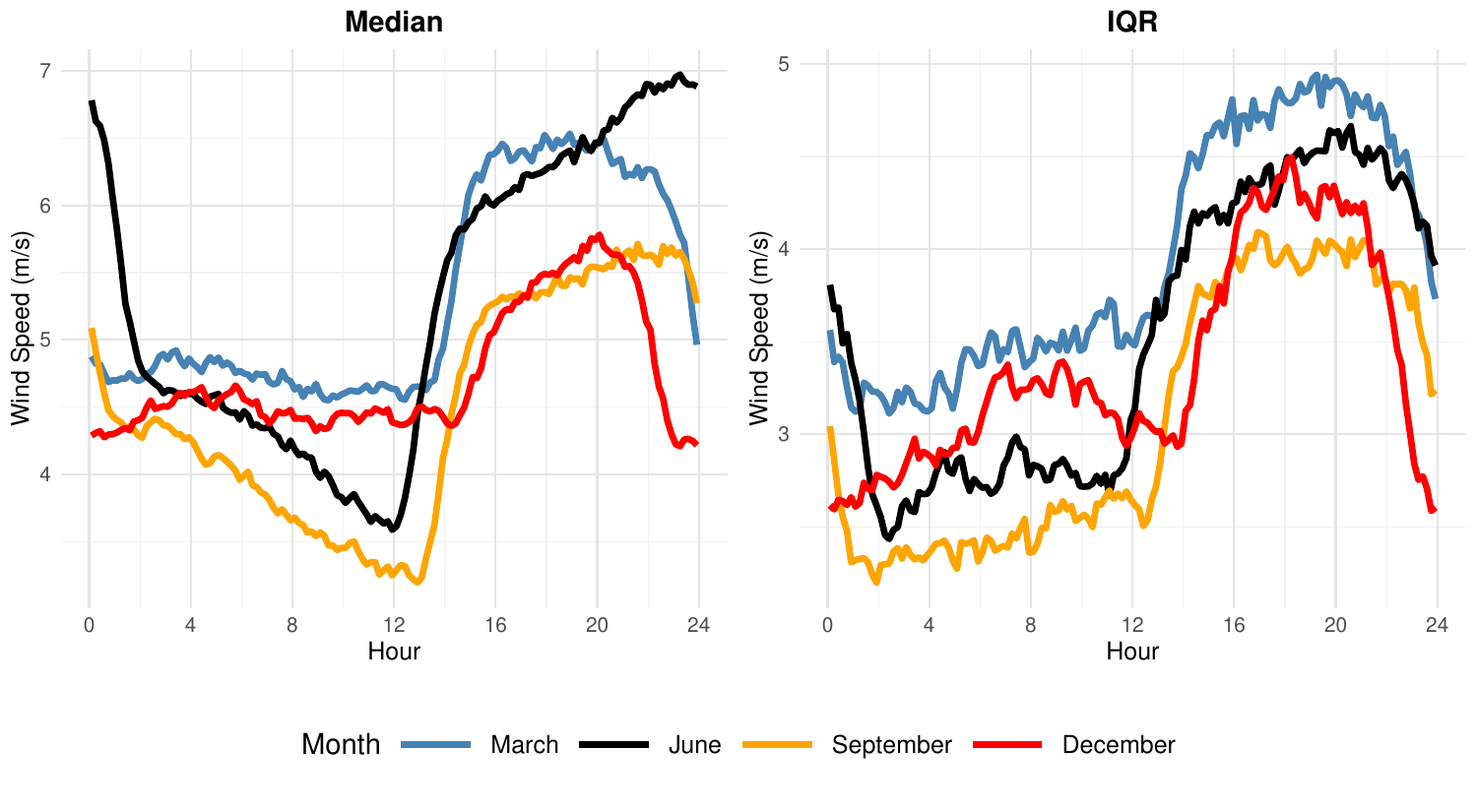}
    \caption{Diurnal cycle in wind speed (m s$^{-1}$) for four months of the year. The left plot shows ten-minute averages of minute-by-minute median wind speeds while the right plot shows the ten-minute averages of minute-by-minute interquartile ranges (IQR).}
    \label{fig:seasonality diurnal plot}
\end{figure}

Figure~\ref{fig:seasonality diurnal plot} displays the diurnal cycle of wind speed across four representative months from 1998 to 2020. To generate this plot, we calculated the median and interquartile range (IQR) for every minute of the day across the entire 22-year period, and then averaged those statistics over consecutive 10-minute intervals. The results reveal a strong diurnal cycle for both metrics. Across all four months, wind speed and its variability are substantially higher during the daytime than at night (note that solar noon in Lamont corresponds to approximately 18:30 UTC).
%}

%\added{
These diurnal patterns largely align with what we would expect if solar radiation is a main driver of variations in wind speed. For example, the stretches of low nighttime wind speeds are longest in December and shortest in June, which coincide with the times of year with the fewest and most hours of sunlight, respectively. More generally, the most rapid changes in both metrics consistently occur in the two hours after sunrise and before sunset. However, sunlight does not explain all seasonal differences. For instance, the IQR is substantially lower throughout the day in September compared to March, even though the amounts of daylight in the two months are roughly the same. There are also more subtle nighttime anomalies: from about 2:00 to 12:00 UTC, the median wind speed remains relatively flat during December and March, whereas it drops noticeably during these same hours in June and September.
%}

% \begin{figure}[!htbp]
%     \centering
%     \includegraphics[width=.88\linewidth]{figures/Data Section/seasonality_lines.pdf}
%     \caption{Diurnal cycle in wind speed (m s$^{-1}$) for four months of the year. The left plot shows ten-minute averages of minute-by-minute median wind speeds while the right plot shows the ten-minute averages of minute-by-minute interquartile ranges (IQR).}
%     \label{fig:seasonality diurnal plot}
% \end{figure}
%\textcolor{red}{I checked the definition of sunset and sunrise, and it is defined in a way that you can't infer the their times at the winter solstice from those at the summer solstice.  I think we should remove these vertical bars and define ``night'', ``day'' and transitions based on Figure 1.}

%It is apparent even from this single figure that there is a strong and fairly complex interaction between seasonality and diurnality in these wind data.
%Thus, for example, the approach used in \citep{hering2015} to remove diurnal and seasonal cycles from wind data using a model that has additive diurnal and seasonal cycles would not be adequate for the Lamont data.
%To reduce both the modeling and computational challenges of working with the full dataset, we have chosen to consider only the month of June in our analyses and we ignore seasonality going forward.
%We also ignore any possible impact of climate change, as some preliminary analyses do not show any clear trend across years in the statistical properties of the wind vectors.

%\added{
  The previous observations suggest the daily wind cycle changes fundamentally depending on the time of year. Because of this complexity, simpler modeling approaches, such as the one proposed by \cite{hering2015}, are inadequate for the Lamont data. Hering’s model assumes that diurnal and seasonal cycles are additive, which implies the daily wind patterns are exactly the same throughout the year except for shifts higher or lower depending on the time of year. Since the actual duration and shape of the daily wind cycles in Lamont change significantly from season to season, an additive model is inadequate.
%}

%\textcolor{red}{MS: This covers the same ground as what I now describe in the previous two paragraphs.} As previously mentioned, we restricted our analysis to the first three weeks of June at the Lamont, Oklahoma E13 ARM station \citeyear{kyrouac2025}. June 1-21 were chosen as they are the days leading to the summer solstice (usually June 21) while not being a long enough time period where seasonality becomes a major concern. Figure \ref{fig:seasonality diurnal plot} illustrates the diurnal $\times$ seasonality interaction. The diurnal cycle generally has lower volatility at night, larger volatility during the day, and sharp transitions in the windows between the two. During periods of the year when the day is shorter, for example December, the period of low volatility (night) is longer than it is in June. Accounting for the diurnal cycle requires modeling the change in the mean and variance structure during the day, night, and two transition periods. Adding how these structures change throughout the year is a worthy exploration, but is left for subsequent work.   

The Lamont ARM station reports data from July 1993 to present, thus our June 1-21 data spans 1994-2025. We use years 1998-2020 for training and the remaining years 1994-1997 and 2021-2025 for testing; Table~\ref{tab:data_summary} summarizes the datasets in detail. 
%\replaced{
We analyzed data from these test years only after all models were finalized.
%}{We did not undertake any analyses of data from these test years until we had finalized all models we planned to use for wind data generation.} %\replaced{
To guide model development and guard against overfitting without compromising the test set, data from June 22–30 for 1998–2020, the same years as for the training data, were used exclusively as a validation set for tuning.
%}{In some analyses, we used data from June 22-30 for the years 1998-2020 as a validation data set so we could look for evidence of overfitting without using any data from the test years. We do not report on any results from these analyses of data from June 22-30.}

%\replaced{
Because training our model requires complete days,
we maximized the size of our training data by imputing days
with only modest missingness. Specifically, for missing segments
spanning up to 20 minutes, we imputed gaps using two independent
and identically distributed Brownian bridges (Appendix A). 
The imputed data were used exclusively to train the VQ-VAE model and were not included in our graphical analyses or the volatility models in Section 8.
%}
%{The machine learning methods we use require complete days of data. For purposes of training the machine model only, we sought to avoid throwing out a whole day of data because of a modest gap in the record. Therefore, we decided to impute gaps under 20 minutes using two independent and identically distributed Brownian bridge processes, see Appendix A for details. This imputed data was only used for training the machine learning models and, unless stated otherwise, was not used in any of the graphical analyses or in the regression models for volatility in} Sect.~\ref{sec:Stochastic-volatility}.

\begin{table}[!htbp]
    \caption{Summary of training and testing data. The missing days are those that were removed after imputation due to having gaps larger than 20 minutes. %\textcolor{red}{MS: Are missing days those with missing AFTER imputation?  Also update days and minutes for test data to include 2025.}
        }
    \label{tab:data_summary}
    \centering
    \begin{tabular}{|l|c|c|c|c|}
        \hline
        \textbf{Dataset} & \textbf{Days} & \textbf{Years} & \textbf{Full Days} & \textbf{Missing Days}  \\
        \hline
        Training & June 1-21 & 1998-2020 & 474 & 9  \\
        \hline
        Test & June 1-21 & 1994-1997, 2021-2025 & 185 & 4  \\
        \hline
    \end{tabular}
    \vspace{8pt}
\end{table}

\section{Exploratory analyses}
\label{sec:Exploratory}

This section presents a range of graphical analyses of the wind vector data from the training period, June 1--21 for the 23 years 1998--2020.
We believe these analyses are independently interesting, identifying a number of features of the data, some of which we hope will be new to the meteorological community and that highlight the opportunities resulting from considering high-frequency observations of the wind vector.
However, for the present work, the main reason for undertaking these exploratory analyses is to provide a series of benchmarks for investigating the quality of the stochastic weather generators we develop.

%Seasonal and diurnal variations are fundamental features of most meteorological variables.
We have already seen large diurnal cycles in wind speed in Figure~\ref{fig:seasonality diurnal plot}; similarly strong diurnal cycles appear in the first differences of wind speed (see Sect.~\ref{sec:differences}).
In particular, during June, a number of properties of the wind vector show sharp changes between 0:00 and 2:00 UTC, roughly the period before sunset, and between 12:00 and 14:00, roughly the period after sunrise.
Many of these properties also clearly vary during other times of day, but, hereafter, in summarizing some of our findings, when we give results separately for what we will call ``nighttime'' and ``daytime,'' we will mean 2:00-12:00 UTC and 14:00-24:00 UTC, respectively.

\begin{figure}[!htbp]
\includegraphics[width=0.75\textwidth]{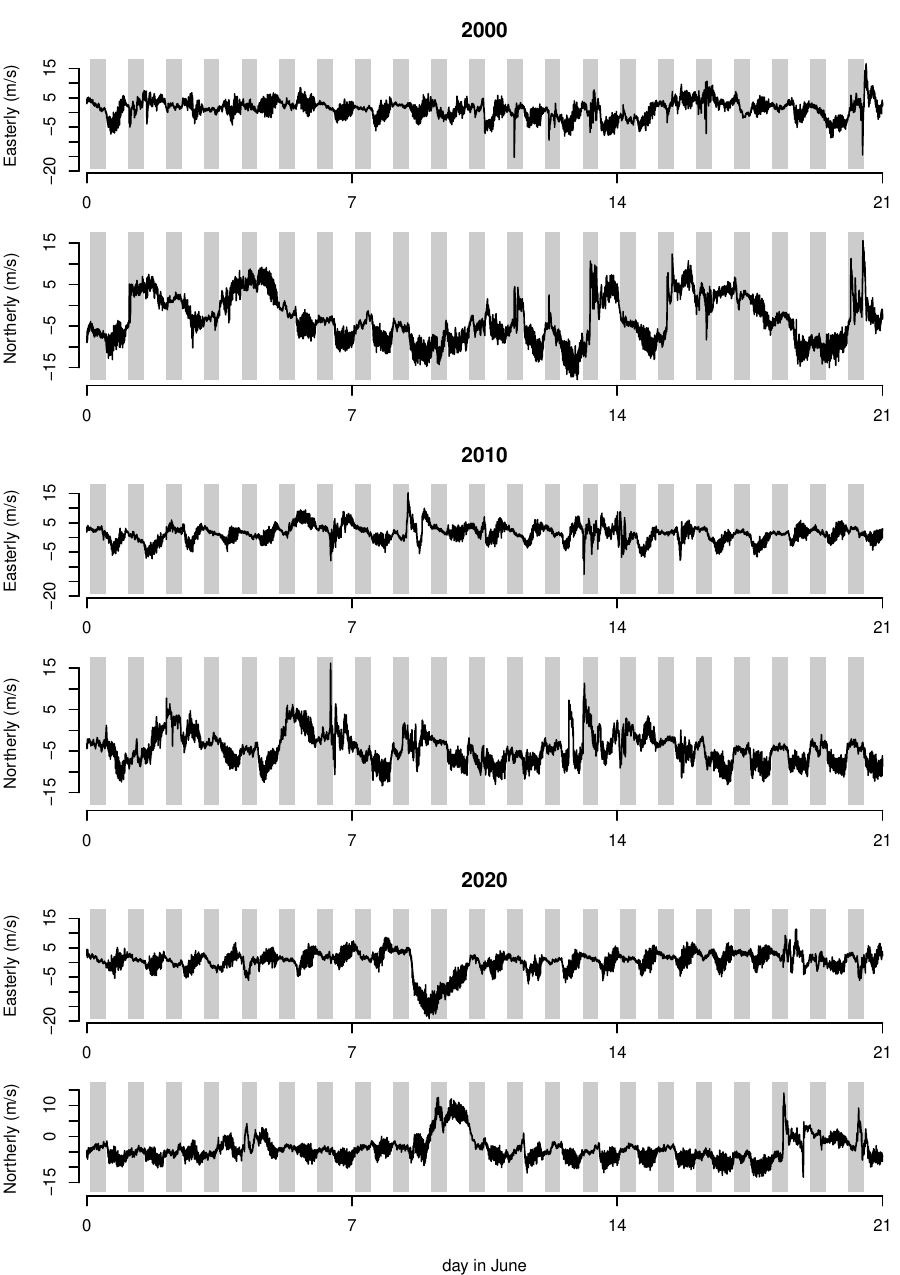}
\caption{Time series of Easterly and Northerly components of wind vector for three years. Gray shaded areas are nighttimes. 
}
\label{fig:three-yearsEN}
\end{figure}

To give some idea as to how the wind vector varies over time, Figure~\ref{fig:three-yearsEN} shows the time series for its northerly and easterly components in the years 2000, 2010, 2020.
The most striking feature of these plots is the much greater high-frequency variation during the daytime than the nighttime for both components of the wind vector.
In contrast to this overall pattern, there are occasionally periods of very high short-term variability in the wind vector during both daytime and nighttime.
Another noteworthy feature of these plots is that the pattern of winds is often similar on consecutive days.
This similarity is most noteworthy during June 11-18 in 2020, where both components of the wind vector show a highly repetitive pattern across these 8 days.
One can also spot approximate repetitions of patterns across days towards the end of 2010 and perhaps June 5-9 in 2000.
Thus, there is evidence of temporal dependence on the scale of at least several days.

% \begin{figure}[!htbp]
% \includegraphics[width=0.75\textwidth]{figures/Exploratory Analysis/three-yearsEN.pdf}
% \caption{Time series of Easterly and Northerly components of wind vector for three years. Gray shaded areas are nighttimes. 
% }
% \label{fig:three-yearsEN}
% \end{figure}

\subsection{Marginal distributions}
\label{sec:marginal}

\begin{figure}[!htbp]
\includegraphics[width=0.9\textwidth]{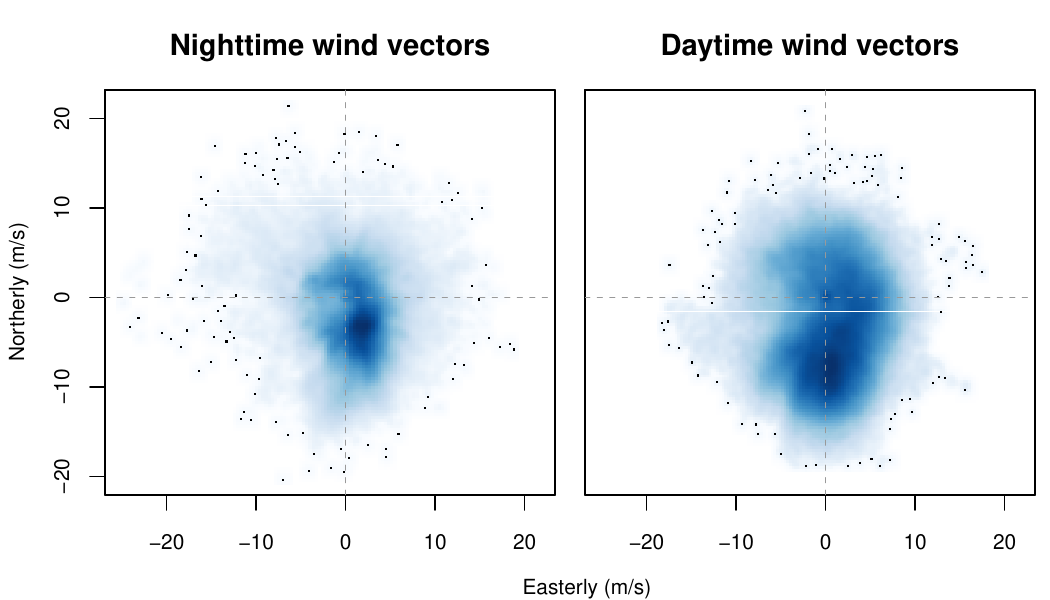}
\caption{Smooth scatter plot of wind vectors during the nighttime (2-12 UTC) and daytime excluding 2 hours after sunrise and 2 hours before sunset (14-24 UTC). Bandwidth parameter in R function \texttt{smoothScatter} set to 0.3 in both plots.
}
\label{fig:wind-scatter}
\end{figure}

This subsection focuses on the marginal distribution of the wind vector and how it varies with time of day.
Using the R command \texttt{smoothScatter}, Figure~\ref{fig:wind-scatter} shows estimated densities for the wind vector during the daytime and nighttime, together with points indicating the wind vector for the 100 observations for which the estimated density is the lowest.
We see that for the bulk of the observations, the variability in the wind vector is substantially greater during the daytime, but the most extreme winds are, if anything, greater in magnitude during the nighttime.
Another noteworthy feature of both of these density plots is that the contours of at least some level sets of the density are clearly not convex.
In particular, both plots show that for wind speeds below, say, 2 m s$^{-1}$, there is a dearth of westerly winds relative to other directions.

% \begin{figure}[!htbp]
% \includegraphics[width=0.9\textwidth]{figures/Exploratory Analysis/wind-scatter2.pdf}
% \caption{Smooth scatter plot of wind vectors during the nighttime (2-12 UTC) and daytime excluding 2 hours after sunrise and 2 hours before sunset (14-24 UTC). Bandwidth parameter in R function \texttt{smoothScatter} set to 0.3 in both plots.
% }
% \label{fig:wind-scatter}
% \end{figure}

\begin{figure}
\begin{center}
    
\includegraphics[width=0.45\textwidth]{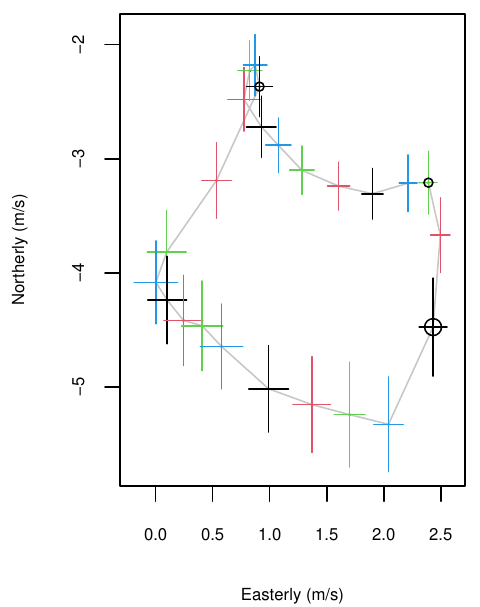}

\end{center}
\caption{Hourly averages of minute-by-minute medians in component-wise median wind vector.  Large black circle corresponds to result for 0:00--0:59 UTC. Results for other hours move in a largely counter clockwise direction. Two small black circles correspond to times of sunset and sunrise. Center of each cross gives the hourly average and each axis of the cross indicates $+/-$ one standard error based on treating results for each year as independent and identically distributed.  Repeating sequence of colors (black, red, green, blue) of crosses are used to enhance visibility. 
}
\label{fig:diurnalUQ}
\end{figure}

Figure \ref{fig:diurnalUQ} shows how the component-wise medians of the wind vector vary with the time of day.
We chose to show medians rather than means because, as shown in the left plot of Figure \ref{fig:wind-scatter}, the marginal distribution of the wind vector is quite heavy-tailed at night.
The component-wise medians are computed separately for every minute of the day and every year and then, for each year, averaged over each hour, yielding 23 values for the northerly and easterly hourly averaged medians.
Write $M_{h,y}$ for this vector of length 2 for hour $h$ (ranging from 0 to 23) and year $y$ (ranging from 1998 to 2020).
If the distribution of $M_{h,y}$ is independent and identically distributed across years with mean vector $\mu_h$, then $\hat\mu_h = \sum_{y=1998}^{2020}M_{h,y}$ is an unbiased estimator of $\mu_h$ and the standard errors of each component of $\hat\mu_h$ can be estimated by treating $M_{h,1998},\ldots,M_{h,2020}$ as independent and identically distributed.
The modest amount of missing data during the training should have only a trivial effect on this assumption of identically distributed averages across years.
The values of $\hat\mu_h$ for each hour are then plotted in Figure \ref{fig:diurnalUQ} as a cross with, for any given hour, the center of the cross at $\hat\mu_h$ and the half-length of each arm of the cross equaling one standard error. 

We see that the diurnal pattern in $\hat\mu_h$ of the component-wise medians makes a fairly substantial (mostly) counterclockwise loop from one day to the next.
These component-wise medians move largely in a southeasterly direction during the daytime and in a northwesterly direction during the nighttime.
During the two hour transition to nighttime, the component-wise medians move rapidly northward and during the two hour transition to daytime, they move rapidly in a southwesterly direction, making a small clockwise loop.
Overall, the typical wind is from a southerly direction during all times of day and nearly always has a positive easterly component except from about 14:00 to 16:00 UTC (mid-morning), where the median easterly component is close to 0 m s$^{-1}$.

Although there are over 680,000 minutes in the training data, the strong dependence in wind vectors over the minute time scale implies that the effective amount of information about marginal distributions in the training data is rather limited.
Indeed, plots (not shown) of the diurnal pattern separately for individual years have patterns that vary greatly across years and, for many years, bear little resemblance to the pattern in Figure \ref{fig:diurnalUQ}.
However, the standard errors in the averages over all 23 years are small enough to support that at least the overall pattern shown in this figure is not a statistical fluke.
Furthermore, because results for nearby hours are generally highly correlated, differences in results for consecutive hours are mostly strongly statistically significant.
For example, the easterly component of $\hat\mu_{20}-\hat\mu_{19}$ equals 0.413 m s$^{-1}$ and its standard error is 0.055 m s$^{-1}$, so this average is over 7 standard errors from 0 even though the standard errors of the easterly components of $\hat\mu_{20}$ and $\hat\mu_{19}$ are both around 0.18 m s$^{-1}$.

% \begin{figure}
% \begin{center}
    
% \includegraphics[width=0.45\textwidth]{figures/Exploratory Analysis/diurnalUQ-2.pdf}

% \end{center}
% \caption{Hourly averages of minute-by-minute medians in component-wise median wind vector.  Large black circle corresponds to result for 0:00--0:59 UTC. Results for other hours move in a largely counter clockwise direction. Two small black circles correspond to times of sunset and sunrise. Center of each cross gives the hourly average and each axis of the cross indicates $+/-$ one standard error based on treating results for each year as independent and identically distributed.  Repeating sequence of colors (black, red, green, blue) of crosses are used to enhance visibility. 
% }
% \label{fig:diurnalUQ}
% \end{figure}

\begin{figure}
\includegraphics[width=0.9\textwidth]{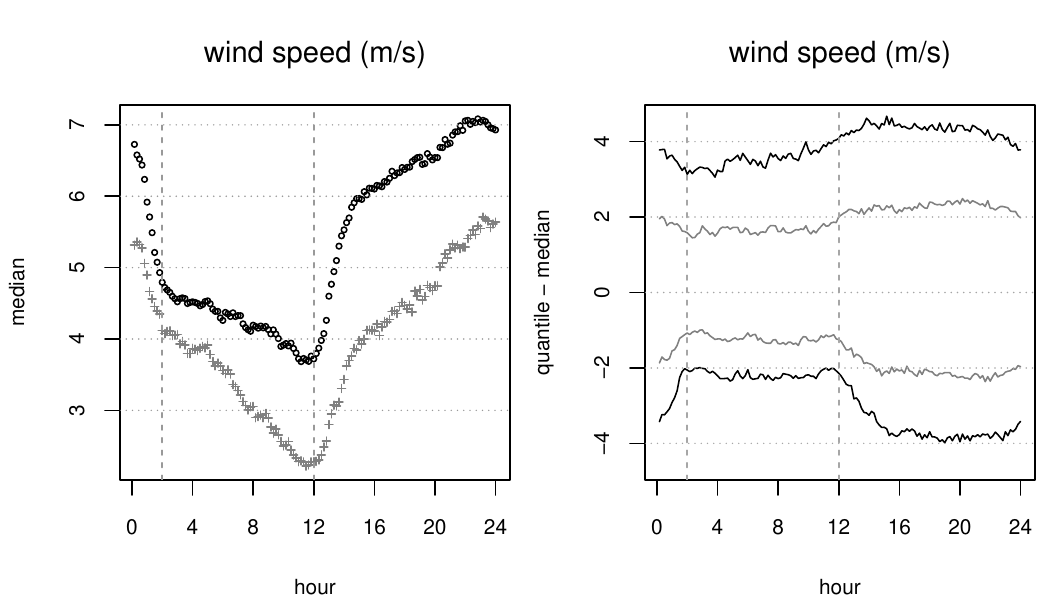}
\caption{Diurnal cycle in wind speed.  Left plot shows ten-minute averages of minute-by-minute medians of wind speeds (black circles) and magnitudes of 10-minute averages of component-wise median wind vectors (gray plus signs).  Right plot shows 10-minute averages of minute-by-minute differences between 0.1, 0.25, 0.75 and 0.9 quantiles and medians of wind speed.
Dashed vertical lines in these and subsequent plots indicate the beginning and end of the ``nighttime'' period.
}
\label{fig:ws-marginal}
\end{figure}

Figure \ref{fig:ws-marginal} shows properties of the diurnal cycle of wind speed.
The left panel of this figure gives the ten-minute averages of the minute-by-minute medians of wind speed, so essentially the black curve in the left panel of Figure \ref{fig:seasonality diurnal plot}.
%showing a steady decrease in speed during the nighttime and a steady increase during most of the daytime except for a sharp increase during the transition to daytime and a sharp decrease during the transition to nighttime.
This plot also shows the wind speeds corresponding to the averaged component-wise medians of the wind vectors (similar to what is shown in Figure \ref{fig:diurnalUQ} except averaging over 10-minute periods rather than hours).
The pattern is qualitatively similar to that of the median wind speeds, showing that the diurnal variation in the median wind speed is largely a function of the diurnal variation in the typical wind vector and not primarily due to differing variability in the wind vector depending on time of day.
The right panel shows deviations of the 0.1, 0.25, 0.75 and 0.9 quantiles from the medians as a function of time of day.
Overall, as we would expect from Figure \ref{fig:wind-scatter}, there is more variation in wind speed during the daytime.
During the nighttime, the distribution of wind speed is noticeably skewed to the right, with this skewness getting stronger during the night due to the difference between the upper quantiles and the median growing.
In the early morning hours, the lower quantiles decrease more rapidly than the upper quantiles increase, so that by 16:00 UTC, the distribution of wind speed is only slightly asymmetric.

% \begin{figure}
% \includegraphics[width=0.9\textwidth]{figures/Exploratory Analysis/ws-marginal.pdf}
% \caption{Diurnal cycle in wind speed.  Left plot shows ten-minute averages of minute by minute medians of wind speeds (black circles) and magnitudes of 10-minute averages of component-wise median wind vectors (gray plus signs).  Right plot shows 10-minute averages of minute-by-minute differences between 0.1, 0.25, 0.75 and 0.9 quantiles and medians of wind speed.
% Dashed vertical lines in these and subsequent plots indicate the beginning and end of the ``nighttime'' period.
% }
% \label{fig:ws-marginal}
% \end{figure}

\begin{figure}
\includegraphics[width=0.8\textwidth]{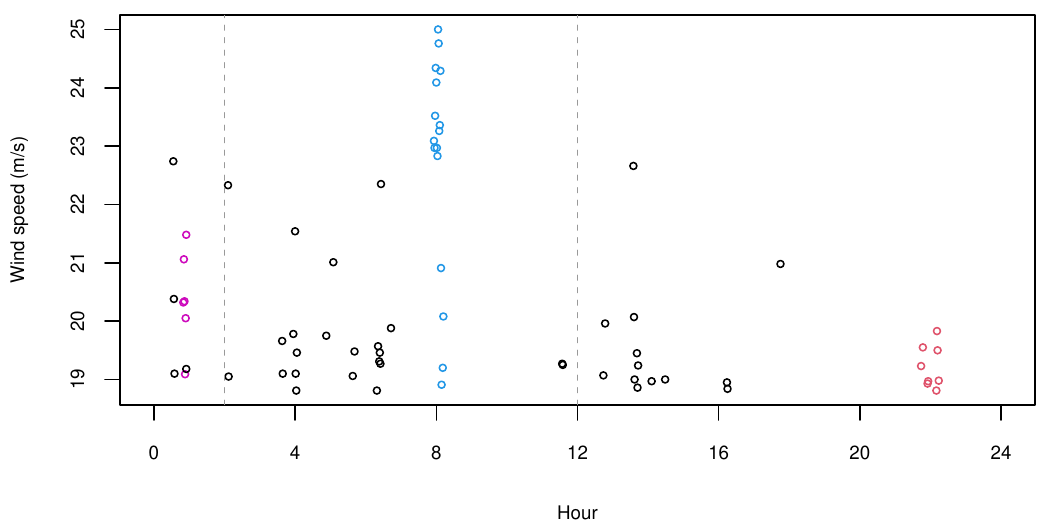}
\caption{Wind speeds greater than the 0.9999 quantile of 18.8 m s$^{-1}$.  
Red points are from a 30-minute period in 2008, magenta points are 6 consecutive minutes about 3 hours after the times of the red points and blue points are 16 points from a 17-minute period in 2011.}
\label{fig:ws-9999}
\end{figure}

Extremes of wind speed are of obvious interest for their impact on human infrastructure.
Figure \ref{fig:ws-9999} shows the wind speed exceedances above the 0.9999 quantile of all wind speeds in the training data.
A substantial fraction of these exceedances come from a small number of weather events; for example, the blue points are from a 17-minute event in 2011 that includes the highest 12 wind speeds during the training period.
The maximum wind speed during this event is 25 m s$^{-1}$, which, while strong, is towards the lower end of what are considered to be damaging winds.
Unfortunately, this event occurred on a day with too many missing observations to include in the training data, which likely had a substantial impact on the most extreme wind speeds our stochastic generators produce.
After eliminating the obviously erroneous wind speeds in the data, the largest recorded wind speed over the entire dataset is 31.29 m s$^{-1}$, which occurred on August 24, 2016.
It is apparent that a larger dataset would be essential for studying extreme winds.
Looking at the entire year would help, as would using data from long-operating weather stations.
However, since the strongest winds in Oklahoma are generally from tornadoes and the chances of a tornado hitting any specific weather station even over many decades are slight, fixed weather stations are probably not the best source of information for the most extreme winds in Oklahoma.
Indeed, 
from 1950-2024, there were 71 tornadoes in the 2600 $\mathrm{km}^2$ of Grant County, OK, which includes Lamont, and only 2 of those have occurred in the month of June, although 37 were in May \citep{Oklahomatornadoes}.
%Thus, we will not put a particular focus on extreme winds in this study.

% \begin{figure}
% \includegraphics[width=0.8\textwidth]{ws-9999.pdf}
% \caption{Wind speeds greater than the 0.9999 quantile of 18.8 m s$^{-1}$.  
% Red points are from a 30-minute period in 2008, magenta points are 6 consecutive minutes about 3 hours after the times of the red points and blue points are 16 points from a 17-minute period in 2011.}
% \label{fig:ws-9999}
% \end{figure}

%Figure \ref{fig:wind-direction-dn} shows the distribution of wind directions for day and nighttimes for various ranges of wind speed. Wind directions for wind speeds less than 1 m{\scslash}s are not given because wind direction at such slow speeds is poorly determined (CITATION) and is not of much practical relevance. The main change in wind directions as wind speeds increase is that they become more concentrated around the dominant southerly wind flow.At lower wind speeds, wind directions are more spread out during the daytime, but for winds greater than 6 m{\scslash}s, the wind direction distributions are very similar for daytime and nighttime.

%\begin{figure}
%\includegraphics[width=0.8\textwidth]{wind-direction-dn.pdf}
%\caption{Wind directions as a function of wind speed and time of day.  Solid gray bars for nighttime and black outlined bars for daytime.
%}
%\label{fig:wind-direction-dn}
%\end{figure}

\subsection{First differences in the wind vector}
\label{sec:differences}

Because of the strong correlation of wind vectors in consecutive minutes, a simple way to investigate the temporal structure of these data is to look at the first differences in the wind vectors.
All of the results shown in the rest of this section consider only first differences when the consecutive observations are one minute apart.

\begin{figure}
\begin{center}
\includegraphics[width=0.6\textwidth]{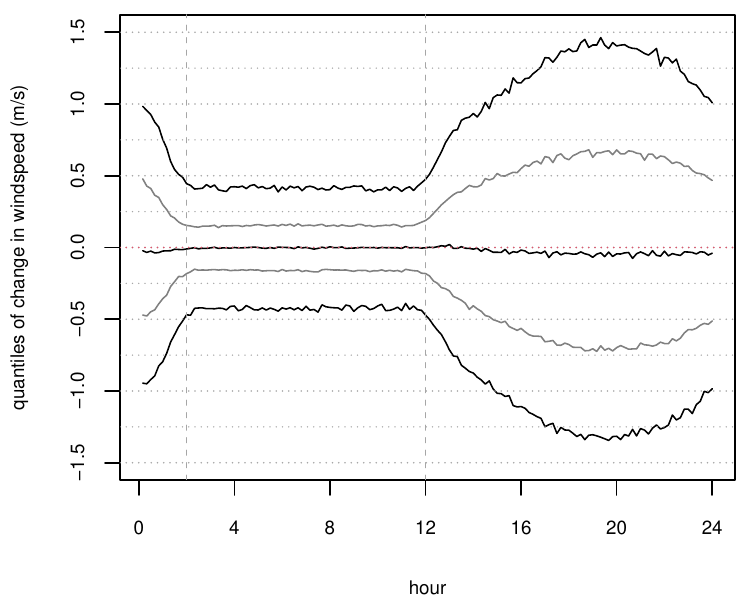}
\end{center}
\caption{Diurnal cycle in one-minute changes in wind speed.  Five curves are for 10-minute averages of minute-by-minute 0.1, 0.25, 0.5, 0.75 and 0.9 quantiles.
}
\label{fig:ws-change}
\end{figure}

Figure \ref{fig:ws-change} shows five quantiles of these first differences in wind speed for each minute of the day that are then averaged over 10-minute time segments.
Not surprisingly, the medians are very close to 0 m s$^{-1}$, although the medians are consistently slightly negative during the daytime, indicating that daytime wind speeds are a bit more likely to decrease from one minute to the next than increase.
All five of these quantiles hardly vary during the nighttime, especially if one restricts to 2:30-11:30 UTC.
These nighttime distributions of the change in wind speed are also very nearly symmetric around 0.
Given that the marginal distribution of wind speed varies substantially during nighttime in its median (left panel of Figure \ref{fig:ws-marginal}) and, to a lesser extent, its spread and shape (right panel of Figure \ref{fig:ws-marginal}), we find these features of the distribution of the changes in wind speed at night rather remarkable.
During the daytime, there is a smooth increase in variability until about 19:30 UTC and then a smooth decrease in variability.
In the middle of the daytime, the wind speed distributions are modestly but consistently skewed to the right.
For example, writing $q_{ds}(\alpha,t)$ for the average of the minute-by-minute $\alpha$ quantile of the change in wind speed for the 10 minutes starting at minute $t$, the measure of skewness $q_{ds}(0.9,t) - 2q_{ds}(0.5,t) + q_{ds}(0.1,t)$ ranges from 0.089 to 0.272 m s$^{-1}$ with a mean of 0.161 m s$^{-1}$ for the 10-minute periods from 14:40 through 22:40 UTC.
In contrast, this statistic ranges from -0.062 to 0.031 m s$^{-1}$ during nighttime with a mean of -0.005 m s$^{-1}$.

% \begin{figure}
% \begin{center}
% \includegraphics[width=0.6\textwidth]{figures/Exploratory Analysis/ws-change.pdf}
% \end{center}
% \caption{Diurnal cycle in one-minute changes in wind speed.  Five curves are for 10-minute averages of minute-by-minute 0.1, 0.25, 0.5, 0.75 and 0.9 quantiles.
% }
% \label{fig:ws-change}
% \end{figure}

By considering changes in the wind vector and not just wind speed, we can see further interesting patterns in the data.
In particular, it is helpful to consider the direction of the change in the wind vector relative to the current wind direction.
Specifically, write $X_t$ for the wind vector at minute $t$ and $H_t$ to be the angle between the vectors $X_{t+1}-X_t$ and $X_t$, measured in degrees and taken to be in the interval $(-180,180]$, where, following the standard convention for measuring wind direction, a positive angle corresponds to the wind vector rotating clockwise and a negative angle counterclockwise.
We can then decompose the minute by minute changes in the wind vector into components parallel and orthogonal to the current wind vector.

\begin{figure}[!htbp]
\centering
\includegraphics[width=0.65\textwidth]{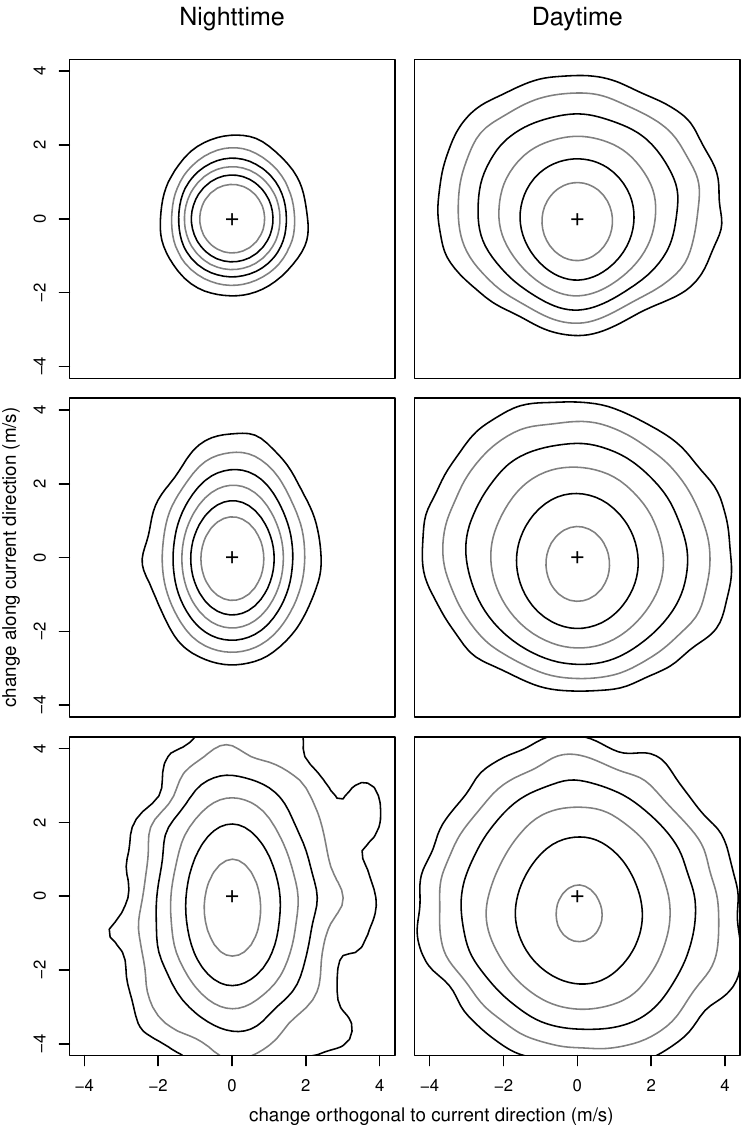}
\caption{Densities for changes in wind vector relative to current wind direction. Top row for current wind speed 1-5 m s$^{-1}$, middle row 5-10 m s$^{-1}$, bottom row greater than 10 m s$^{-1}$. Contour levels are for $\log_{10}$ of the densities at values from -1 (innermost contours) to -3.5 (outermost) with increments of 0.5.  Changes in the wind vector outside the ranges shown in these plots occur in about 1 in 1000 minutes. $+$ signs indicate the origin.
}
\label{fig:intro-vol}
\end{figure}

Figure \ref{fig:intro-vol} shows density plots for these two components of the change in the wind vector for each of three ranges of wind speed for daytime and nighttime.
%Figures \ref{fig:H1-5}-\ref{fig:H10} show estimated densities for these components of $w_{t+1}-w_t$ for three ranges of the current wind speed for the two times of day. In all three figures, the scales in the left-hand plots are all the same and cover the entire range of outcomes, whereas all of the right-hand plots focus on a narrower range of outcomes in which both components of $w_{t+1}-w_t$ are, roughly, at most 3.2 m{\scslash}s in magnitude. In these six right-hand side plots, the fraction of the observations covered by this range of values is at least 0.979.
There are several interesting differences between the results for daytime and nighttime.
First, during the nighttime, the densities change much more across the three wind speed ranges.
In particular, the upper left-hand plot in Figure~\ref{fig:intro-vol} shows much less variability in both directions for wind speeds from 1 to 5 m s$^{-1}$ than the corresponding plot in the right column for the daytime, whereas, for wind speeds greater than 10 m s$^{-1}$, the bottom row of plots in Figure \ref{fig:intro-vol} shows the variability in the parallel direction is very similar during the two times of day.
%As we might expect given the histograms for $H_t$ at night in Figure \ref{fig:dhead31},
During nighttime, for the two higher ranges of wind speed (left column, second and third rows of figure), the variation in changes in the orthogonal direction are considerably smaller than those in the parallel direction.
In contrast, these distributions are much closer to isotropic during the daytime, although for the lower wind speeds in the upper right plot in Figure~\ref{fig:intro-vol}, the contours get noticeably farther apart as one moves up the vertical axis.
Sect.~\ref{sec:Stochastic-volatility} further explores this relationship between volatility in winds as a function of recent behavior of the wind vector.
%For the lower wind speeds in Figure \ref{fig:H1-5}, the modes of the density are very close to the origin, whereas for the other two ranges of wind speed the modes of the densities are somewhat below the origin very nearly on the vertical axis.
%The tendency for higher wind speeds to diminish should be expected since there needs to be some overall tendency for strong winds to weaken to keep wind speeds within a reasonable range.

% \begin{figure}[!htbp]
% \centering
% \includegraphics[width=0.65\textwidth]{figures/Exploratory Analysis/intro-vol.pdf}
% \caption{Densities for changes in wind vector relative to current wind direction. Top row for current wind speed 1-5 m s$^{-1}$, middle row 5-10 m s$^{-1}$, bottom row greater than 10 m s$^{-1}$. Contour levels are for $\log_{10}$ of the densities at values from -1 (innermost contours) to -3.5 (outermost) with increments of 0.5.  Changes in the wind vector outside the ranges shown in these plots occur in about 1 in 1000 minutes. $+$ signs indicate the origin.
% }
% \label{fig:intro-vol}
% \end{figure}

\FloatBarrier
%\begin{figure}
%\includegraphics[width=0.93\textwidth]{H1-5.pdf}
%\caption{Density plots for changes in wind vector (m{\scslash}s) relative to current wind direction for present wind speed between 1 and 5 m{\scslash}s.  Top plots for nighttime and bottom plots for daytime.  Left plots show full range of changes and right plot gives contours for densities over a limited range.  Contour levels are for $\log_{10}$ of the densities.
%}
%\label{fig:H1-5}
%\end{figure}

%\begin{figure}
%\includegraphics[width=0.93\textwidth]{H5-10.pdf}
%\caption{Same as Figure \ref{fig:H1-5} for present wind %speed between 5 and 10 m{\scslash}s
%}
%\label{fig:H5-10}
%\end{figure}

%\begin{figure}
%\includegraphics[width=0.93\textwidth]{H10.pdf}
%\caption{Same as Figure \ref{fig:H1-5} for present wind %speed greater than 10 m{\scslash}s.
%}
%\label{fig:H10}
%\end{figure}

\section{Methodology}
\label{sec:Methodology}

%Machine learning approaches for time series generation include adversarial training (e.g. generative adversarial networks, GANs), variational inference (e.g. VAEs) and, more recently, discrete tokenization and flow-matching foundation models. TimeGAN \citep{yoon2019time} introduced adversarial training to time series. Unlike standard GANs, TimeGAN learns an embedding space using an autoencoder and trains the adversary within that latent space. This allows it to enforce step-wise temporal dynamics more effectively than a standard GAN. TimeVAE \citep{desai2021timevae} introduced a decomposable decoder that explicitly splits the generation into trend (polynomial) and seasonality (sinusoidal) blocks, demonstrating that VAEs can match GAN performance with significantly faster and more stable training. More recently, \citep{liu2025sundial} introduced Sundial, a generative time series foundation model that combines a transformer backbone \citep{vaswani2017attention} with a flow-matching objective \citep{lipman2023flow} to sample the next set of time points, conditional on the past. Sundial is trained on TimeBench \citep{liu2025sundial} which comprises over one trillion time points from multiple sources. 

The primary challenge in synthetic time series generation lies in capturing complex temporal dependencies, a problem that has driven continuous architectural improvements in deep learning. Early approaches primarily adapted adversarial training to sequence data; for example, TimeGAN \citep{yoon2019time} embedded a standard GAN within an autoencoder's latent space to better enforce step-wise temporal dynamics. Subsequent methods shifted toward variational inference to improve training stability. Notably, TimeVAE \citep{desai2021timevae} demonstrated that variational autoencoders with decomposable decoders that explicitly separate trend and seasonality can match GAN performance while remaining significantly easier to train. Following the success of discrete tokenization in natural language processing, models like the Time Vector-Quantized Variational Autoencoder \citep[Time VQ-VAE;][]{lee2023vector} were introduced to encode continuous time series into discrete tokens and demonstrated strong performance on benchmark datasets. Most recently, the field has seen the emergence of massive foundation models, such as Sundial \citep{liu2025sundial}, which leverages a transformer architecture \citep{vaswani2017attention} and flow-matching \citep{lipman2023flow} to sample time points conditional on past data.

%In this paper, we focus on task-specific deep generative models based on the Time Vector-Quantized Variational Autoencoder \citep[Time VQ-VAE;][]{lee2023vector}, which demonstrated good performance on small datasets from the UCR archive \citep{UCRArchive2018}. Time VQ-VAE has three stages. In stage one, the continuous data is tokenized into discrete values. Specifically, a VQ-VAE \citep{van2017neural} is trained to encode the time series as discrete tokens (Sect.~\ref{subsec:tokenization}). In stage two, a generative model is trained on the discrete tokens. Specifically, a bidirectional transformer \citep{chang2022maskgit} is trained to predict tokens (Sect.~\ref{subsec:generator-training}). In stage 3, an iterative decoding strategy is used to generate new token sequences \citep{chang2022maskgit}. These generated token sequences are then decoded into a synthetic time series using the VQ-VAE decoder (Sect.~\ref{subsec:iterative-decoding}).

While large-scale foundation models like Sundial are trained on generalized, multi-source datasets, capturing the highly localized, minute-by-minute stochastic volatility of wind vectors requires a more specialized architecture. Therefore, in this paper, we focus on task-specific deep generative models built upon the Time VQ-VAE framework \citep[Time VQ-VAE;][]{lee2023vector}. This framework consists of three stages. In stage one, the continuous data is tokenized into discrete values using a VQ-VAE \citep{van2017neural} trained to encode the time series (Sect. ~\ref{subsec:tokenization}). In stage two, a generative model built on a bidirectional transformer \citep{chang2022maskgit} is trained to predict these discrete tokens (Sect.~\ref{subsec:generator-training}). Finally, in stage three, an iterative decoding strategy \citep{chang2022maskgit} generates new token sequences, which are then converted back into a synthetic continuous time series using the VQ-VAE decoder (Sect.~\ref{subsec:iterative-decoding}).

We consider two scenarios: (1) independent one-day time series generation; (2)  consecutive multi-day time series generation that is conditioned on the previous day. For both scenarios, we further consider whether weather-state information is incorporated into the model. When weather states are included, we study two conditioning strategies: (i) by including weather states in the first tokenization stage as an additional feature; and (ii) by including weather states in the second transformer learning stage (Sect.~\ref{Subsection:conditionalon-Weather-States}). For scenario (1), without weather states information, we train the original unconditional Time VQ-VAE model, omitting the pre-processing Fourier transform step. For the remaining cases, we propose new strategies to condition on the weather states and,  in scenario (2), previous-day wind vectors. This is distinct from \citet{lee2023vector,chang2022maskgit} who only consider conditioning on a class label.

%\textcolor{red}{As we discussed, a relatively short paragraph mentioning other approaches that were tried but found wanting would be worthwhile.}

 We initially explored a TimeVAE framework \citep{desai2021timevae}, a variational autoencoder designed for multivariate time-series generation that combines convolutional encoder-decoder architectures with a structured latent space to capture temporal dependencies. The model follows the standard VAE formulation, learning an approximate posterior $q_{\phi}(z \mid x)$ and a decoder $p_{\theta}(x \mid z)$ under an ELBO objective, while allowing incorporation of domain-specific temporal components such as trend and seasonality through structured decoder modules. However, despite its ability to capture temporal structure in reconstruction, we found that TimeVAE exhibits fundamental limitations in the generative setting: samples drawn from the learned posterior $q_{\phi}(z \mid x)$ tend to exhibit memorization behavior, while samples drawn from the prior $p(z)$ often fail to produce realistic wind dynamics unless the KL regularization is heavily constrained, revealing a trade-off between latent space regularization and generative fidelity. To mitigate the memorization issue, we further explored augmenting the latent sampling procedure by fitting a Gaussian mixture model (GMM) to the learned latent representations and sampling from this mixture distribution. This approach is motivated by the observation that the aggregated posterior often deviates from the imposed Gaussian prior, and more flexible density estimators such as GMMs can better approximate the true latent distribution \citep{10715230}. However, despite improving the flexibility of the sampling distribution, this strategy did not resolve the core issue in our setting: the generated samples continued to exhibit memorization behavior and lacked sufficient diversity in temporal dynamics.

We also investigated frequency-domain modeling strategies inspired by the original Time-VQ-VAE framework \citep{lee2023vector}, where the time series is transformed via the short-time Fourier transform (STFT) and decomposed into low-frequency (LF) and high-frequency (HF) components that are modeled separately. In this framework, LF components are modeled with higher compression and larger receptive fields to capture global structure, while HF components are generated subsequently as refinements conditioned on LF. While this design improves reconstruction by allowing different treatment of global structure and fine details, it implicitly assumes that LF components are more predictable and carry the primary signal, whereas HF components are less structured and can be treated as residual details. However, in the context of minute-by-minute wind vector data, we found this assumption to be problematic: the HF components do not merely correspond to noise, but capture essential short-term variability, stochastic fluctuations, and rapid changes in wind dynamics. As a result, treating HF components as secondary or suppressing them produces overly smooth synthetic sequences that fail to capture realistic high-frequency variations. This suggests that preserving the full spectral structure is essential in this setting.

These limitations motivated our adoption of a VQ-VAE-based framework which leverages discrete latent tokenization to model time series without imposing restrictive distributional assumptions or explicit frequency decomposition. By learning a prior directly over discrete latent tokens, our approach enables stable generation while preserving both global temporal structure and high-frequency variability essential for realistic wind dynamics.

\subsection{Tokenizing Wind Vectors}
\label{subsec:tokenization}

In the first step, following \citet{lee2023vector}, we employ a VQ-VAE to learn discrete latent representations from input data.  The VQ-VAE consists of an encoder, a quantization module based on a learned codebook, and a decoder. The VQ-VAE  learns to encode the high dimensional inputs as compressed discrete codes and then decode these discrete codes back to the original time series.

\begin{figure}[!htbp]
    \centering
    \includegraphics[width=\linewidth]{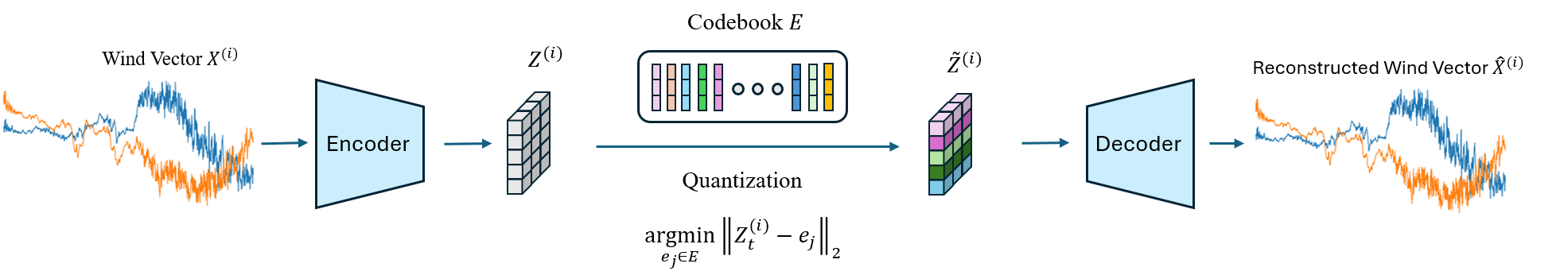}
    \caption{Stage 1: Time VQ-VAE. Learn a compact, discrete representation of high-resolution wind speed vectors using VQ-VAE. The encoder maps $\bm{X}^{(i)} \in \mathbb{R}^{1440 \times 2}$ to a continuous latent embedding $\bm{Z}^{(i)}\in \mathbb{R}^{1440 \times 64}$, where $T=1440$ represents $60 \times 24=1440$ minutes for day $i$ and $D=64$ matches the length of each codebook vector $\bm{e}_k$, where $\bm{E} = \{\bm{e}_k\}_{k=1}^{512}$. }
    \label{fig:stage1}
\end{figure}

Figure~\ref{fig:stage1} presents an overview of our wind vectors tokenization method. The observed data on day $i$ is $\bm{X}^{(i)} \in \mathbb{R}^{T \times P}$, where $T$ is the time sequence length ($T=60\times 24=1440$ for minute-by-minute data) and $P=2$ are the two components of the wind vector.  For models including weather state, $P=2$ or 3 depending on the specific way they are included in the model (see Sect.~\ref{Subsection:conditionalon-Weather-States}). 

The encoder first maps $\bm{X}^{(i)}$ to a continuous latent embedding $\bm{Z}^{(i)} = [(\bm{z}_1^{(i)})^\top \cdots (\bm{z}_T^{(i)})^\top] \in \mathbb{R}^{T \times D}$, where $D=64$ is the latent dimension. The encoder architecture consists of convolutional layers. In the encoder, there is no downsampling for time steps; this helps maintain high frequency characteristics \citep{rombach2022high}.

The continuous latent space is transformed into a discrete latent space via vector quantization. In this process, the continuous latent embedding at minute $t$ on day $i$, denoted $\bm{z}_t^{(i)} \in \mathbb{R}^D$, is assigned to the closest vector $\bm{e}_k\in\mathbb{R}^D$ from a learned codebook $\bm{E} = \{\bm{e}_k\}_{k=1}^K$, where $K=512$ is the number of codebook vectors, each of length of $D=64$. More formally, the quantization step maps $\bm{z}_t^{(i)}$ to a quantized vector $\widetilde{\bm{z}}_t^{(i)}$ as:
\begin{equation}
    \widetilde{\bm{z}}_t^{(i)} = \bm{e}_k
    \text{ where }k = \operatorname*{argmin}_j ||\bm{z}_t^{(i)} - \bm{e}_j||_2.
\end{equation}
The codebook loss is:
\begin{equation}
    L_{\text{codebook}}(\bm{z}_t^{(i)}) = \lVert sg[\bm{z}_t^{(i)}] - \widetilde{\bm{z}}_t^{(i)} \rVert^2_2 + \beta\lVert \bm{z}_t^{(i)} - sg[\widetilde{\bm{z}}_t^{(i)}]\rVert^2_2
\end{equation}
where $sg[\cdot]$ denotes the stop-gradient operator and $\beta$ controls the strength of the commitment term to encourage encoder outputs to stay close to the codebook embeddings. Here, we set $\beta=1$. 

Then the quantized latent vectors $[\widetilde{\bm{z}}_1^{(i)} \cdots \widetilde{\bm{z}}_T^{(i)}]^\top$ are passed through the decoder, producing $\widehat{\bm{X}}^{(i)}$, where the decoder mirrors the encoder's structure. Overall, the  optimization objective is:

\begin{align}
    L(\psi, \bm{E}; \mathcal{D}_{\mathrm{train}}) &= \sum_{i=1}^{n_{\mathrm{train}}} \left[ L_{\text{recon}}(\bm{X}^{(i)}) + \sum_{t=1}^T L_{\text{codebook}}(\bm{z}_t^{(i)}) \right] \\
      &= \sum_{i=1}^{n_\mathrm{train}} \Bigg[ \lVert \bm{X}^{(i)} - \widehat{\bm{X}}^{(i)} \rVert ^2_2 + \sum_{t=1}^T \lVert sg[\bm{z}_t^{(i)}] - \widetilde{\bm{z}}_t^{(i)} \rVert^2_2   + \beta\lVert \bm{z}_t^{(i)} - sg[\widetilde{\bm{z}}_t^{(i)}]\rVert^2_2 \Bigg], 
\end{align}
where the vector $\psi$ contains the parameters of the encoder and decoder networks and $\mathcal{D}_{\mathrm{train}} = \{\bm{X}^{(i)}\}_{i=1}^{n_\mathrm{train}}$ is the training data.  We use mini-batch stochastic gradient descent for optimization. In this case, the codebook vectors $\bm{E}$ are updated using exponential moving averages to stabilize training and avoid codebook collapse.

% \begin{figure}[!htbp]
%     \centering
%     \includegraphics[width=\linewidth]{figures/Methodology/stage1-.png}
%     \caption{Stage 1: Time VQ-VAE. Learn a compact, discrete representation of high-resolution wind speed vectors using VQ-VAE. The encoder maps $\bm{X}^{(i)} \in \mathbb{R}^{1440 \times 2}$ to a continuous latent embedding $\bm{Z}^{(i)}\in \mathbb{R}^{1440 \times 64}$, where $T=1440$ represents $60 \times 24=1440$ minutes for day $i$ and $D=64$ matches the length of each codebook vector $\bm{e}_k$, where $\bm{E} = \{\bm{e}_k\}_{k=1}^{512}$. }
%     \label{fig:stage1}
% \end{figure}

\subsection{Training the Generator}
\label{subsec:generator-training}

\begin{figure} [!htbp]
    \centering
    \includegraphics[width=\linewidth]{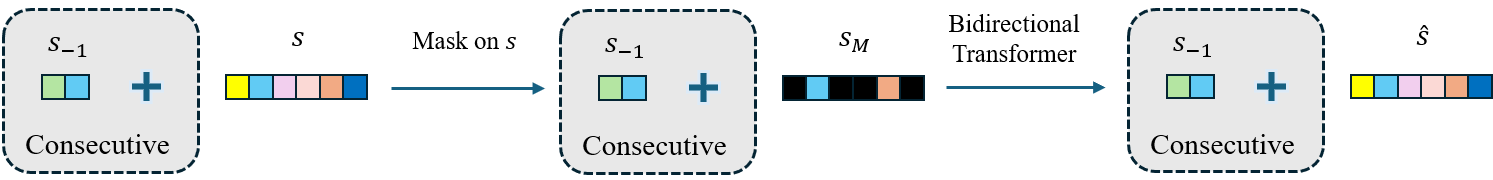}
    \caption{MaskGIT training. For day $i$, $\bm{s}_{-1}$ denotes $\bm{s}^{(i-1)}_{1:C}$, $\bm{s}$ denotes $\bm{s}^{(i)}_{1:T}$, $\bm{s}_M$ denotes $\bm{s}^{(i)}_{\bm{m}(1:T)}$.}
    \label{fig:maskgit}
\end{figure}

After training the VQ-VAE, we have discrete representations of the wind sequences for each day.
To capture the temporal dependencies and underlying patterns within the wind sequences, in the second step, we train a model that learns the distribution of the tokenized sequences. Following \citet{lee2023vector}, we use MaskGIT, a bidirectional transformer trained to predict randomly masked tokens \citep{chang2022maskgit}; Figure~\ref{fig:maskgit} shows the structure of this generator training process. 
Since the tokens can be represented as indices of the codebook, we denote the tokens as $\bm{s} \in \{1,\dots, K\}$, where $\bm{s}_t^{(i)} = k \iff \widetilde{\bm{z}}_{t}^{(i)} = \bm{e}_k$.
We use $\bm{s}_t^{(i)}$ in the following to represent the token at time $t$ for day $i$.

\textbf{Bidirectional transformers.} Bidirectional transformers process entire sequences simultaneously, in contrast to traditional sequence models, which process data autoregressively. This simultaneous processing is accomplished using
 self-attention heads, which allow each element in the sequence to ``attend'' to all other elements in the sequence, not just previous elements \citep{vaswani2017attention}. Bidirectional transformers can better capture global consistency and accelerate sampling \citep{chang2022maskgit}.

\textbf{MaskGIT.} For the wind vector data, we train the bidirectional transformer to generate an entire day at a time, in contrast to autoregressive token-by-token generation.  To train the bidirectional transformer, we follow MaskGIT \citep{chang2022maskgit}. The key idea of MaskGIT is to randomly mask tokens in a sequence and train a bidirectional transformer to predict these masked tokens. 

% \begin{figure} [!htbp]
%     \centering
%     \includegraphics[width=\linewidth]{figures/Methodology/maskgit.png}
%     \caption{MaskGIT training. For day $i$, $s_{-1}$ denotes $\bm{s}^{(i-1)}_{1:C}$, $s$ denotes $\bm{s}^{(i)}_{1:T}$, $s_M$ denotes $\bm{s}^{(i)}_{\bm{m}(1:T)}$.}
%     \label{fig:maskgit}
% \end{figure}

Denote the bidirectional transformer as $f_{\theta}: \{1,\dots,K\}^{C+T} \to \mathbb{R}^{T\times K}$. That is, $f_{\theta}$ takes in a $C+T$ dimensional sequence, and outputs a vector of unnormalized probabilities over $K$ for each time point $t=1,\dots, T$. Here, $C$ is the dimension of a conditioning vector; in the independent one-day generative model, $C=0$, similarly to \citet{chang2022maskgit,lee2023vector}. For the consecutive multi-day generative model, we take $C$ tokens from the wind vector sequence of the previous day (typically, the final 60 minutes). 
We use $\bm{s}_{1:C}^{(i-1)} \in \{1,\dots, K\}^C$ to denote this sequence (with slight abuse of notation). The bidirectional transformer has parameters $\theta\in\Theta$, which comprise the neural network weights (from self-attention heads and feedforward layers) and positional embeddings.   

\textbf{Training.} The goal is to train $f_{\theta}$ to take in $\{\bm{s}_{1:C}^{(i-1)}, [\mathrm{MASK}], \dots, [\mathrm{MASK}]\}$, and output unnormalized probabilities for each of the $T$ masked positions, corresponding to $\bm{s}_{1:T}^{(i)}$. Note that for the independent generative model, we take $\bm{s}_{1:C}^{(i-1)}=\emptyset$.

During training, we mask a random subset of $\bm{s}_{1:T}^{(i)}$, similarly to MaskGIT. Specifically, a masking vector is randomly generated as
$
\bm{m}_{1:T}^{(i)}
=
(m_t^{(i)})_{t=1}^T
\in
\{0,1\}^T $.
The masked sequence is denoted as $\bm{s}_{\bm{m}(1:T)}^{(i)}$ where for each $t$, we have:
\begin{align}
\bm{s}_{\bm{m}(t)}^{(i)} =
\begin{cases}
    \bm{s}_t^{(i)} &\text{if } m_t^{(i)} = 1 \\
    [\mathrm{MASK}] &\text{if } m_t^{(i)} = 0.
\end{cases} 
\end{align}
That is, for the wind tokens on the $i$th day, we model the token-wise conditional distribution as:
\begin{align}
\label{eq:maskgit}
p(\bm{s}_t^{(i)}|\bm{s}_{1:C}^{(i-1)}, \bm{s}_{\bm{m}(1:T)}^{(i)}) = \mathrm{Categorical}(\mathrm{Softmax}(f_{\theta}([\bm{s}_{1:C}^{(i-1)}, \bm{s}_{\bm{m}(1:T)}^{(i)}])_{t})).
\end{align}
The Softmax operation is applied row-wise. Specifically, given the $f_{\theta}$ output matrix $\bm{A} \in \mathbb{R}^{T\times K}$, we have:
\begin{align}
    (\text{Softmax}(\bm{A}))_t = \left(\frac{\exp(A_{t1})}{\sum_{k=1}^K \exp(A_{tk})},\dots, \frac{\exp(A_{tK})}{\sum_{k=1}^K \exp(A_{tk})} \right).
\end{align}

We train the model by minimizing the masked token-wise negative log-likelihood:
\begin{align}
\mathcal{L}(\theta) = -\sum_{i=1}^{n_{\mathrm{train}}} \sum_{t: m_t^{(i)}=1} \sum_{k=1}^K 1(\bm{s}_t^{(i)}=k) \log p(\bm{s}_t^{(i)}=k|\bm{s}_{1:C}^{(i-1)}, \bm{s}_{\bm{m}(1:T)}^{(i)}).
\end{align}
Although tokens within a day are treated as conditionally independent in the training objective, the per-token predictions are coupled through the shared transformer representation $f_{\theta}$. This objective can be interpreted as an approximation to maximum likelihood for a joint sequence model. At generation time, we use iterative decoding (Sect.~\ref{subsec:iterative-decoding}), which progressively enlarges the conditioning set of tokens.
Further implementation details are given in the Appendix~\ref{app:imp-detail}.

\subsection{Sampling synthetic days via iterative decoding}\label{subsec:iterative-decoding}

\begin{figure}[!htbp]
    \centering
    \includegraphics[width=\linewidth]{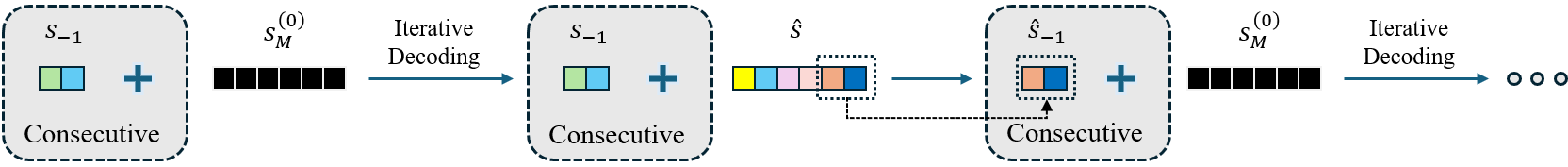}
    \caption{Iterative Decoding to sample synthetic wind data. We can sample consecutive synthetic days by sampling a new day, conditioned on the last 60 minutes of the previously sampled day.}
    \label{fig:iterative-decode}
\end{figure}

We now describe how to generate a synthetic day, having trained the bidirectional transformer $f_{\theta}$. 
In the third step, instead of generating all tokens at once, an iterative generative scheme is used, following MaskGIT \citep{chang2022maskgit}.
More specifically, instead of sampling a sequence using the probabilities estimated from $f_{\theta}(\bm{s}_{1:C}^{(i-1)}, \{[\mathrm{MASK}]\}^{T})$ directly, we generate $\bm{s}^{(i)}_{1:T}$ by iteratively unmasking over $B$ refinement steps, as shown in Figure~\ref{fig:iterative-decode}, and Algorithm~\ref{alg:iterative-decode}, which contains details of the iterative decoding algorithm.   

% \begin{figure}[!htbp]
%     \centering
%     \includegraphics[width=\linewidth]{figures/Methodology/iterative decode.png}
%     \caption{Iterative Decoding to sample synthetic wind data. We can sample consecutive synthetic days by sampling a new day, conditioned on the last 60 minutes of the previously sampled day.}
%     \label{fig:iterative-decode}
% \end{figure}

At each refinement step, the model outputs probabilities for every currently masked token. MaskGIT assigns each token a confidence score (a function of the probabilities, perturbed by Gumbel noise for stochasticity). The most confident subset of tokens is unmasked and used as additional conditioning in the next refinement step.

\smallskip
\begin{algorithm}[!htbp]
\caption{Iterative decoding with MaskGIT}
\label{alg:iterative-decode}

\begingroup
\raggedright

\textbf{Initialization:}\par
Create initial masked sequence for day $i$: $$\bm{s}^{(i)}_{\bm{m}(1:T)} \leftarrow \{[\mathrm{MASK}]\}^{T}$$

\smallskip
\textbf{For} $b=1,\ldots,B$ \textbf{do}\par

\begingroup
\leftskip=1.5em

Compute number of tokens to keep masked:$$n_M \leftarrow \left\lfloor \cos\left(\frac{\pi b}{2B}\right)T \right\rfloor$$

\smallskip
\textbf{Model prediction:}\par
Compute probabilities:
$$
\bm{P}
\leftarrow
\operatorname{Softmax}
\left(
f_{\theta}
\left(
\bm{s}_{1:C}^{(i-1)},
\bm{s}^{(i)}_{\bm{m}(1:T)}
\right)
\right).
$$

Sample tokens based on the probabilities to obtain $\widehat{\bm{s}}^{(i)}_{1:T}$.

\smallskip
\textbf{Compute confidence:}\par
Assign infinite confidence to already unmasked tokens in $\bm{s}^{(i)}_{\bm{m}(1:T)}$.

Extract probabilities corresponding to the sampled class. \\
For tokens $t=1,\ldots,T$,
$$
P_{t,\mathrm{selected}}
=
P_{t k'},
\quad
\text{where }
k'=\widehat{\bm{s}}^{(i)}_t .
$$

Compute confidence with Gumbel noise. For tokens $t=1,\ldots,T$,
$$
c_t
\leftarrow
\log\left(P_{t,\mathrm{selected}}\right)
+
\mathrm{temperature}
\times
\mathrm{Gumbel\ noise}.
$$

\smallskip
\textbf{Mask uncertain tokens:}\par
Keep masked the $n_M$ tokens with the smallest confidence scores.

Unmasked tokens are set to the sampled value $\widehat{\bm{s}}^{(i)}_t$.

Update
$$
\bm{s}^{(i)}_{\bm{m}(1:T)}
\leftarrow
\widehat{\bm{s}}^{(i)}_{\bm{m}(1:T)} .
$$

\endgroup

\smallskip
\textbf{End for}\par

\smallskip
\textbf{Return} final generated token sequence $\bm{s}^{(i)}_{1:T}$.

\endgroup

\end{algorithm}

\subsection{Conditioning on Weather States}
\label{Subsection:conditionalon-Weather-States}
Training the model to capture periods of high volatility is difficult, in part because these periods tend to be short and
infrequent. Thunderstorms, for example, often generate high wind speeds and usually last under an hour. The high frequency behavior of wind is also likely affected by cloud coverage because of the important role convection plays in wind dynamics. The correlation between wind behavior and measurable weather conditions suggest incorporating weather state covariates to improve learning high volatility events. We constructed a discrete set of weather states with $2^4$ levels to label the minute-by-minute data. While continuous variables might offer higher precision, this simplified discrete approach allows us to isolate and demonstrate the value of adding weather information without shifting our research focus away from the wind vector modeling itself.

In this paper, we consider two approaches to include this weather state variable. The first approach treats the weather state as another feature (channel) in the tokenization stage, in addition to the easterly and northerly wind vectors; consequently, $P=3$. Then, the model is trained as before, allowing the model to  simultaneously capture the weather states and wind vector information. 
The second approach is to incorporate the weather states during stage 2. That is, weather state embeddings are appended during the training of the MaskGIT model. The weather state is a categorical variable with 16 levels,  denoted as $\bm{W}^{(i)} \in \{1,\dots, 16\}^{T}$. The weather state is mapped to a real-valued learnable embedding, which is easier for the transformer to utilize. Specifically, the weather state is embedded via $g_{\phi}:\{1, \dots
, 16\}^T \rightarrow \mathbb{R}^{T \times D'}$. We denote $\widetilde{\bm{W}}^{(i)} = g_{\phi}(\bm{W}^{(i)}) \in \mathbb{R}^{T\times D'}$, where $D'=128$. 

We train the MaskGIT model to predict wind tokens and weather states, both randomly masked at the same time period, conditional on both wind tokens and weather embeddings from the previous day. 
Specifically, for the wind tokens, we model the token-wise conditional distributions as
     \begin{align}
        & p({\bm{s}}_{t}^{(i)}|\bm{s}_{1:C}^{(i-1)}, \bm{s}_{\bm{m}(1:T)}^{(i)}, \widetilde{\bm{W}}^{(i-1)}_{1:C}, \widetilde{\bm{W}}^{(i)}_{\bm{m}(1:T)}) \nonumber\\
         &\quad = \mathrm{Categorical}(\mathrm{Softmax}(f_{\theta}([\bm{s}_{1:C}^{(i-1)}, \bm{s}_{\bm{m}(1:T)}^{(i)}, \widetilde{\bm{W}}^{(i-1)}_{1:C}, \widetilde{\bm{W}}^{(i)}_{\bm{m}(1:T)}])_t)).
    \end{align}
For the weather states, we add an additional layer onto  $f_{\theta}$, denoted $h_{\phi}:\mathbb{R}^{T\times K} \to \mathbb{R}^{T\times 16}$. The token-wise conditional distributions are:
    \begin{align}
        &p(\bm{w}_t^{(i)}|\bm{s}_{1:C}^{(i-1)}, \bm{s}_{\bm{m}(1:T)}^{(i)}, \widetilde{\bm{W}}^{(i-1)}_{1:C}, \widetilde{\bm{W}}^{(i)}_{\bm{m}(1:T)}) \\
         &\quad= \mathrm{Categorical}(\mathrm{Softmax}((h_{\phi} \circ f_{\theta}([\bm{s}_{1:C}^{(i-1)}, \bm{s}_{\bm{m}(1:T)}^{(i)}, \widetilde{\bm{W}}^{(i-1)}_{1:C}, \widetilde{\bm{W}}^{(i)}_{\bm{m}(1:T)}]))_t)). \nonumber
    \end{align}
We train the model by minimizing the negative log-likelihood:
\begin{align}
\mathcal{L}(\theta,\phi) &= -\sum_{i=1}^{n_{\mathrm{train}}} \sum_{t: m^{(i)}_t=1} \left[\sum_{k=1}^K 1(s_t^{(i)}=k)  \right.
%\right.  \\ 
%&\qquad\qquad\qquad\qquad \times 
\log p(\bm{s}_t^{(i)}=k|\bm{s}_{1:C}^{(i-1)}, \bm{s}_{\bm{m}(1:T)}^{(i)},\widetilde{\bm{W}}_{1:C}^{(i-1)}, \widetilde{\bm{W}}_{\bm{m}(1:T)}^{(i)}) \nonumber\\
&\qquad\qquad\qquad\qquad + \left. \sum_{j=1}^{16} 1(\bm{w}_t^{(i)}=j) \log p(\bm{w}_t^{(i)}=j|\bm{s}_{1:C}^{(i-1)}, \bm{s}_{\bm{m}(1:T)}^{(i)},\widetilde{\bm{W}}_{1:C}^{(i-1)}, \widetilde{\bm{W}}_{\bm{m}(1:T)}^{(i)})
\right],
\nonumber
\end{align}
where $1(\cdot)$ is the indicator function.
While this likelihood factorizes across time, wind and weather, the predictions again remain coupled through the shared transformer representation. This objective can be interpreted as an approximation to maximum likelihood for the joint wind and weather sequence model. At generation time, we follow the same iterative decoding strategy as before (Sect.~\ref{subsec:iterative-decoding}).

%\textcolor{red}{MS: So here we need to (briefly) discuss how we determined that generators aren't just reproducing the observed data.  This needs to be done for all the generators in Table 3 I would think.}

%\textcolor{blue}{Justin: This seems like the natural place for the boxplots showing the distribution of minimum Euclidean distances. For now, I will have it as a subsection of section 4 since part of the methodology was sanity checking. I will try to incorporate some of Zhiqiu's and Professor Stein's setups of the synthetic results so that it is defined once. Sorry that this will cause some rewriting.}

%\textcolor{blue}{Justin (April 12, 2026, after the two above comments) The full details for the no memorization discussion are being moved to the supplement, however there should be a brief paragraph that reports this analysis being done and points to the supplement for full details. }

In all, we have six models that we use to generate synthetic data. The models vary in two ways: first, how the weather state is incorporated into the model and, second, the temporal conditioning strategy. For weather incorporation, ``No Weather'' will refer to no incorporation, ``Features'' denotes weather state treated as a feature, and ``Embedded'' refers to adding weather state as an additional component of the state vector. For temporal conditioning, ``Independent'' refers to generating individual days independently and ``Consecutive'' starts with an independently generated day and generates the remaining 20-days of the sequence conditional on the previous days' final hour. 

To ensure that our generative models are not merely reproducing the observed patterns in our training data, we compute the minimum Euclidean distance between each training day and the closest day in a series of comparison sets: the other 473 training days and 474 synthetic days from each of our six models. If the synthetic data were near-identical reproductions of training days, we would expect these minimum distances to be close to zero. However, upon inspection, we find that the distributions of minimum distances between training and synthetic days are comparable to those observed between training days themselves, indicating the models generate novel patterns rather than memorized copies. Additionally, we plotted the time series of the Easterly and Northerly components of the wind vector for the closest training-synthetic pair across all generators, which showed that while the pair shared similarities in general shape, there were distinct features present in the synthetic day that are not present in the training. Additional implementation details and results are provided in Appendix~\ref{app:min_dist_boxplots}.

\section{Discriminative Evaluation}
\label{sec:discriminative}

A good all-purpose weather generator should be able to reproduce all of the main features of the observations.
This section describes an approach to assess the realism of the simulations via a classifier that attempts to distinguish real and synthetic data.
We tried many generators, but here we only report on the six generators described in Sect.\ \ref{Subsection:conditionalon-Weather-States}.
%Specifically, we consider not including the weather states, treating the weather state as a feature or embedding the weather state together with the wind data into the state vector.
%For each of these three treatments of the weather state, we generate days of wind vectors one at a time or consecutively for periods of 21 days, so a total of six stochastic wind generators.
In training the discriminator, we use two different units: either one day at a time or periods of three consecutive days.
When using three consecutive days, we allow overlap between the periods, so any one day can appear in up to three 3-day periods.
It does not make sense to use the generators that do not condition on the previous day when discriminating 3-day periods, so we only give results for the three generators that do this conditioning.

%for both the one day at a time approach (denoted as \textbf{Uncond.}) and the consecutive days approach (denoted as \textbf{Consec.}).
% For the approaches generating consecutive days, we evaluate them using a classifier that takes each day as the unit and a classifier that considers three consecutive days as the unit.
%\textcolor{red}{Maybe in this section only, worthwhile to give results for generators that don't use weather states so we can quantify their impact?}  \textcolor{orange}{Zhiqiu: It is included in the next subsection. Let me know if any fix is needed.}

\subsection{Experimental Setup}

To evaluate the realism of the generated time-series data, we employ a post-hoc classification metric following the methodology of prior work by \citet{yoon2019time}. This approach trains a discriminative classifier to differentiate between real observations and synthetic samples. Unlike previous works that calculate a ``discriminative score'', we report the raw classification accuracy. Since we always evaluate the classifier using an equal number of real and synthetic days, an accuracy of 0.5 indicates that the classifier performs no better than random guessing. 
%This implies the synthetic data is indistinguishable from real data, signifying a high-quality generator. 
On the other hand, an accuracy close to 1.0 indicates that the classifier can easily distinguish synthetic samples from real observations, signifying lower generation quality.
Note, though, even if the accuracy is near 0.5, that only shows that our discriminator could not distinguish real and synthetic days and does not demonstrate that reliable discrimination is impossible.

The characteristics of the real dataset are detailed in Table~\ref{tab:data_summary}. Additionally, the validation subset consists of 205 samples. Regarding synthetic data, the total generation count depends on the method used. We generated 50,000 independent samples when producing wind vectors one day at a time, and a total of 4,830 days when generating consecutive 21-day sequences. 
Synthetic days for this exercise were sampled from these larger sets of simulations to match the size of the observational record.
In particular, to maintain consistency with the observational record, we artificially applied a missing-data mask to these simulations to replicate the patterns of missingness observed in the real data. For the analysis utilizing 3-day periods, we ensured that all segments consisted of consecutive days. Consequently, for 3-day periods, the training, validation, and testing set sizes for this experiment were 422, 127, and 162 periods, respectively. 
%\textcolor{red}{Shouldn't these numbers be slightly smaller for 3-day periods because there are fewer of them even when allowing overlapping periods?} \textcolor{orange}{Zhiqiu: Yes, they are. As you can see from Table~\ref{tab:data summary} and the second sentence of this paragraph. The numbers 422, 127, 162 correspond to 3-day periods.}
To match these sizes, we sampled the corresponding number of synthetic examples from a pool of 4,220 available 3-day sequences, which is derived from the 4,830 generated days.

Our discriminative classifier utilizes a bidirectional Long Short-Term Memory network, followed by a feed-forward network, to classify time-series sequences as either real or synthetic. The input to the network consists of two-dimensional data representing the easterly and northerly wind components. The sequence length is determined by the number of consecutive days being evaluated. For the training, validation, and testing datasets, we sample an equal number of synthetic and real samples to ensure balanced evaluation, guaranteeing that the baseline random guessing accuracy is exactly 0.5. To ensure robust statistical evaluation, we generate 50 different sets of synthetic samples for training, validation, and testing using different random seeds, and report the average and standard deviation of the accuracy across these runs. Complete architectural specifications and training hyperparameters are provided in Appendix~\ref{app:disc}.

\subsection{Results}

Table~\ref{tab:discriminativescores_50} presents the classification accuracy for all evaluated methods on one-day sequences. 
When discriminating a day at a time, the generator that embeds weather states and produces each day conditional on the previous day does best with minimizing the probabilities of the classifier identifying both real and synthetic days.
Interestingly, when not incorporating weather states, using consecutive days actually increases these probabilities.
Among the three generators that produce consecutive days, embedding weather states is again the winner when discriminating three-day periods.

Note that for all three of these generators, the accuracies of detecting real weather periods are largely the same whether one considers 1 or 3 day periods, whereas the discriminator does substantially better at detecting 3-day synthetic periods than 1-day synthetic periods.
Thus, there appears to be a weakness in our approach to producing consecutive days that the discriminator is able to detect.
One obvious possibility is that the transitions between days in the consecutive generators are unrealistic.
Indeed, for the observational data, the root mean squared difference in the change in wind speed between the last minute of one day and the first minute of the next day is 0.73 m s$^{-1}$, whereas it is 1.17 m s$^{-1}$ using embedded weather states and 1.34 m s$^{-1}$ using weather states as a feature.
Whether or not the discriminator is making use of this difference between observations and the generators to classify 3-day periods, there is clear scope for improving how we generate one day of winds given the previous day's winds.

%To quantify the benefit of conditioning on weather states, we included ``NoW'' baselines.
%The results demonstrate that incorporating weather states generally improves the realism of the generated data. Notably, the \textit{W-Emb Consec.} method achieves the lowest classification accuracy of $\mathbf{0.755 \pm 0.052}$. It indicates that this configuration produces the most realistic time-series data among the models tested.

\begin{table}[!htbp]
\caption{
Accuracies in detecting real and synthetic wind vector time series using discriminative classifier for six stochastic wind generators.
Weather column indicates how weather states are included in the generator and Consecutive column is Yes if each day after the first in a year is simulated conditional on the previous day.
Period is length of time periods used as individual cases by the classifier.
Accuracies are averages over 50 sets of synthetic training and testing days and numbers in parentheses are the standard deviations of these 50 results.
For each period, best stochastic wind generator (lowest accuracy for discriminator) is given in bold.
}
\label{tab:discriminativescores_50}
\centering
\begin{tabular}{ccccc}
%\hline
\multicolumn{2}{c}{Method} & &  \multicolumn{2}{c}{Accuracy} \\
\cmidrule(lr){1-2} \cmidrule(lr){4-5}
Weather & Consecutive & Period & Real & Synthetic  \\ 
\hline
None & No & 1-day & $.814 (.048)$ & $.789 (.044)$  \\
None & Yes & 1-day &  $.845 (.048)$ & $.799 (.069)$ \\
Feature & No & 1-day &  $.843 (.033)$ & $.795 (.033$) \\
Feature & Yes & 1-day & $.811 (.056)$ & $.794 (.066)$ \\
Embed & No & 1-day & $.791 (.058)$ & $.791 (.059)$ \\
Embed & Yes & 1-day & $\bm{.766 (.052)}$ & $\bm{.744 (.081)}$ \\
\hline
None & Yes & 3-day & $.851 (.039)$ & $.885 (.045)$ \\
Feature & Yes & 3-day & $.819 (.054)$ & $.865 (.052)$ \\
Embed & Yes & 3-day & $\bm{.759 (.055)}$ & $\bm{.857 (.044)}$
%\hline
\end{tabular}
\vspace{8pt}
\end{table}

% We further evaluate the model's capability to generate longer sequences by extending the generation window to three days. As shown in Table \ref{tab:discriminativescores_three_days}, the difficulty of the generation task increases with sequence length, resulting in generally higher discriminative scores across all methods compared to the one-day results.

%We further assessed the models' capabilities to maintain realism over longer temporal horizons by extending the classification window to three consecutive days. As shown in Table \ref{tab:discriminativescores_50}, the classification accuracy generally increases for the three-day window compared to the one-day window across all methods.

%This rise in accuracy suggests that the generated sequences fail to fully capture the underlying temporal dependencies present in real data. Over longer horizons, inconsistencies or implausible transitions between days become more apparent, making synthetic samples easier for the classifier to distinguish from real ones.

Even using our best generator and 1-day periods, the discriminator does much better than just guessing, so it is worthwhile to understand better how the discriminator is making decisions.
As a first step in this direction, we analyzed whether the classifier relies more on high-frequency variation or low-frequency behavior to distinguish observations from simulations. We applied high-pass and low-pass filters to datasets generated by our best-performing generator (embedded weather state and consecutive days). To distinguish high-frequency noise from low-frequency behavior, we applied 4th-order Butterworth filters to the generated datasets and utilized a zero-phase forward-backward application. The cutoff frequency was set to 48, corresponding to 30-minute intervals in a 1440-minute daily cycle, to separate short-term fluctuations from longer-term patterns.
%\textcolor{red}{MS: Please say exactly what these filters were.} \textcolor{orange}{Zhiqiu: have added, from ``To distinguish high-frequency noise from ... ''}

The results in Table~\ref{tab:freq_analysis_50} reveal distinct spectral behaviors. On low-pass filtered data, the discriminator is noticeably less accurate than when using the unfiltered data for both synthetic and real days. This relatively low accuracy suggests the generator is relatively effective at capturing macro-level trends and general wind patterns, making them difficult to distinguish from reality.
In contrast, the discriminator based on high-pass filtered data performs substantially better than the discriminator using unfiltered data, especially for synthetic days, yielding an accuracy of 0.896 as opposed to 0.744.
Since all of the information in the high-pass filtered data is contained in the unfiltered data, this indicates a lack of optimality of our discriminator when applied to the unfiltered data.
Apparently the best information for discriminating between real and synthetic days is in the high frequencies and, by first applying a high-pass filter, we focus the discriminator's attention on to this high-frequency information and end up with a better discriminator.

Another noteworthy result in Table~\ref{tab:freq_analysis_50} is that, for the high-pass filtered data, the standard deviations in accuracies across the 50 repetitions of the procedures are much higher: 0.215 for real data and 0.160 for synthetic data, whereas the next largest standard deviation in this table is 0.081.
Figure \ref{fig:gen-real} in Appendix E plots the accuracy results for the discriminator for all 50 repetitions of the procedure when applied to both the unfiltered and high-pass filtered data.
We see that when applied to the filtered data, the discriminator is often nearly perfect but sometimes performs poorly, whereas, when applied to the unfiltered data, the discriminator performs fairly consistently across repetitions but is never close to perfect.
These results suggest that a better discriminator might be able to consistently perform nearly perfectly even for our best synthetic wind generator.

%significantly higher overall accuracy of $0.857$ , which . This indicates the generator struggles to reproduce realistic high-frequency noise, providing a clear signal for the classifier. Furthermore, the high standard deviation ($\pm 0.150$) in the high-pass regime indicates instability. In some training runs, the generator's noise is distinguishable, while in others, it effectively mimics the stochastic nature of the observations.

\begin{table}[!htbp]
\caption{Accuracies for classifying 1-day periods for frequency-filtered data using the consecutive generator with embedded weather states.}
\label{tab:freq_analysis_50}
\centering
\begin{tabular}{cccc}
%\hline
 && \multicolumn{2}{c}{Accuracy} \\
\cmidrule(lr){3-4}
Filter && Real & Synthetic  \\ 
\hline
None  && $.766(.052)$ & $.744(.081)$  \\ 
Low-pass  && $.729(.052)$ & $.683(.058)$ \\
High-pass && $.817(.215)$ & $.896(.160)$  \\
\end{tabular}
\vspace{8pt}
\end{table}

\section{Energy scores} 
\label{sec:energy_scores}

% \textcolor{blue}{Justin: Not sure if the paper would flow better having the energy scores somewhere else, but for now I will make the edits where it has been.}

Another way to evaluate the accuracy of the estimated distributions of our generators is through scoring rules.
Here, we use the multivariate version of continuous ranked probability score (CRPS), called an \textit{energy score} \citep{gneiting2007, gneiting2008}.

Let $\bm{y} = (y_1,\dots, y_d) \in \mathbb{R}^d$ be a real observation and let $F$ denote a forecast distribution on $\mathbb{R}^d$ characterized by $m$ discrete samples $\bm{X}_1,\dots, \bm{X}_m$ from $F$ with $\bm{X}_i \in \mathbb{R}^d$ for $i=1,\dots, m$. The energy score is calculated as
\begin{equation}
  ES(F,\bm{y}) = \frac{1}{m}\sum_{i=1}^m ||\bm{X}_i - \bm{y} || - \frac{1}{2m^2}\sum_{i=1}^m \sum_{j=1}^m ||\bm{X}_i - \bm{X}_j|| \label{eq:energy_score}
\end{equation}
where $||\cdot ||$ denotes the Euclidean norm on $\mathbb{R}^d$ \citep{jordan2019}. Lower energy scores signify the distribution $F$ being a better forecast. We use the \verb|es_sample()| function from the package \verb|scoringRules| for efficient computation of energy scores \citep{jordan2019}.
% The second term on the right-hand side of (\ref{eq:energy_score}) is known as the energy norm.
% It does not depend on $\bm{y}$, which saves on computations.

For our evaluation, we calculate energy scores for each of the 185 testing days using the distributions characterized by the 474 training days and 474 synthetic days from each generator with the same full-day missing pattern as observed in the training days. For each day $i$, we have $\bm{X}^{(i)} \in \mathbb{R}^{T \times P}$ where $T = 1440$ minutes and $P = 2$ are the easterly and northerly components of the wind vector. We reshape this matrix into a vector by concatenating all columns:
\begin{equation}
    \bm{v}^{(i)} = \text{vec}(\bm{X}^{(i)}) \in \mathbb{R}^{TP} \label{eq:concat_east_north1}
\end{equation}
where $\text{vec}(\cdot)$ stacks the columns of $\bm{X}^{(i)}$ into a single vector of dimension $TP = 2880$. Each day is represented as $\bm{v}^{(i)} \in \mathbb{R}^{2880}$ for day-level evaluation. For hourly evaluation, we similarly vectorize each hour of day $i$ as $\bm{v}_h^{(i)} \in \mathbb{R}^{120}$ containing the 60 easterly followed by 60 northerly components for hour $h$. 

For each test day $j$, we compute two energy scores $ES(V_{\text{train}}, \bm{v}_{\text{test}}^{(i)})$ and $ES(V_{\text{synth}}, \bm{v}_{\text{test}}^{(i)})$, where $V_{\text{test}}$ and $V_{\text{synth}}$ are the sets of vectorized training and synthetic days respectively, with $m = 474$ in both cases. If the generators produce realistic wind patterns, we expect these two scores to be similar for each test day.

\begin{figure}[!htbp]
    \centering
    \includegraphics[width=0.55\linewidth]{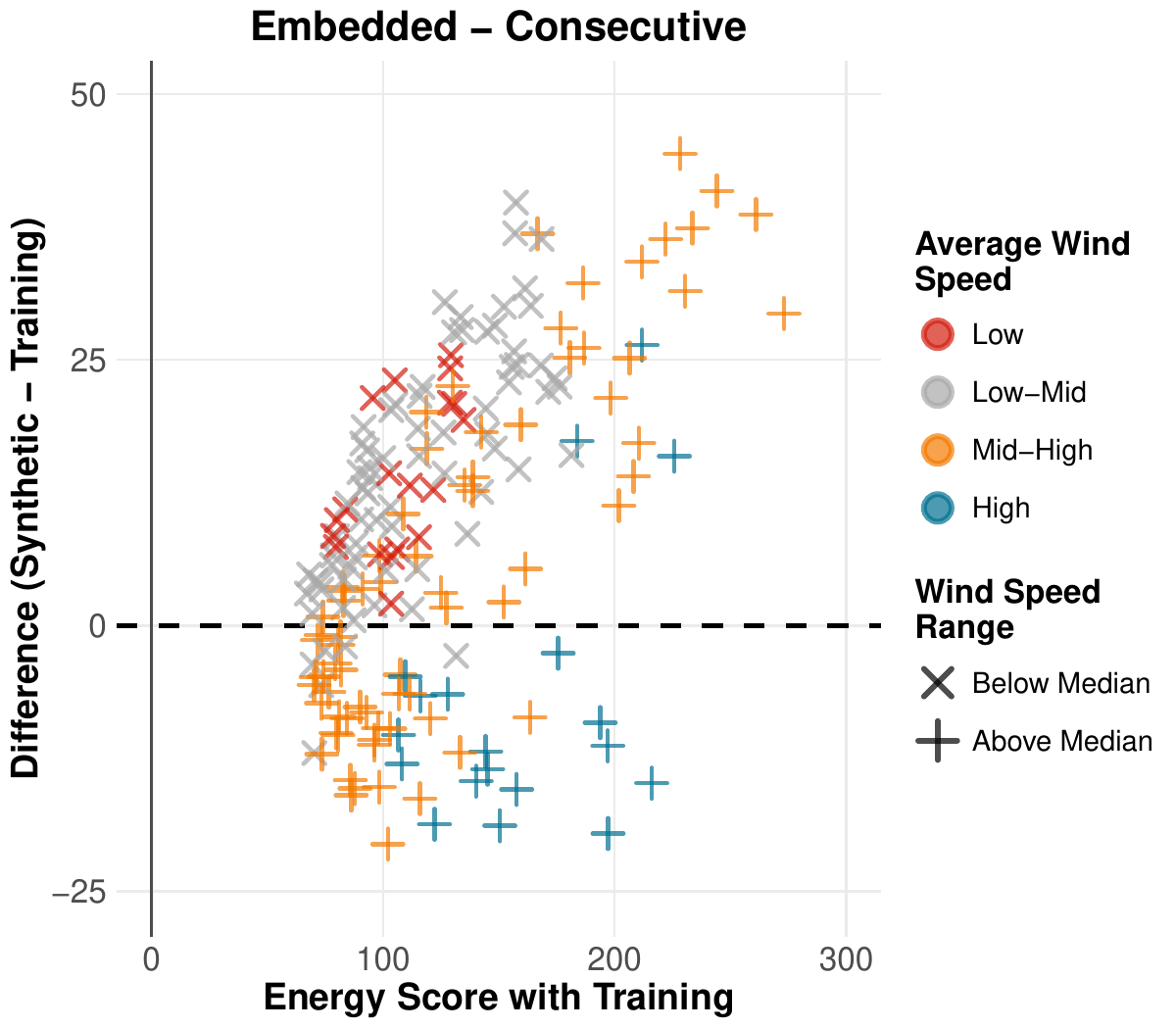}
    \caption{Day-level energy score differences for Embedded generator with consecutive sampling. Each point represents one of the 185 test days colored by average wind speed. The vertical axis shows difference between synthetic and training energy scores, with the horizontal dashed line ($y=0$) indicating perfect agreement. Color indicates average wind speed over the test day: Low (0-0.1 quantile), Low-Mid (0.1-0.5 quantile), Mid-High (0.5-0.9 quantile), and High (0.9-1 quantile).}
    \label{fig:es_day_emb_con}
\end{figure}

Figure~\ref{fig:es_day_emb_con} shows the differences between the day-level energy score performance for the Embedded generator using consecutive sampling and the energy score based on the training data; that is,  $ES(V_{\text{synth}}, \bm{v}_{\text{test}}^{(i)})-ES(V_{\text{train}}, \bm{v}_{\text{test}}^{(i)})$ versus $ES(V_{\text{train}}, \bm{v}_{\text{test}}^{(i)})$ for each test day $i=1,\dots, 185$. 

% \begin{figure}[!htbp]
%     \centering
%     \includegraphics[width=0.55\linewidth]{figures/Energy Scores/Embedded/embedded_es_scatter_day_diff_consecutive.pdf}
%     \caption{Day-level energy score differences for Embedded generator with consecutive sampling. Each point represents one of the 185 test days colored by average wind speed. The vertical axis shows difference between synthetic and training energy scores versus training energy, with the horizontal dashed line ($y=0$) indicating perfect agreement. Color indicates average wind speed over the test day: Low (0-0.1 quantile), Low-Mid (0.1-0.5 quantile), Mid-High (0.5-0.9 quantile), and High (0.9-1 quantile).}
%     \label{fig:es_day_emb_con}
% \end{figure}

\begin{figure}[!htbp]
    \centering
    \includegraphics[width=0.92\linewidth]{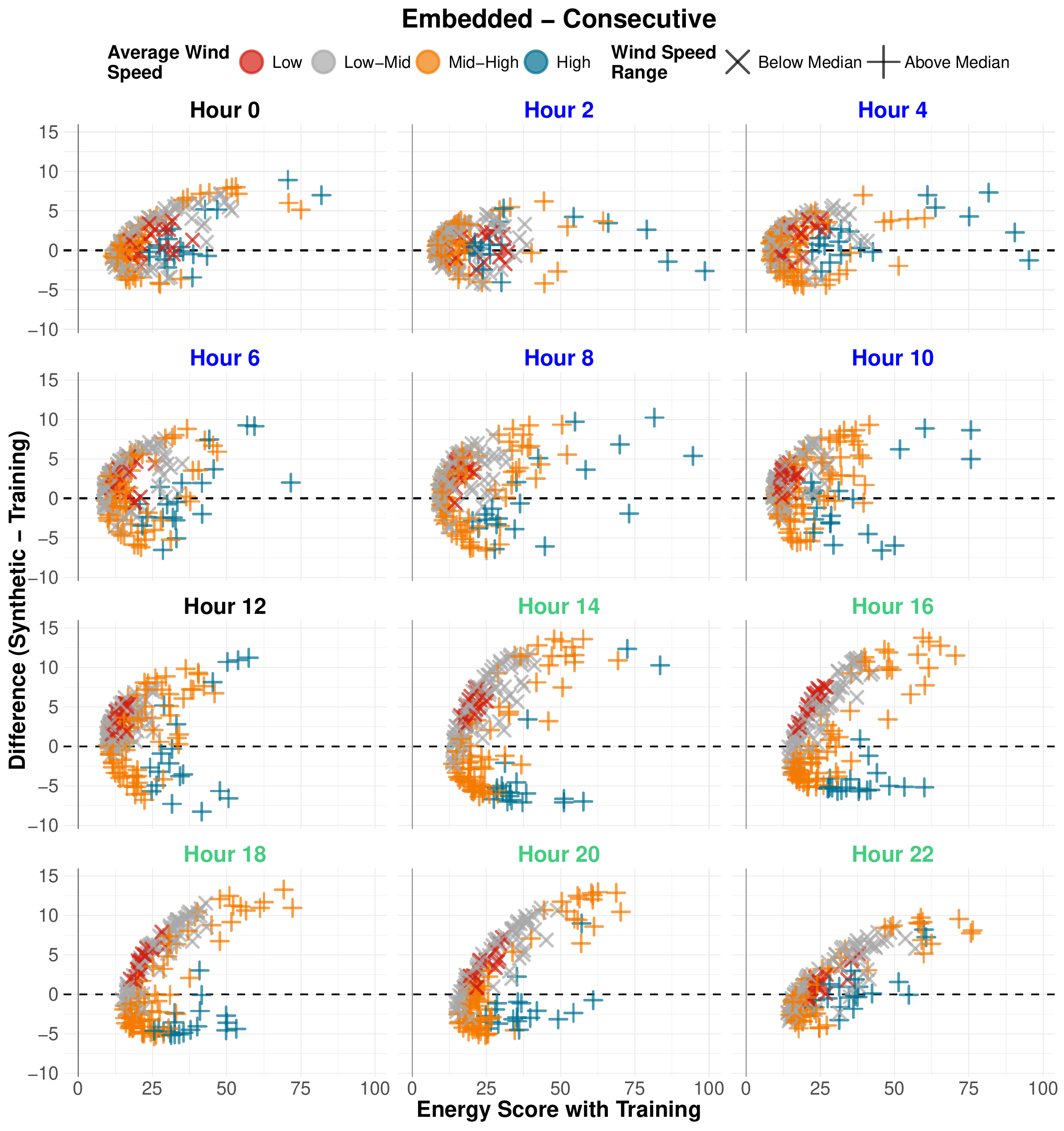}
    \caption{ Hourly energy score differences for Embedded generator with consecutive sampling for every other hour of the day. Each subplot shows one hour of the day with 185 (one per test day) colored by average wind speed over that hour in the testing data. The x-axis shows energy score with training data as the forecast distribution, the y-axis shows the difference between synthetic and training energy scores, and the horizontal dashed line $y=0$ indicates comparable performance. Colored subplot titles indicate nighttime hours (blue for 2-11), daytime hours (green for 14-24), and transition hours (black for 0,1,12,13). Symbols same as Figure \ref{fig:es_day_emb_con}. }
    \label{fig:es_hour_emb_con}
\end{figure}

%Points clustering near the diagonal line indicate that the synthetic distribution performs similarly to the training distribution in characterizing the test days \textcolor{blue}{Justin: Might need to reword this sentence, feels slightly clunky}. 
%The right panel shows the same energy scores but as differences, plotting $ES(V_{\text{synth}}, \bm{v}_{\text{test}}^{(i)}) - ES(V_{\text{train}}, \bm{v}_{\text{test}}^{(i)})$ versus $ES(V_{\text{train}}, \bm{v}_{\text{test}}^{(i)})$. 
%his residual representation makes it easier to assess systematic patterns. 
If the generator is working well, then the differences on the vertical axis should be close to 0.
Differences well above zero indicate the synthetic distribution yields higher (worse) energy scores than the training distribution. 
Expressed as percentages of training energy scores (i.e. $100 \times \frac{\text{Diff}}{\text{Train}}$), the differences range from -20\% to +25\%, with a median of +7.3\% and an interquartile range of $[-4.4\%, +14.4\%]$. This indicates that while synthetic performance is generally comparable to training performance, synthetic data tend to produce moderately higher energy scores. 
%An interesting feature of Figure \ref{fig:es_day_emb_con}
Figure~\ref{fig:es_hour_emb_con} extends the day-level analysis to individual hours. We only show results for every other hour; the other hours show patterns similar to their neighboring hours. 
We see that the differences in energy scores tend to be larger during the daytime hours.
%Each of the 12 subplots displays energy score differences $ES(V_{\text{synth}}, \bm{v}_{h,\text{test}}^{(i)}) - ES(V_{\text{train}}, \bm{v}_{h,\text{test}}^{(i)})$ versus energy scores $ES(V_{\text{train}}, \bm{v}_{h,\text{test}}^{(i)})$ for hour $h$ across all 185 test days. 

% \begin{figure}[!htbp]
%     \centering
%     \includegraphics[width=0.92\linewidth]{figures/Energy Scores/Embedded/embedded_es_scatter_hourly_diff_consecutive.pdf}
%     \caption{ Hourly energy score differences for Embedded generator with consecutive sampling for every other hour of the day. Each subplot shows one hour of the day with 185 (one per test day) colored by average wind speed over that hour in the testing data. The x-axis shows energy score with training data as the forecast distribution, the y-axis shows the difference between synthetic and training energy scores, and the horizontal dashed line $y=0$ indicates comparable performance. Colored subplot titles indicate nighttime hours (blue for 2-11), daytime hours (green for 14-24), and transition hours (black for 0,1,12,13). Symbols same as Figure \ref{fig:es_day_emb_con}. }
%     \label{fig:es_hour_emb_con}
% \end{figure}

In both the day-level differences in Figure~\ref{fig:es_day_emb_con} and hour-level differences in Figure~\ref{fig:es_hour_emb_con}, there is a bifurcation for higher energy scores using the training data as the forecast distribution, creating a ``C'' shaped pattern. Larger values on the horizontal axis represent test days (or hours) that are less similar to typical training days. For these time periods with higher scores, the differences in the two energy scores tend to show a bimodal distribution.
%synthetic scores split into two clusters: one group with positive differences, i.e. the test day/hour is further from the synthetic distribution than training and another with negative differences, i.e. synthetic distribution is closer. 
As seen in Figure~\ref{fig:es_hour_emb_con}, this bimodality is the strongest for the daytime hours 14-20. 
Note further that for these hours, times in which the average wind speeds are above the 0.9 quantile for that hour tend to have negative values for the difference in energy scores whereas the reverse is true for hours with average wind speeds below the 0.1 quantile.
We are not sure how to interpret this result since, as we will see in the next section, all of our generators fail to reproduce the strongest wind speeds in the observations.

Test data with high energy scores using training samples as the forecast distribution represent conditions that deviate from typical training patterns, potentially caused by unusual weather events, transitional periods, or rare atmospheric conditions. For these atypical data, the synthetic distribution sometimes captures features that happen to match the test data better than the training distribution (negative differences), while other times it fails to capture aspects present in training (positive differences). For consecutive sampling, the largest average wind speeds tend to fall below $y = 0$, however they tend to be above $y=0$ for independent sampling. This difference between the sampling types persists when looking at results for the Features and No Weather generators, suggesting this relationship does have a connection to the sampling approach, with the exact cause still unclear. Additional details regarding the energy scores for the other generator and sampling types are available in Appendix~\ref{app:Add-En}. 

\section{Graphical summaries for two generators}
\label{sec:graph-simulated}

A good all-purpose weather generator should be able to reproduce all of the main features of the observations.
In this section, we return to the graphical summaries in Sect.~\ref{sec:Exploratory} to evaluate the two generators that produce consecutive days and treat the weather state as a feature or as another component of the state vector.
For convenience, this first approach is called ``Features'' and the second ``Embedded'' throughout this section.
From any simulated 23-year dataset, all plots in this section artificially remove those minutes that are missing in the observational data to make the comparisons between real and synthetic data as fair as possible.
Because these comparisons are informal, we compare synthetic results to the larger training dataset rather than the testing dataset.

\begin{figure}[!htbp]
    \centering
    \includegraphics[width=.7\linewidth]{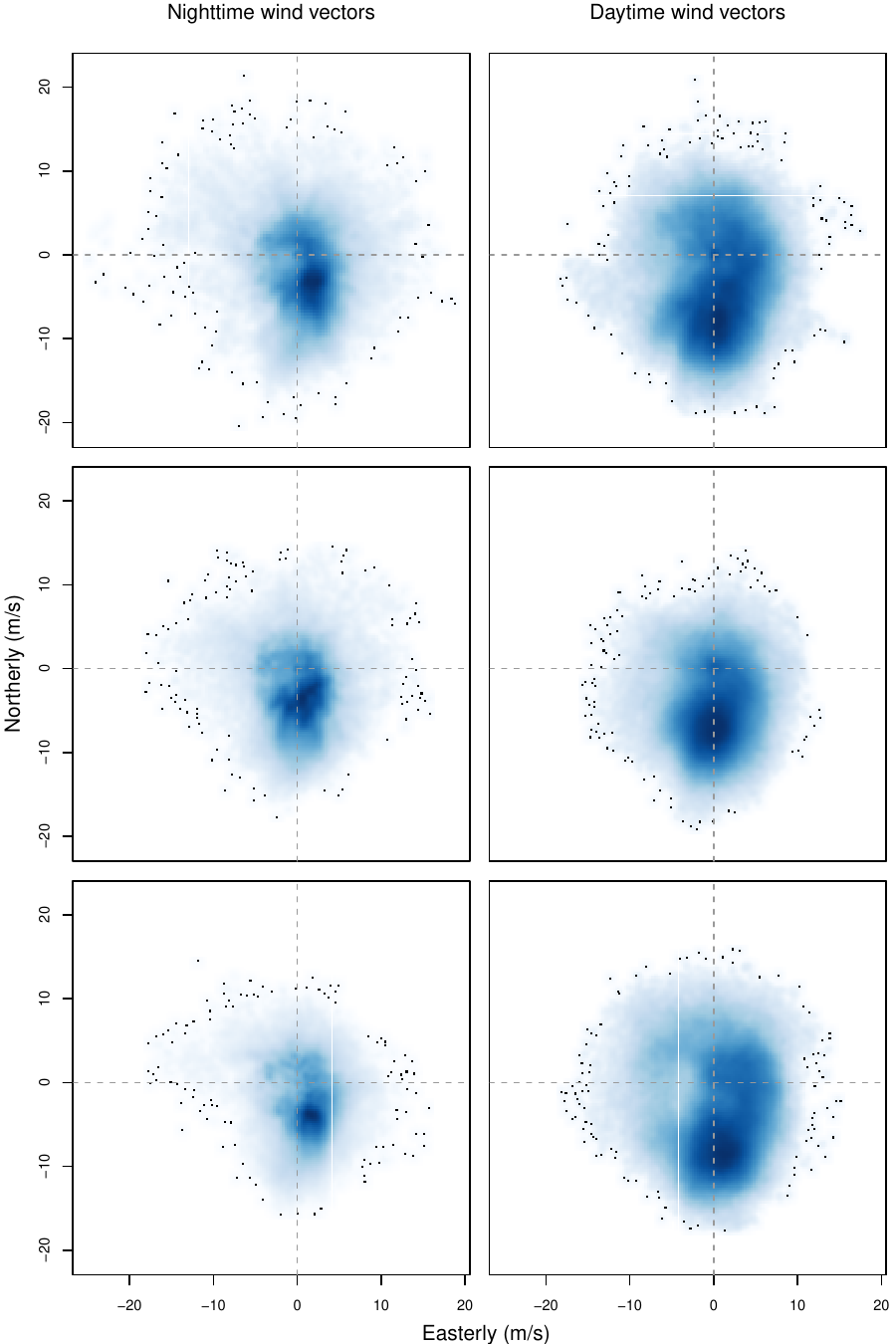}
    \caption{Density plots of wind vector for daytime and nighttime.  Top row is for training data as in Figure \ref{fig:wind-scatter}.  Middle row is for the Embedded generator.  Bottom row is for the Features generator.}
    \label{fig:wind-scatter4}
\end{figure}

When considering characteristics of the marginal distributions of the wind vectors, we need to keep in mind that these characteristics may not be estimated all that well from the observations because of the strong temporal dependence in the data.
Figure~\ref{fig:wind-scatter4} shows wind vector densities for nighttime and daytime, with the top row  giving results for the observational data (so reproducing Figure~\ref{fig:wind-scatter}), the middle row for the Embedded generator and the bottom row for the Features generator.
Results for further 23-year simulations from these two generators are given in Figures~\ref{fig:wind-scatter-cf} and \ref{fig:wind-scatter-em} in Appendix C.
The two additional Embedded simulations show this generator struggling to capture some of the apparent non-convexity in the density of the training data during the daytime.
Overall, both generators arguably perform fairly well, although neither can capture more extreme winds, especially during the nighttime.

% \begin{figure}[!htbp]
%     \centering
%     \includegraphics[width=.7\linewidth]{figures/graphical summaries/wind-scatter4.pdf}
%     \caption{Density plots of wind vector for daytime and nighttime.  Top row is for training data as in Figure \ref{fig:wind-scatter}.  Middle row is for the Embedded generator.  Bottom row is for the Features generator.}
%     \label{fig:wind-scatter4}
% \end{figure}

\begin{figure}[!htbp]
    \centering
    \includegraphics[width=.82\linewidth]{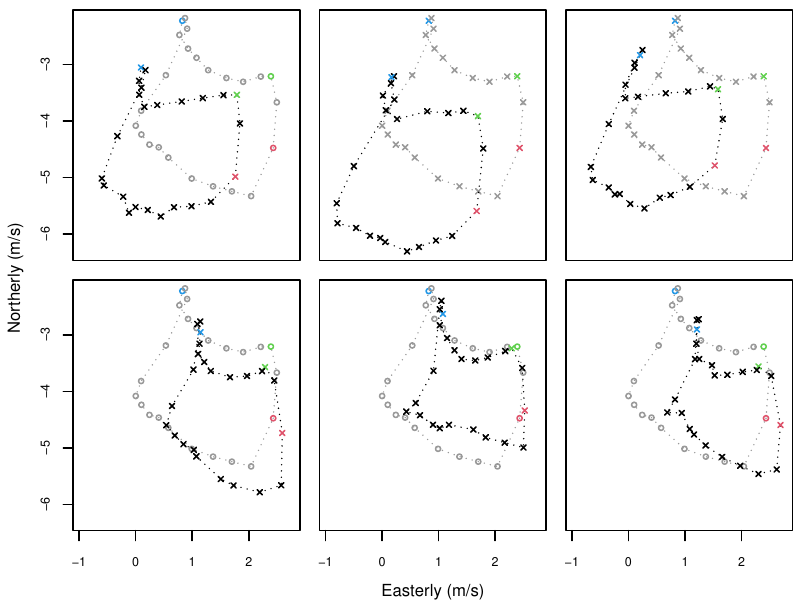}
    \caption{Comparisons of observed to simulated diurnal cycles in median winds as in Figure \ref{fig:diurnalUQ}.  Gray circles are observed values.  Black $\times$'s are simulated values.  Top row is for three Embedded simulations, bottom row is for three Features simulations.  Red symbols indicate hour starting at 0:00 UTC, green symbols for hour starting at 2:00 UTC (near sunset) and blue symbols for hour starting at 10:00 UTC (near sunrise).}
    \label{fig:diurnal3com}
\end{figure}

Figure~\ref{fig:diurnal3com} shows the hourly average of minute-by-minute medians as in Figure~\ref{fig:diurnalUQ} for three simulations from each of the two generators with the results from the observational data included for comparison.
As we should expect, results can vary substantially across simulations from either generator.
Nevertheless, for both generators, all three simulations show fairly consistent deviations from the observational results.
In particular, the Embedded generator shifts the entire diurnal towards the southwest and the Features generator mostly shifts the diurnal cycle to the southeast.
Overall, the Features generator gives results somewhat closer to those of the observations.

% \begin{figure}[!htbp]
%     \centering
%     \includegraphics[width=.82\linewidth]{figures/graphical summaries/diurnal3com.pdf}
%     \caption{Comparisons of observed to simulated diurnal cycles in median winds as in Figure \ref{fig:diurnalUQ}.  Gray circles are observed values.  Black $\times$'s are simulated values.  Top row is for three Embedded simulations, bottom row is for three Features simulations.  Red symbols indicate hour starting at 0:00 UTC, green symbols for hour starting at 2:00 UTC (near sunset) and blue symbols for hour starting at 10:00 UTC (near sunrise).}
%     \label{fig:diurnal3com}
% \end{figure}

\begin{figure}[!htbp]
    \centering
    \includegraphics[width=0.85\linewidth]{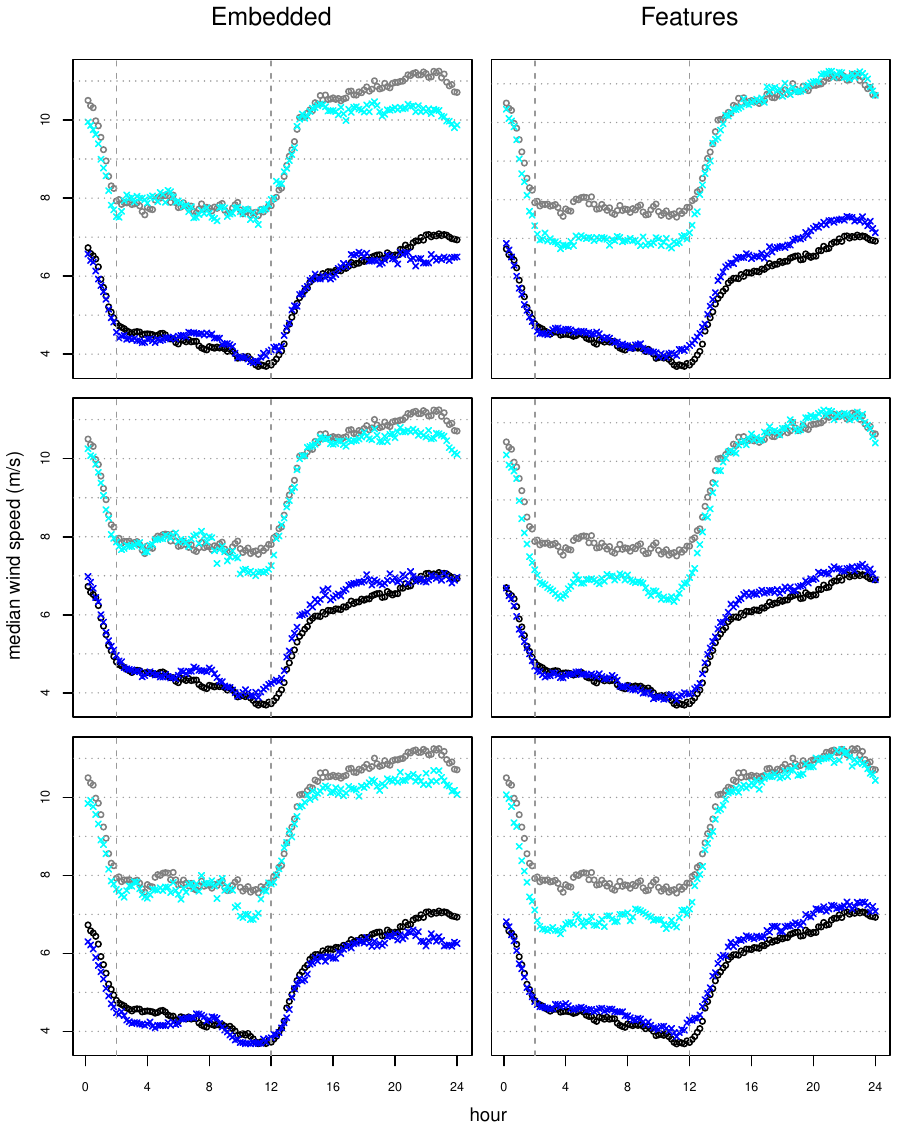}
    \caption{Ten-minute averages of minute-by-minute quantiles.  Circles for observations, $\times$'s for simulated data with left column for Embedded generator and right column for Features generator.  Dark blue and black for 0.5 quantiles and light blue and gray for 0.9 quantiles.}
    \label{fig:wsemcf}
\end{figure}

Figure~\ref{fig:wsemcf} shows, for three simulations each of the two generators, comparisons of the diurnal cycles in the 0.5 and 0.9 quantiles of wind speed.
We see that the Features generator clearly underestimates the 0.9 quantile during the nighttime for all three simulations.
In contrast, the Embedded simulation does well during the nighttime but modestly underestimates the 0.9 quantile during the late afternoon.

% \begin{figure}[!htbp]
%     \centering
%     \includegraphics[width=0.85\linewidth]{figures/graphical summaries/wsemcf-marginal.pdf}
%     \caption{Ten-minute averages of minute-by-minute quantiles.  Circles for observations, $\times$'s for simulated data with left column for Embedded generator and right column for Features generator.  Dark blue and black for 0.9 quantiles and light blue and gray for 0.5 quantiles.}
%     \label{fig:wsemcf}
% \end{figure}

\begin{figure}[!htbp]
\centering
\includegraphics[width=0.8\textwidth]{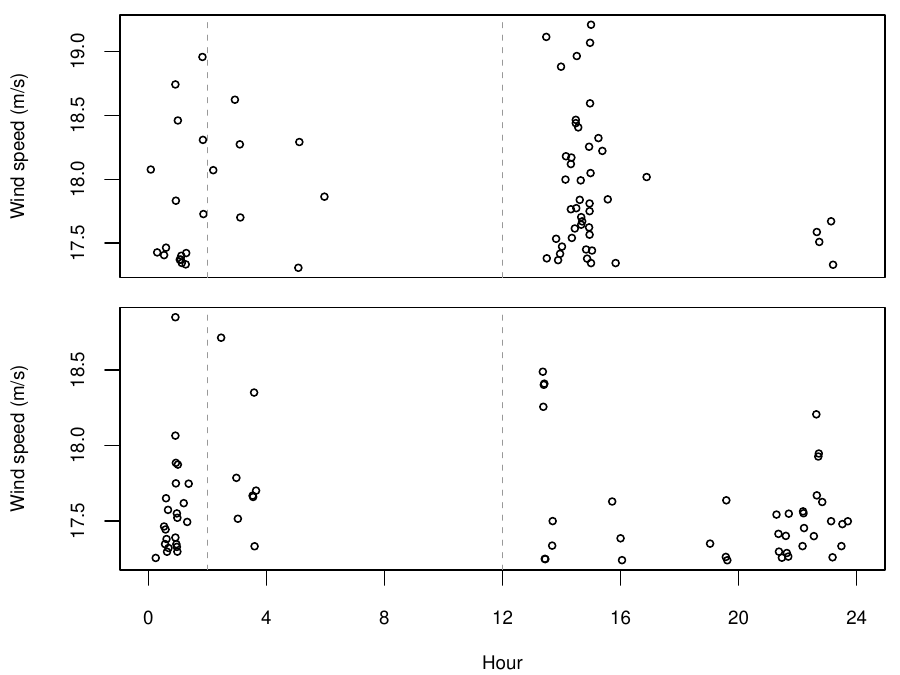}
\caption{Wind speeds above 0.9999 quantile for a simulation from Embedded (top) and Features (bottom) generators.  Compare to Figure~\ref{fig:ws-9999}.
}
\label{fig:ext-comp}
\end{figure}

Figure~\ref{fig:ext-comp} shows wind speeds above the 0.9999 quantile for one simulation each from the Embedded and Features generators.
We see that neither generator produces any wind speeds greater than 20 m s$^{-1}$, whereas the observations exceed that level 28 times.
The problem is partly due to the methodology and partly due to bad luck.
The bad luck is due to the fact that the day in 2011 with the 12 highest wind speeds, including all 10 wind speeds exceeding 23 m s$^{-1}$, contains a substantial stretch of missing data, and thus this day is excluded from the training data.
More consequentially, even across a large number of synthetic realizations, the maximum generated wind speed rarely exceeds the maximum observed in the training data. For example, there are only 3 times that the generated wind speed exceeds the maximum of the wind speed in training data (22.7 m s$^{-1}$) within 50,000 generated days using \textit{Weather as Embeddings}, and no exceedances using \textit{Weather as Features}.  This behavior can be further illustrated through high quantile analysis (e.g., 98\% or 99\% quantiles), where the upper-tail estimates of generated wind speeds remain systematically bounded by, and often less stable than, those of the observed data (results not shown). 
%Specifically, \textcolor{red}{MS: complete this sentence: what are the maximum wind speeds in the simulations?}

% \begin{figure}[!htbp]
% \centering
% \includegraphics[width=0.8\textwidth]{figures/graphical summaries/ext-comp.pdf}
% \caption{Wind speeds above 0.9999 quantile for a simulation from Embedded (top) and Features (bottom) generators.  Compare to Figure~\ref{fig:ws-9999}.
% }
% \label{fig:ext-comp}
% \end{figure}

\begin{figure}[!htbp]
\centering
\includegraphics[width=0.6\textwidth]{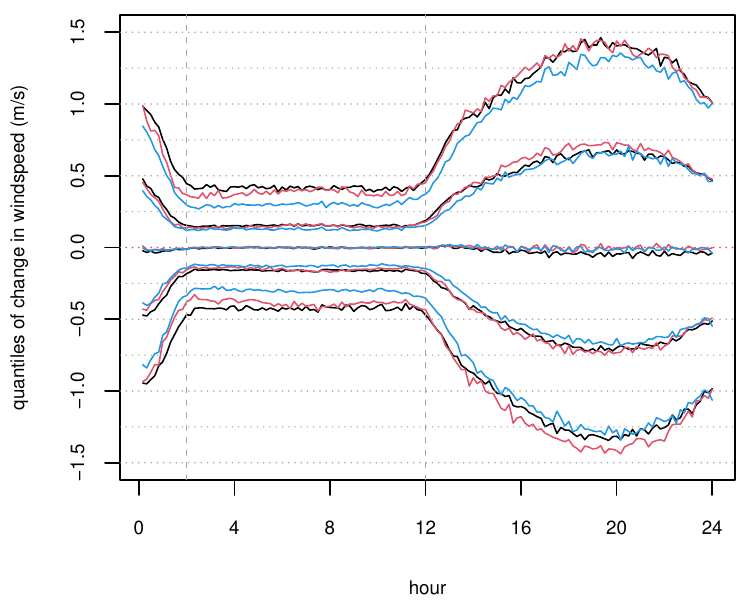}
\caption{Same as Figure~\ref{fig:ws-change} except now including results for a simulation from the Embedded (red) and Features (blue) generators.
}
\label{fig:ws-change-comp}
\end{figure}

Changes in the wind vector, while certainly not independent in time, are much closer to independent than the wind vectors themselves, so it is not critical to look at how properties in these changes vary across simulations for any given generator.
Figure~\ref{fig:ws-change-comp} compares the diurnal cycles for changes in wind speed for the observations and the Embedded and Features generators.
We see that the Embedded generator fits the pattern in the observations quite well, whereas the Features generator gets the magnitudes of the 0.1 and 0.9 quantiles substantially too small during the nighttime.

% \begin{figure}[!htbp]
% \centering
% \includegraphics[width=0.6\textwidth]{ws-change-comp.pdf}
% \caption{Same as Figure~\ref{fig:ws-change} except now including results for a simulation from the Embedded (red) and Features (blue) generators.
% }
% \label{fig:ws-change-comp}
% \end{figure}

Figure~\ref{fig:intro-vol} showed the densities of changes in the wind vector relative to the current wind direction for daytime and nighttime and three ranges of wind speed.
Figures~\ref{fig:obs-vcf} and~\ref{fig:obs-vem} in Appendix C compare similar results for simulations from the Features and Embedded simulations to the observational results.
Overall, both generators do a respectable job of reproducing the patterns in the observations.
However, neither generator captures the asymmetry of the observational results for the change along the current wind direction for daytime and wind speeds between 1 and 5 m s$^{-1}$.

In principle, a dominant southerly flow could at least in part explain the tendency for changes in nighttime wind vectors to be stronger parallel to the present wind direction than in the orthogonal direction.
However, Figure~\ref{fig:ew-ns-ch} shows that even when the wind vector is predominantly in an east-west direction, the changes in the observational wind vectors parallel to the current wind vector tend to be larger than in the orthogonal direction. 

Figures~\ref{fig:obs-vcf} and~\ref{fig:obs-vem} do not display the largest changes in the wind vector.
For the Euclidean norm of the changes in the wind vector, the Features generator performs fairly well, with 0.9999 quantiles of this norm of 6.39, 6.03, 5.79, 6.59 and 5.44 m s$^{-1}$ in five simulations compared to the observed value of 6.29 m s$^{-1}$.
In contrast, five simulations under the Embedded generator yield 0.9999 quantiles of 7.63, 8.21, 7.54, 7.31 and 7.58 m s$^{-1}$, which  are consistently too large.

%However, the absence of this effect during the daytime, during which a southerly flow is also evident, makes this explanation doubtful.
%For wind speeds greater than 7.5 m s$^{-1}$$ during the nighttime, 
%Figure \ref{fig:ew-ns-ch} separates out the changes in the wind vector into times when the wind vector is predominantly in a north-south direction and predominantly in an east-west direction and shows that the changes in the observational wind vectors parallel to the current wind vector tend to be larger in both cases.
%This result confirms the value of decomposing the changes in the wind vector into components parallel and orthogonal to the current wind vector.
%Both generators qualitatively capture the greater variation in the direction parallel to the current wind vector for both wind directions, although the Features generator somewhat underestimates the anisotropy for the east-west winds.
%It is interesting to note, though, that, when wind speed is greater than 7.5 m/s, the changes in the wind vector show greater variation when the current winds are predominantly along the east-west axis.

\section{Stochastic volatility}
\label{sec:Stochastic-volatility}

Another way to evaluate our stochastic wind generators is to compare the fits of parametric models to the observations with the fits of the same parametric models to simulated data.
A comparison of a parametric model fitted to real data and synthetic data can be used to assess the quality of a stochastic generator even if the parametric model is not correct.
In particular, if we use a synthetic dataset that has the same availability in time as the training data (the same number of years, days, and missingness pattern), then a good wind generator should yield similar summary statistics as the observational data when fitted to a misspecified parametric model.

There are many aspects of the wind vector time series that we could use in this fashion, but here we will just consider some models for the stochastic volatility noted in Figure~\ref{fig:intro-vol}, where the volatility of the wind vector depends on the present state of the winds even after controlling for time of day.
In particular, Figure~\ref{fig:intro-vol} shows that the volatility of the wind vector increases with the current wind speed.
However, the volatility of the wind vector depends not just on the time of day and the current wind speed, but also recent volatility of the wind vector.
We will examine this phenomenon by fitting quantile regressions \citep{koenker2005, koenker2017} to the absolute change in wind speeds during nighttime.
We restrict to nighttime hours because the variations in volatility are greater then.
We use quantile regressions because of the heavy upper tail in the distribution of changes in wind speed.
Within this restricted time window, we do not let the model depend on the time of night even though there is some evidence that the relationship between wind volatility and recent winds varies somewhat during the nighttime.
%In doing so, we are not assuming that the volatility process is the same throughout the night, only that giving results averaged over nighttime provides a useful summary of the process and a valid tool for assessing the realism of our stochastic wind generators by comparing fitted coefficients to the quantile regressions based on observations and simulations.

Writing $s_t = \|X_t\|$ for the wind speed at minute $t$, we consider predicting the absolute change in wind speed, $d_{t+1} = |s_{t+1}-s_t|$, using various functions of $\{X_l\}_{l\le t}$. The quantile regression objective is to find $\beta$ which minimizes: $\sum_{t\in \mathcal{T}} \rho_{\tau}(d_t - \beta^\top f(\{X_l\}_{l\le t})),$ where $\rho_\tau(u) = u(\tau - 1_{\{u < 0\} })$ and $\mathcal{T}$ is the set of time indices where all values $\{X_{t-9}, \dots, X_t\}$ are observed. Call the minimized value of the criterion function $C_\tau$. Note that values of $X_{t+1}-X_t$ are unreliable when $s_t$ is small; consequently, times $t$ when $s_t \le 1$ m s$^{-1}$ are omitted from these analyses. The regressions were fitted using the R package \texttt{quantreg} \citep{quantreg2025}.

Table~\ref{tab:rq.vol} reports $C_\tau$ values for the various sets of covariates and $\tau$ equal to 0.5, 0.75 and 0.9. Let us first focus on those models whose covariates are functions of the three most recent wind vectors. For the three values of $\tau$, when comparing using no covariates to using $s_t$, the relative reduction of $C_\tau$ increases substantially with $\tau$ (from 11.6\% for $\tau=0.5$ to 26.1\% for $\tau=0.9$). This pattern of greater relative reductions in $C_\tau$ as $\tau$ increases holds for every model considered in Table \ref{tab:rq.vol} and indicates that varying volatility has its largest impact on more extreme quantiles. Adding $s_{t-1}$ as a second covariate results in negligible further reductions in $C_\tau$. In contrast, adding $|s_t-s_{t-1}|$ as a second covariate substantially improves the fit for all three values of $\tau$. Interestingly, adding $\|X_t-X_{t-1}\|$ as the second covariate improves the fit modestly more in all three cases. The statistical significance of these greater improvements can be demonstrated by fitting the quantile regressions separately for each of the 23 years in the training data. For all three $\tau$ values and all 23 years, the value of $C_\tau$ is smaller when using $\|X_t-X_{t-1}\|$ as the second covariate than when using $|s_t-s_{t-1}|$. Thus, even if one were only interested in wind speed, there can be advantages in using the wind vector information.

Adding values of either the absolute change in wind speed or the norm of the change in the wind vector from further in the past noticeably reduces $C_\tau$ for all three values of $\tau$. This phenomenon is illustrated in Table \ref{tab:rq.vol} by adding absolute changes in wind speed or in the norm of the change in wind vector for eight additional minutes. Again, the fits are moderately better when one uses the norm of the change in the wind vector than the absolute change in wind speed. The last row of Table \ref{tab:rq.vol} gives results when only using $\|X_l-X_{l-1}\|$ for $l=t-8,\ldots,t$ and not $s_t$. We see that excluding $s_t$ modestly degrades the quality of fit. Thus, it appears that the current wind speed provides information about future volatility that is not contained in measures of recent volatility. 
%Is there something regarding the comparison to stochastic volatility models in finance?

\begin{table}
 \caption{Minimized criterion function values for quantile regressions of change in wind speed. Relative (to model with just intercept) reductions in $C_\tau$ in parentheses as percentages. All models include an intercept.}
\label{tab:rq.vol}
\begin{center}
\begin{tabular}{ c c c c}
& \multicolumn{3}{c}{$\tau$} \\
%\vspace{0.05cm}
 \cline{2-4} 
 \vspace{0.1cm}
regressors & 0.5 & 0.75 & 0.9 \\
\hline
intercept & 29043 & 31777 & 23325 \\
$s_{t}$ & 25687(11.6) & 25451(19.9) & 17228(26.1) \\
\vspace{0.1cm}
$s_{t}, s_{t-1}$ & 25681(11.6) & 25432(20.0) & 17196(26.3) \\
\vspace{0.1cm}
$s_{t}, |s_{t}-s_{t-1}|$ &  24596(15.3) & 23659(25.8) & 15586(33.2) \\
\vspace{0.1cm}
$s_{t}, \|X_{t}-X_{t-1}\|$& 24277(16.4) & 23042(27.5) & 14912(36.1) \\
$s_{t}, |s_{t}-s_{t-1}|$, &  23718(18.3) & 22035(30.7) & 13878(40.5) \\
\vspace{0.1cm}
$\cdots, |s_{t-8}-s_{t-9}|$ & & & \\
$s_{t}, \|X_{t}-X_{t-1}\|,$&  23596(18.7) & 21757(31.5) & 13517(42.0) \\
\vspace{0.1cm}
$\cdots, \|X_{t-8}-X_{t-9}\|$& & & \\
$\|X_{t}-X_{t-1}\|$,&  23839(17.9) & 22180(30.2) & 13877(40.5) \\
$\cdots, \|X_{t-8}-X_{t-9}\|$& & & \\
 \end{tabular}
 \end{center}
 \vspace{8pt}
\end{table}

\begin{figure}[!htbp]
    \centering
    \includegraphics[width=1.0\linewidth]{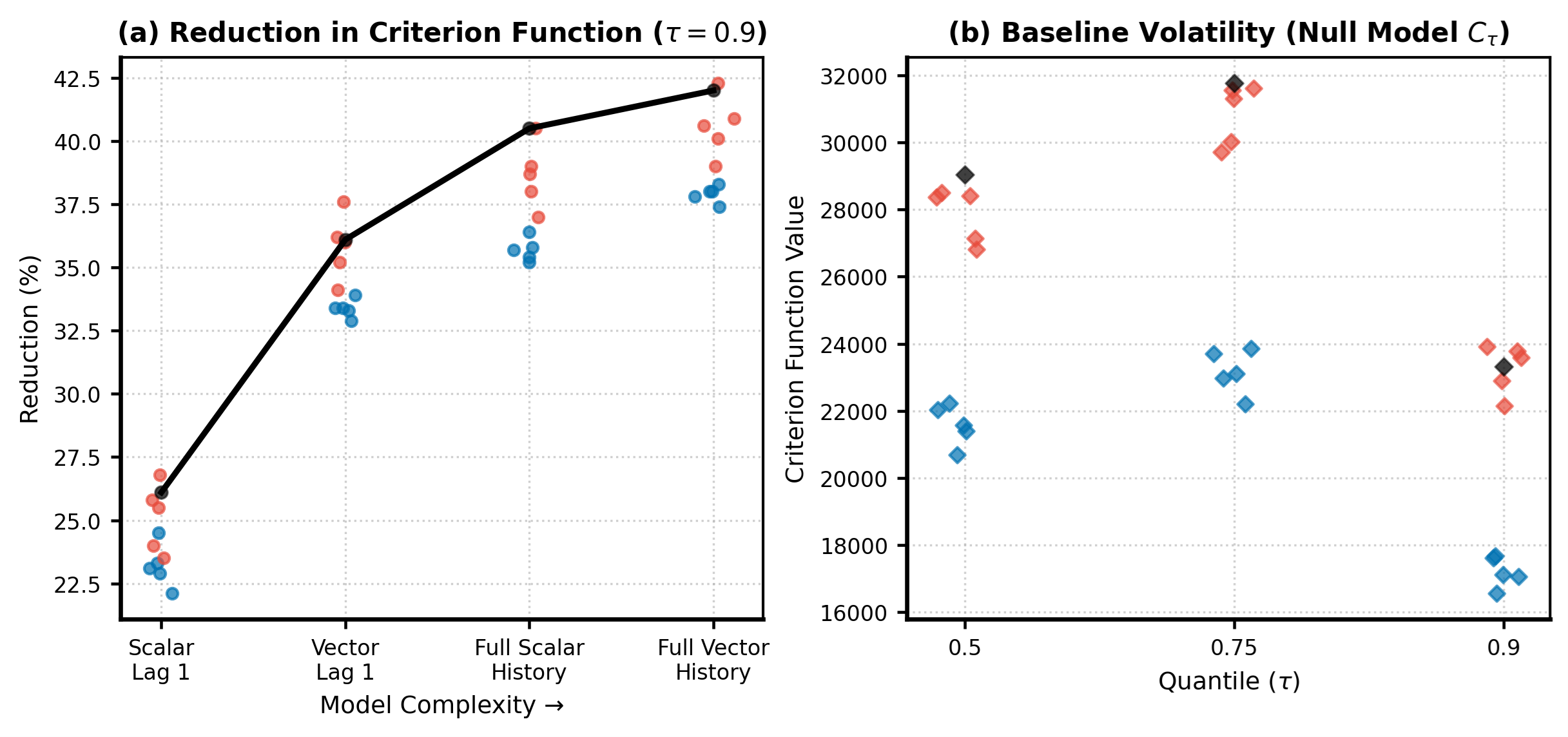}
    \caption{Quantile regression validation of stochastic volatility dynamics at $\tau=0.9$. 
\textbf{(a)} Relative reduction in the minimized criterion value $C_{0.9}$ as predictive complexity increases from scalar lag-1 to full vector history. 
\textbf{(b)} Baseline volatility under the null (intercept-only) model, quantified by $C_{0.9}$. 
Black markers denote observed data; red markers denote the Weather as Embeddings generator; blue markers denote the Weather as Features generator ($n=5$ runs each). 
Synthetic replicate points are jittered slightly in the horizontal direction to reduce overlap (no jitter is applied to the observed series). 
The Embeddings generator closely tracks the observational record in both structure and scale, while the Features generator systematically underestimates volatility.}
    \label{fig:quantile_validation}
\end{figure}

To validate the realism of our stochastic generators, we apply this same quantile regression framework to the simulated time series. This allows us to directly compare the learned volatility structure against the empirical benchmark. We focus our analysis on upper tail of the volatility distribution ($\tau=0.9$), where heteroskedastic effects are most pronounced and are more critical for energy applications. For each generator, Figure~\ref{fig:quantile_validation} presents the results across 5 independent simulation runs of 23 years, each with 21 days and with minutes corresponding to missing minutes in the observational record removed.

% \begin{figure}[!htbp]
%     \centering
%     \includegraphics[width=1.0\linewidth]{figures/stochastic volatility/quantile.png}
%     \caption{Quantile regression validation of stochastic volatility dynamics at $\tau=0.9$. 
% \textbf{(a)} Relative reduction in the minimized criterion value $C_{0.9}$ as predictive complexity increases from scalar lag-1 to full vector history. 
% \textbf{(b)} Baseline volatility under the null (intercept-only) model, quantified by $C_{0.9}$. 
% Black markers denote observed data; red markers denote the Weather as Embeddings generator; blue markers denote the Weather as Features generator ($n=5$ runs each). 
% Synthetic replicate points are jittered slightly in the horizontal direction to reduce overlap (no jitter is applied to the observed series). 
% The Embeddings generator closely tracks the observational record in both structure and scale, while the Features generator systematically underestimates volatility.}
%     \label{fig:quantile_validation}
% \end{figure}

Panel~(a) compares the relative reduction in the criterion function $C_{0.9}$ as predictive complexity increases. The Weather as Embeddings generator (red) replicates the empirical trajectory almost exactly, achieving a 40.6\% mean reduction for the full vector history model compared to 42.0\% in the observed data—a gap of only 1.4 percentage points. In contrast, the Weather as Features generator (blue) produces values of $C_{0.9}$ that are systematically smaller, reaching only a 37.9\% reduction and leaving a persistent 4.1 percentage point gap. 
%This indicates that conditioning on raw meteorological features is insufficient for the model to learn the complex, state-dependent volatility structure present in real wind data.

Panel (b) reveals a more fundamental discrepancy in the scale of generated volatility. The intercept-only model measures baseline variance—the intrinsic ``noise floor'' of minute-to-minute wind fluctuations before any predictive structure is exploited. The Weather as Features generator produces a baseline criterion of $C_{0.9} = 17{,}200$, roughly 26\% below the observed value of $C_{0.9} = 23{,}325$. This confirms that the Features generator suffers from systematic over-smoothing, producing synthetic time series that lack the natural magnitude of high-frequency turbulence. The Weather as Embeddings generator, by contrast, achieves $C_{0.9} = 23{,}264$—within 0.3\% of the observational benchmark—demonstrating that learned embeddings successfully preserve both the structure and intensity of stochastic volatility.

These results highlight that matching marginal distributions of wind vectors and their first differences is not sufficient for evaluating a high-frequency stochastic generator.  A realistic generator should also reproduce the \emph{conditional} variability of the wind process---both its overall scale and its dependence on the current wind state and recent history.  The quantile-regression diagnostic provides a simple and interpretable way to assess this aspect of realism, and in our experiments it clearly distinguishes between generators that preserve minute-scale dependence and those that tend to over-smooth the dynamics.

\FloatBarrier

\section{Discussion}
\label{sec:Discussion}

A stochastic weather generator should ideally provide realistic output that could be used as a substitute for actual observations in a broad range of applications, including applications not anticipated by the developers of the weather generator.
Thus, in evaluating the quality of such a generator, it is important to consider a diverse set of assessment tools that explore various aspects of the generator's output.

We have applied a number of approaches to assessing a range of machine-learning based stochastic wind generators, some numerical but many graphical.
We have found the graphical assessments very helpful in identifying features in the observations that standard time series models would struggle to capture.
Many of these features are accurately captured by our best stochastic generator, but others are not.
In particular, all of our approaches are nonparametric and are thus understandably unable to reproduce extreme winds accurately.

If one were interested in a stochastic wind generator that could reproduce both extreme winds and accurate temporal dependence across many days, then it would be advisable to use a broader range of data, possibly from other sites and/or from different times of year than the Lamont site in June.
Even with more data, changes in our methodology would be needed to generate realistic extreme behavior.
A well-known limitation of many generative models is that their outputs tend to concentrate within the support of the training data, making it difficult to generate samples that exceed historically observed extremes. 
%In our setting, this phenomenon is evident when examining the distribution of generated wind speeds: even across a large number of synthetic realizations, the maximum generated wind speed rarely exceeds the maximum observed in the training data. For examples, there are only 3 times that the generated wind speed goes beyond the maximum of the wind speed in training data (22.7 m s$^{-1}$) within 50,000 generated days using \textit{Weather as Embeddings}, and never happens using \textit{Weather as Features} .  This behavior can be further illustrated through high quantile analysis (e.g., 98\% or 99\% quantiles), where the upper-tail estimates of generated wind speeds remain systematically bounded by, and often less stable than, those of the observed data. 
%This limitation is particularly concerning for applications involving risk assessment and extreme event analysis. For example, when generating wind trajectories for a future year, it is unrealistic that repeated simulations almost never produce wind speeds exceeding historical maxima, as real-world extreme events are inherently stochastic and may surpass past observations. Moreover, under global warming, extreme wind events are expected to become more frequent and severe over time. Therefore, 
Developing generative models that can accurately capture and extrapolate tail behavior beyond the observed data represents a critical and challenging direction for future research. One possible direction is to design more robust and task-specific loss functions. In this paper, we use standard loss functions commonly adopted in VAE and transformer models, which mainly focus on overall reconstruction accuracy. While these losses capture general patterns in the data, they may not adequately reflect extreme values or short-term variability in wind dynamics. In particular, we observe that the current loss formulation does not explicitly account for extreme wind speeds or rapid temporal changes. This may contribute to the difficulty in reproducing high quantile behavior and fine-scale fluctuations in the generated sequences. These observations suggest that the choice of loss function plays an important role in modeling both the distributional properties and the temporal dynamics of wind data, and deserves further investigation.

Given that our data is at the minute scale, we have naturally focused on dependence on shorter time scales.
In particular, by using measurements taken every minute, we have been able to see patterns in the data that would be much harder to detect with measurements taken even every 10 minutes, which would often be considered high-frequency.
In particular, we find clear evidence of stochastic volatility in the wind vectors and that this volatility process has a diurnal cycle.
We believe there is much more that could be learned exploring this volatility process, but we hope that the findings here could be of some meteorological interest beyond their role in assessing our stochastic wind generator.
%We have investigated a number of machine learning approaches to synthesizing high frequency wind vector time series at a single site and evaluated them using a number

Furthermore, our method for generating one day of winds conditional on the previous day's winds does not adequately capture the transition from one day to the next.
Another aspect that deserves further attention is the trade-off between temporal continuity and diversity in the generated sequences. In the current framework, the continuity between consecutive simulated days depends on how much information from the previous day is used as conditioning input. For example, conditioning on a longer window (e.g., the final 240 minutes of the previous day) generally improves the smooth transition between days, while conditioning on a shorter window (e.g., the final 60 minutes) may lead to less consistent transitions. However, increasing the conditioning length also reduces the diversity of the generated sequences. When a large portion of the previous day is fixed, the generated next day becomes more constrained, and repeated simulations tend to produce more similar outcomes. This highlights an inherent trade-off between maintaining temporal continuity and preserving variability across simulated trajectories, which is an important consideration for future work.

A further problem with our approach to modeling dependence across days is that it has no mechanism for capturing the kind of recurring patterns across multiple days shown in Figure~\ref{fig:three-yearsEN} for 2020.
Perhaps a better model for weather states could help with this problem.
\citet{bessac2016} study a range of more sophisticated models for weather states in the analysis of wind vectors than we have used here, but all of them treat the regime as a discrete Markov chain, which would likely also struggle with reproducing patterns like we see in 2020 for which Figure~\ref{fig:three-yearsEN} shows there is a fairly strong repetitive diurnal pattern early in the month, a strong break in this pattern around June 12 and then another strong repeated diurnal pattern for around seven days that is somewhat similar to the pattern earlier in the month.
Our machine learning approach does not assume the regime is Markov, but we would need to find an effective way to condition on more than just the last hour of the previous day to capture this kind of pattern.

Beyond the current generative framework, several promising directions for future work naturally arise. One particularly compelling extension is to develop predictive models for wind or weather dynamics based on the proposed stochastic generator. Given a well-trained generative model, it is natural to leverage its learned temporal structure for forecasting tasks. In particular, the prior learning stage in our framework is built upon a transformer architecture, which is well-suited for sequence modeling and time series prediction. 

However, substantial methodological challenges remain in transitioning from stochastic generation to conditional prediction. For example, constructing valid joint prediction intervals for multivariate wind vectors is nontrivial and requires careful treatment of temporal and cross-variable dependencies. In addition, improving prediction flexibility is essential: unlike the current generation setting, which assumes a fixed starting time (e.g., midnight), a practical forecasting model should be capable of producing predictions conditioned on observations from arbitrary time points.

Other areas for future work include further development of our assessment efforts.
In particular, the fact that our classifier of real versus synthetic wind data worked better after the data was processed through a high-pass filter indicates room for improvement in this classifier.
It would also be valuable to have a better understanding why the \textit{Weather as Embeddings} generator overall works better than the \textit{Weather as Features} approach.
We suspect that embedding discrete weather states provides a more effective conditioning signal for learning nonlinear interactions between atmospheric regimes and minute-scale turbulence, whereas using raw features yields over-smoothed simulations with systematically reduced volatility.

For many applications, including wind energy and wildfire spread, one would often be interested in generating wind vector time series simultaneously at multiple sites.
The machine-learning approaches used here might extend reasonably well to a small number of locations, but we believe generators for a large number of locations should directly incorporate physical knowledge of meteorology.
However, even if one were interested in only a modest number of locations, which could be the case in some wind energy applications, developing appropriate and informative graphical diagnostics for dependencies in the data is much more challenging than when considering a single site.
Despite this difficulty, we believe that innovative and well-chosen graphical tools for exploring space-time dependencies will be essential for the proper evaluation of stochastic wind generators. 

A more fundamental constraint of extending our approach to other locations is that high-quality wind vector measurements at the minute frequency are not routinely collected.
The Advanced Radiation Measurement facility in the Southern Great Plains has collected similar data at other sites in the region over the years, but as of February 2026, such data is only collected at five other sites in the region.
At least for winds, we maintain that there is much to be learned by observing winds every minute or perhaps even more frequently and we strongly advocate for maintaining and expanding these facilities.

% Appendix here-----------------------------------------------------------------------------------
\appendix
\section{Data issues}
\label{supsec:data}

\subsection{Zero wind speeds}
\label{supsec:zeroes}

Throughout June for all reported years (1994-2025), there were 1848 minutes with recorded wind speed of 0 m s$^{-1}$. Interestingly, there are also 401 total minutes in this time frame for which temperature and relative humidity are both missing, and in all but one of them (June 30, 1995 21:27), the wind speed is reported as 0 m s$^{-1}$. In addition, wind speeds just before and after such stretches of missing temperature and relative humidity were generally not close to 0 m s$^{-1}$, leading us to treat all such recorded zeroes as missing values. 
%We could not see any obvious problem with the other observed wind speeds of 0 m s$^{-1}$ and left them in the database.
Another anomaly in the data is that all of the wind speeds of 0 m s$^{-1}$ in the data we used occurred after 2006, so there must have been some change in either the instrumentation or the recording of very low wind speeds that changed after 2006, although we could not find any documentation to this effect.
Because the wind speeds immediately preceding and following  sequences of zeroes not associated with missing temperature readings were mostly under 0.2 m s$^{-1}$ (and always less than 2 m s$^{-1}$), we chose to retain those recorded 0 m s$^{-1}$ wind speeds in our dataset.

%we refer to figure \ref{M-fig:seasonality diurnal plot}

%\textcolor{red}{MS: Justin, please update these numbers as they are based just on June 1-21 data and not including 2025. Maybe switch to all of June and include all years.}
%One odd feature in the recorded data is that there is not a single recorded wind speed of 0 through 2006 and then 1183 wind speeds of 0 in the subsequent years.
%For 354 of the minutes with 0 reported wind speeds, the temperature and relative humidity are listed as missing and there are no other missing values for these variables during this time frame.
%We found this coincidence disturbing and chose to treat the wind speeds for these minutes as also missing.
%We could not say any obvious problem with the other observed wind speeds of 0 and left them in the database, even though clearly there must have been some change in instrumentation or recording that led to no zeroes through 2006 and so many afterwards.

\subsection{Filling in missing data}
\label{app:filling}

As noted in Sect.~\ref{sec:Data}, for the purposes of expanding the number of training days only, we impute gaps in the observational record of at most 20 minutes for days during the training period (June 1--21 for the years 1998--2020).
We fill in the gaps by acting as if the wind vector process during a gap is made up of two independent and identically distributed Brownian bridge processes that are tied down to match the last observed wind vector before the gap and the first observed wind vector after the gap.
The only parameter that needs to be specified to carry out the imputation is the common scale parameter $\sigma^2$ for the Brownian bridges.
Because the volatility of the wind vector is constantly changing, we estimate $\sigma^2$ separately for each gap using the 15 minutes before and after the gap.
Write $j_0$ for the last minute before a gap and $j_1$ for the first minute after a gap and $X_j$ for the wind vector at minute $j$.
Then define
\begin{equation}
    \hat\sigma^2 = \frac{1}{60}\bigg[\sum_{j = j_0-14}^{j_0} ||X_j - X_{j-1}||^2 + \sum_{j = j_1}^{j_1+14} ||X_{j+1} - X_{j} ||^2  \bigg].
\end{equation}
Let $\bm{Y}_1,\ldots,\bm{Y}_{j_1-j_0}$ be independent and identically distributed bivariate normal random vectors with common mean 0 and common covariance matrix $\hat\sigma^2 I_2$.
Then for $j_0  < j < j_1$ and $\bar{\bm{Y}} = (\bm{Y}_1+\cdots + \bm{Y}_{j_1-j_0})/(j_1-j_0)$, we impute
\begin{equation}
    X_j = X_{j_0} + \frac{j-j_0}{j_1-j_0}(X_{j_1}-X_{j_0}) + \sum_{k=1}^{j-j_0} (\bm{Y}_k-\bar{\bm{Y}}).
\end{equation}
Undoubtedly one could develop a more sophisticated and realistic imputation scheme, but given that there are only 16 such gaps in the training days for a total of 48 minutes, we expect that the effect on our results would be negligible.

\begin{figure}
    \centering
    \includegraphics[width=0.82\linewidth]{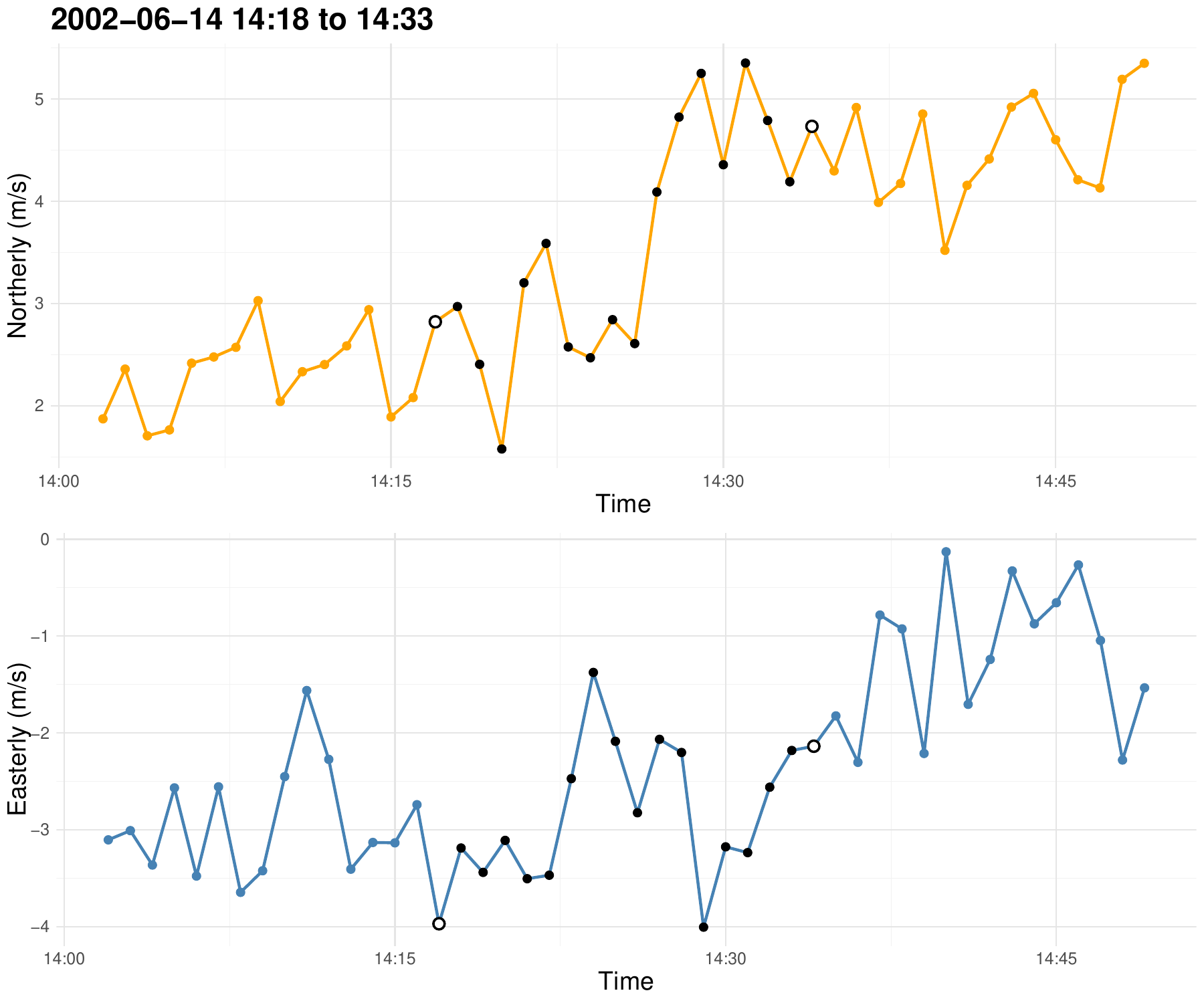}
    \caption{Example of the Brownian bridge imputation method for our longest imputed gap of sixteen minutes. The black points represent the imputed minutes, open circles are the last minute before the gap ($j_0$) and first minute after the gap ($j_1$), colorful points are the additional minutes surrounding the gap used to estimate local variability. }
    \label{fig:impute_example}
\end{figure}

Figure~\ref{fig:impute_example} shows an example of this imputation procedure.
During the 16-minute period between the open circles on the plot, recorded wind speeds were 0 m s$^{-1}$ and relative humidity and temperature were missing.
Given the wind speeds in the surrounding minutes, it is evident that these zeroes are erroneous.
The black points show the imputed wind vectors we used based on the Brownian bridge approach described here.

% \begin{figure}
%     \centering
%     \includegraphics[width=0.82\linewidth]{figures/Imputation Plots/20020614_length_16.pdf}
%     \caption{Example of the Brownian bridge imputation method for our longest imputed gap of sixteen minutes. The black points represent the imputed minutes, open circles are the last minute before the gap ($j_0$) and first minute after the gap ($j_1$), colorful points are the additional minutes surrounding the gap used to estimate local variability. }
%     \label{fig:impute_example}
% \end{figure}

\section{Ensuring Generators are not Memorizing Training Data} 
\label{app:min_dist_boxplots}

One possible failure mode for synthetically generated output is simply replicating the training samples without learning the underlying distributional rules. Before any evaluation of our synthetic results, we first ensure memorization is not a problem for output from each of our stochastic generators as described in Sect.\ \ref{Subsection:conditionalon-Weather-States}.

For each day $i$, we have $\bm{X}^{(i)} \in \mathbb{R}^{T \times P}$ where $T = 1440$ minutes and $P = 2$ are the easterly and northerly components of the wind vector. We reshape this matrix into a vector by concatenating all columns:
    $\bm{v}^{(i)} = \text{vec}(\bm{X}^{(i)}) \in \mathbb{R}^{TP}$,
where $\text{vec}(\cdot)$ stacks the columns of $\bm{X}^{(i)}$ into a single vector of dimension $TP = 2880$. Let $V_{\text{train}}$ and $V_{\text{synth}}$ denote the sets of vectorized training and synthetic days respectively. For any day vector $\bm{v}$ and set $V$, the minimum Euclidean distance is
    $d_{\min}(\bm{v}, V) = \min_{\bm{u} \in V} || \bm{v} - \bm{u} ||_2.$
This metric quantifies how close $\bm{v}$ is to its nearest neighbor in $V$. 

% \begin{figure}
%     \centering
%     \includegraphics[width=0.85\linewidth]{figures/Euclidean Distance Between Days/min_distance_combined.pdf}
%     \caption{The distribution of Euclidean distance from each training day (474 days) to its closest
% other training (473) and each set of synthetic (474) days. All box plots show 474 minimum distances, one for each training day.}
%     \label{fig:min_dist_boxplots}
% \end{figure}

\begin{figure}
    \centering
    \includegraphics[width=0.85\linewidth]{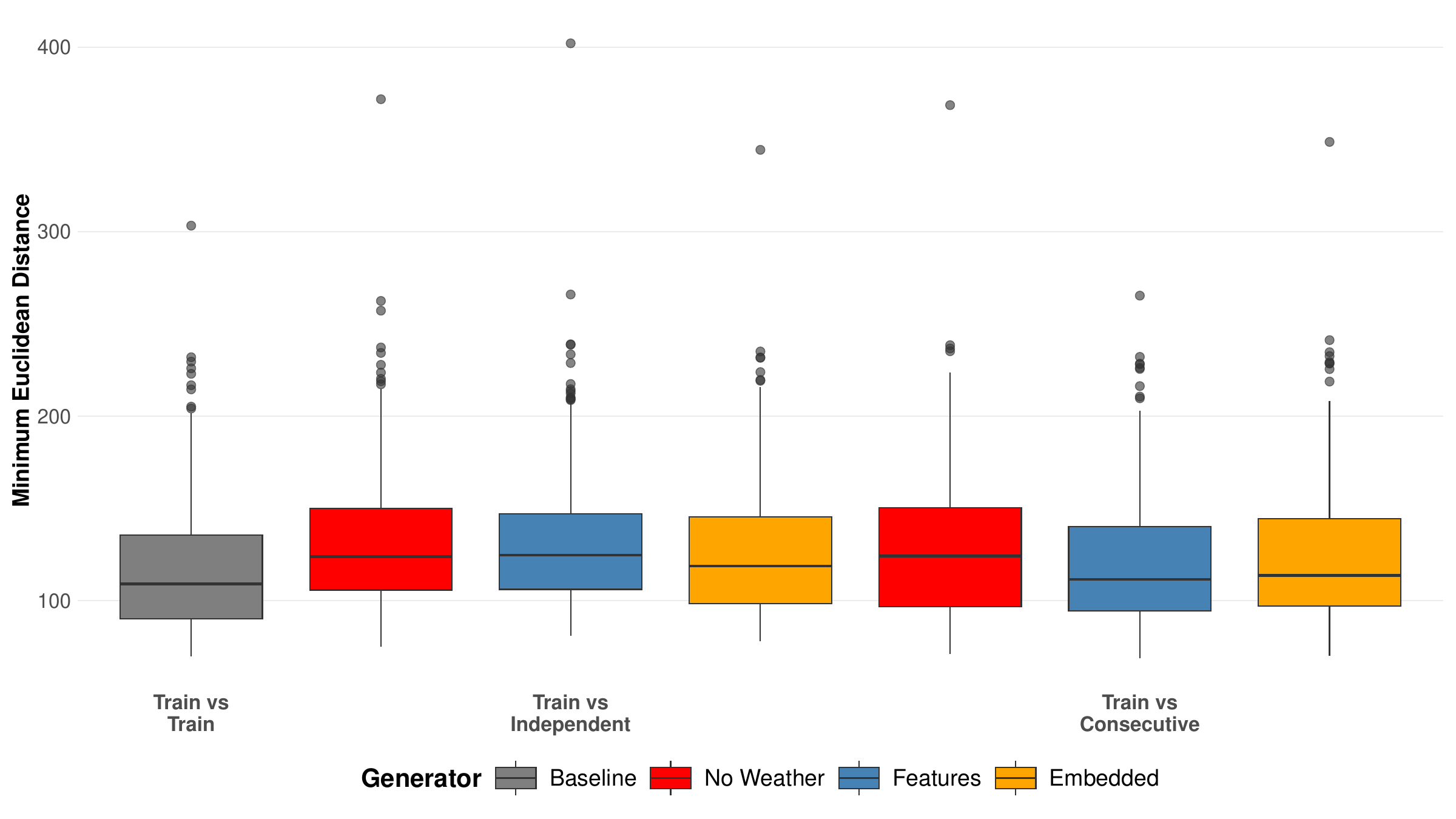}
    \caption{The distribution of Euclidean distance from each training day (474 days) to its closest
other training (473) and each set of synthetic (474) days. All box plots show 474 minimum distances, one for each training day.}
    \label{fig:min_dist_boxplots}
\end{figure}

Figure~\ref{fig:min_dist_boxplots} shows the distribution of how close each training day is from different sets of days. For each training day $i=1,\dots, 474$, we compute $d(\bm{v}_{\text{train}}^{(i)}, V)$ where $V$ is either the other 473 training days or 474 synthetic days from each generator. For synthetic days, we match the missing pattern of the training data by removing the fully missing days, but to maintain as many full days as possible do not remove imputed minutes. Each box plot in Figure \ref{fig:min_dist_boxplots} shows the distribution of $d_{\min}(\bm{v}_{\text{train}}^{(i)}, V)$ for $i=1,\dots, 474$ and each set of days $V$. 

The key diagnostic is comparing the synthetic distances to the baseline training distances. If the generators merely copied the training data, synthetic distances would cluster near zero and be well below the ``Train vs Train'' baseline. Instead, all six synthetic distributions show slightly greater distances compared to the real data baselines. 

\begin{figure}[!htbp]
    \centering
    \includegraphics[width=0.75\linewidth]{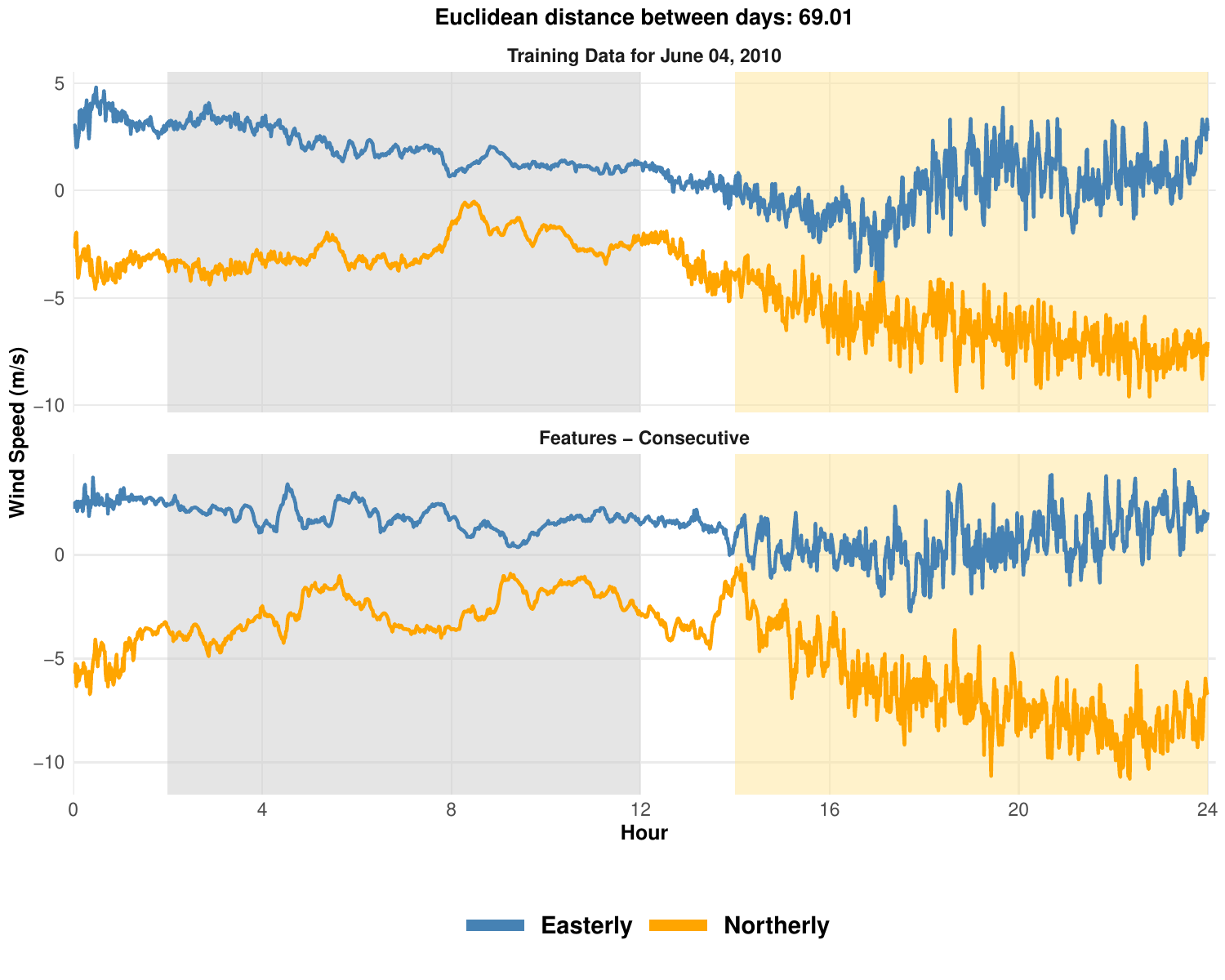}
    \caption{Time series of Easterly and Northerly components of wind vector for June 4, 2010 (top) and its nearest synthetic day across all generators (bottom). The synthetic day using the ``Features'' generator with independent sampling. }
    \label{fig:min_euc_dist_pair_time_series}
\end{figure}

To further illustrate that the generators produce similar but distinct patterns rather than near exact reproductions, Figure~\ref{fig:min_euc_dist_pair_time_series} displays the closest training-synthetic pair across all generators. The top panel shows the training day (June 4, 2010), and the bottom panel its nearest neighbor from the synthetic data generated by the Features generator using independent sampling. Despite being the most similar pair we identified, the two days are clearly not identical. While both days exhibit similar overall patterns, there are clear differences. Most notably, the synthetic day shows wind speed spikes between hours 4-5 and 13-14 that are not present in the training day, along with variations in the timing and specific magnitudes of wind changes. These differences confirm that the generators produce novel, realistic patterns rather than memorized copies. 

% \begin{figure}[!htbp]
%     \centering
%     \includegraphics[width=0.75\linewidth]{figures/Euclidean Distance Between Days/Features/Train v Con/1_69.01.pdf}
%     \caption{Time series of Easterly and Northerly components of wind vector for June 4, 2010 (top) and its nearest synthetic day across all generators (bottom). The synthetic day using the ``Features'' generator with independent sampling. }
%     \label{fig:min_euc_dist_pair_time_series}
% \end{figure}

\section{Graphical comparisons of simulations and observations}

\begin{center}
\begin{figure}[!htbp]
\includegraphics[width=0.7\linewidth]{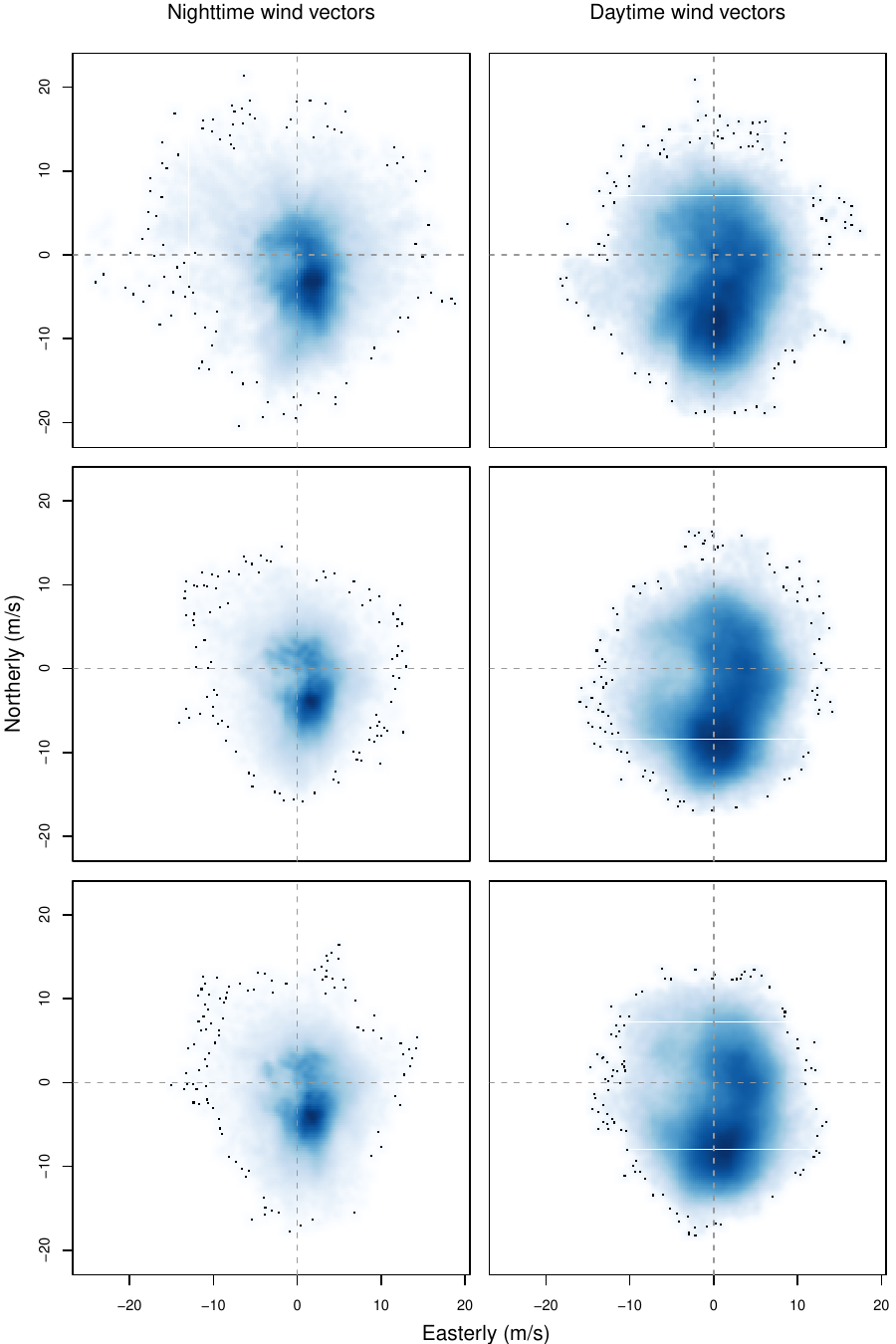}
\caption{Top row same as Figure \ref{fig:wind-scatter}.  Bottom two rows give similar density estimates for two simulations from the Features generator.
}
\label{fig:wind-scatter-cf}
\end{figure}
\end{center}

\begin{center}
\begin{figure}[!htbp]
\includegraphics[width=0.7\linewidth]{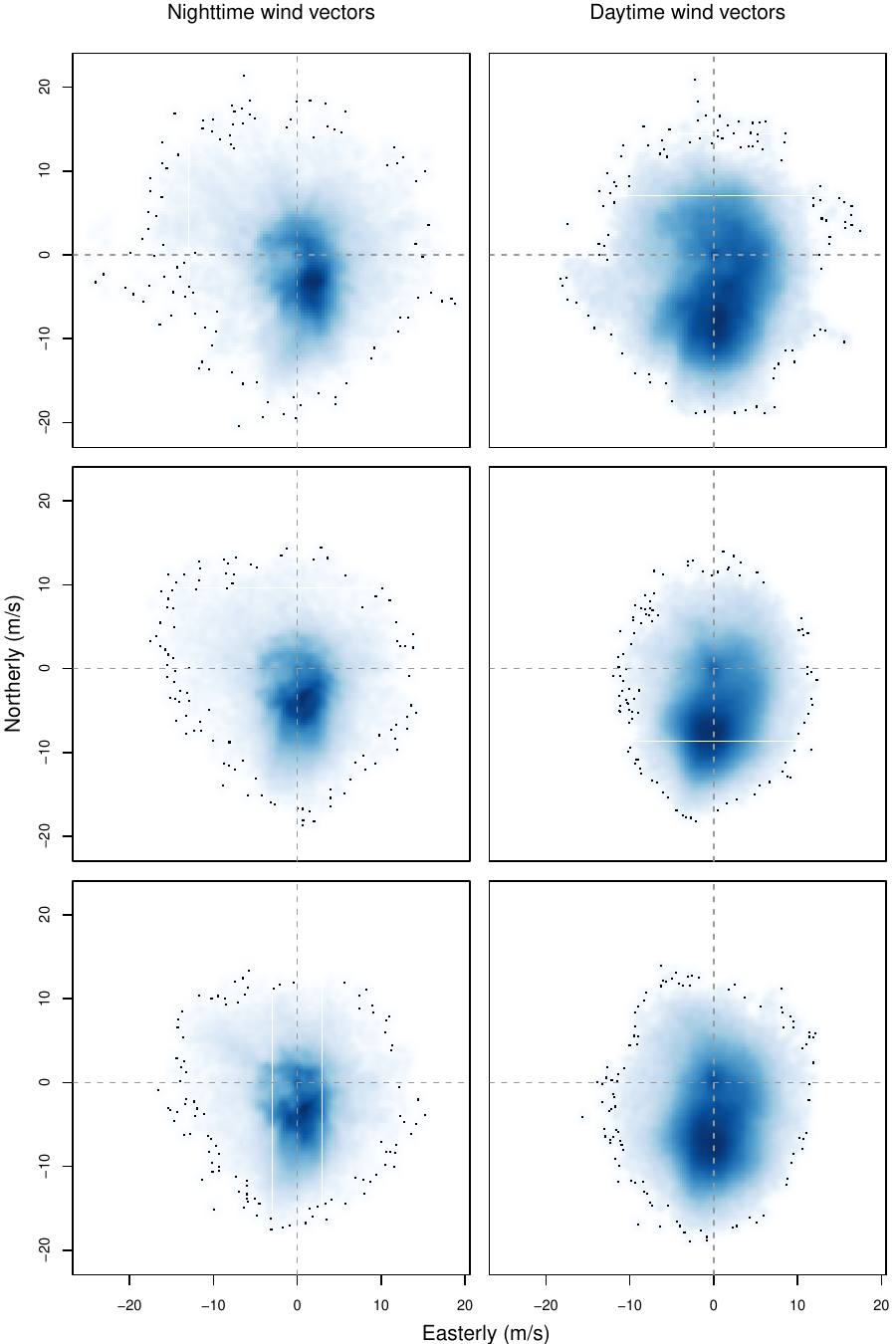}
\caption{Same as Figure \ref{fig:wind-scatter-cf} except for two simulations from the Embedded generator.
}
\label{fig:wind-scatter-em}
\end{figure}
\end{center}

We show results for two generators: weather states as features and consecutive days (hereafter ``Features") and embedded weather states and consecutive days (hereafter ``Embedded'').
Figure~\ref{fig:wind-scatter} showed marginal density estimates during daytime and nighttime for the training data and for one simulation from the Features generator and one simulation from the Embedded generator.
Figure~\ref{fig:wind-scatter-cf} shows similar marginal density estimates for two more simulations from the Features generator and
Figure~\ref{fig:wind-scatter-em} does the same for the Embedded generator.

\begin{center}
%\makebox[0pt]{[0pt][0pt]
\begin{figure}[!htbp]
\includegraphics[width=0.65\linewidth]{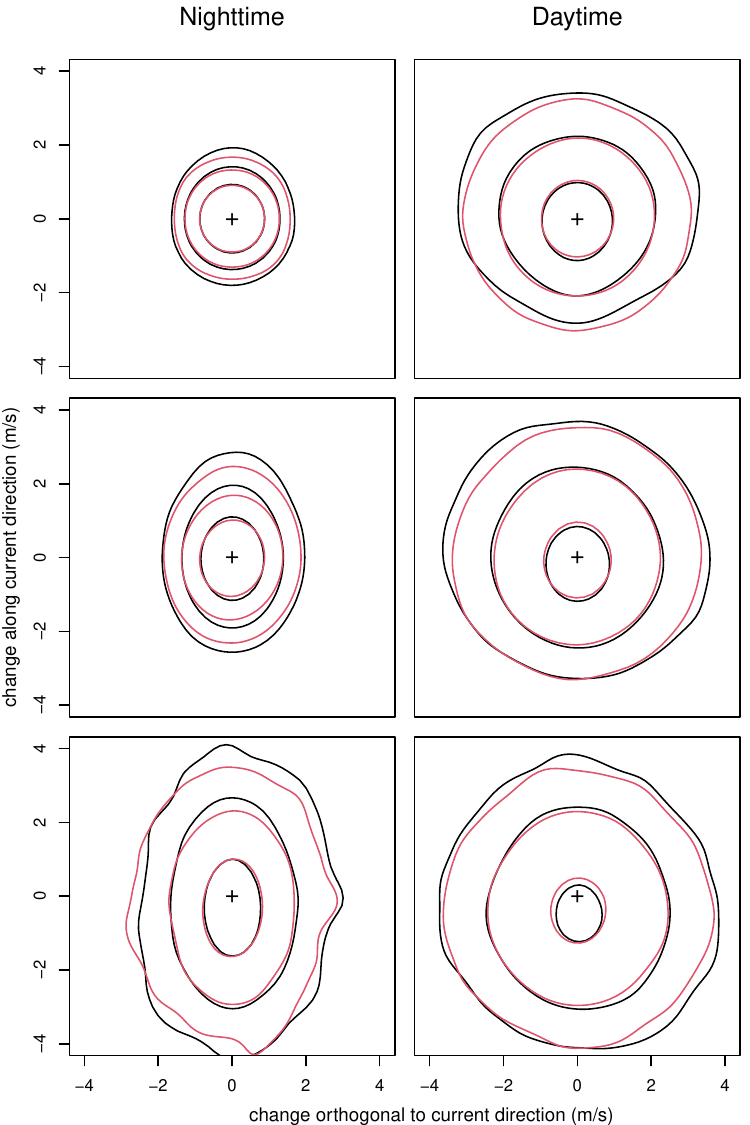}
\caption{Contour plots for densities of changes in wind vector relative to current wind vector. Top row for wind speed between 1 and 5 m s$^{-1}$, middle between 5 and 10 m s$^{-1}$ and
bottom greater than 10 m s$^{-1}$.  Black contours for observations, red for Features generator at levels $0.1,0.01$ and $0.001$.
}
\label{fig:obs-vcf}
\end{figure}

%\clearpage

\end{center}

Figure~\ref{fig:intro-vol} shows the densities of changes in the wind vector relative to the current wind direction for daytime and nighttime and three ranges of wind speed.
Figures~\ref{fig:obs-vcf} and~\ref{fig:obs-vem} compare similar results for simulations from the Features and Embedded simulations to the observational results.
Overall, both generators do a respectable job of reproducing the patterns in the observations.
However, neither generator captures the asymmetry in the observational results for the change along the current wind direction for daytime wind speeds between 1 and 5 m s$^{-1}$.

\begin{figure}[!htbp]
\includegraphics[width=0.9\linewidth]{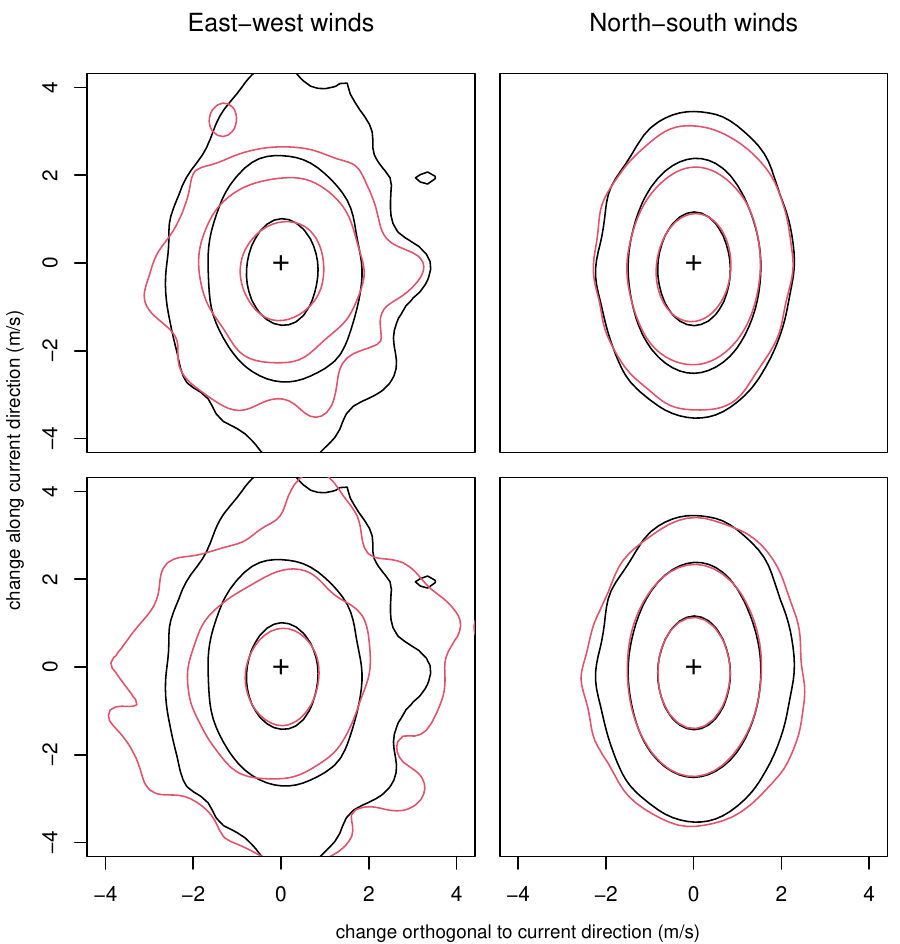}
\caption{Contour plots for nighttime densities of changes in wind vector relative to current wind vector for current wind speed greater than 7.5 m s$^{-1}$.  Left column for east-west winds (current wind direction within $45^\circ$ of east-west axis) and right column for north-south winds (current wind direction within $45^\circ$ of north-south axis). Top row for Features generator, bottom row for Embedded generator.  Colors and levels of contours as in Figure \ref{fig:obs-vcf}.
}
\label{fig:ew-ns-ch}
\end{figure}

%In principle, a dominant southerly flow could at least in part explain the tendency for changes in nighttime wind vectors to be stronger parallel to the present wind direction than in the orthogonal direction.
%However, the absence of this effect during the daytime, during which a southerly flow is also evident, makes this explanation doubtful.
For wind speeds greater than 7.5 m s$^{-1}$ during the nighttime, Figure~\ref{fig:ew-ns-ch} separates out the changes in the wind vector into times when the wind vector is predominantly in a north-south direction and predominantly in an east-west direction and shows that the changes in the observational wind vectors parallel to the current wind vector tend to be larger in both cases.
This result confirms the value of decomposing the changes in the wind vector into components parallel and orthogonal to the current wind vector.
Both generators qualitatively capture the greater variation in the direction parallel to the current wind vector for both wind directions, although the Features generator somewhat underestimates the anisotropy for the east-west winds.
It is interesting to note that when wind speed is greater than 7.5 m s$^{-1}$, the changes in the observed wind vectors show greater variation when the current winds are predominantly along the east-west axis.
The Embedded generator appears to capture this feature in the data better than the Features generator.

% \begin{center}
% \begin{figure}[!htbp]
% \includegraphics[width=0.7\linewidth]{figures/App graphical comparison/Weather States/wind-scatter-cf.pdf}
% \caption{Top row same as Figure \ref{fig:wind-scatter}.  Bottom two rows give similar density estimates for two simulations from the Features generator.
% }
% \label{fig:wind-scatter-cf}
% \end{figure}
% \end{center}

% \begin{center}
% \begin{figure}[!htbp]
% \includegraphics[width=0.7\linewidth]{figures/App graphical comparison/Weather States/wind-scatter-em.pdf}
% \caption{Same as Figure \ref{fig:wind-scatter-cf} except for two simulations from the Embedded generator.
% }
% \label{fig:wind-scatter-em}
% \end{figure}
% \end{center}

% \begin{center}
% %\makebox[0pt]{[0pt][0pt]
% \begin{figure}[!htbp]
% \includegraphics[width=0.65\linewidth]{figures/App graphical comparison/obs-vcf.pdf}
% \caption{Contour plots for densities of changes in wind vector relative to current wind vector. Top row for wind speed between 1 and 5 m s$^{-1}$, middle between 5 and 10 m s$^{-1}$ and
% bottom greater than 10 m s$^{-1}$.  Black contours for observations, red for Features generator at levels $0.1,0.01$ and $0.001$.
% }
% \label{fig:obs-vcf}
% \end{figure}

% %\clearpage

% \end{center}

% \begin{center}
% \begin{figure}[!htbp]
% \includegraphics[width=0.65\linewidth]{figures/App graphical comparison/obs-vem.pdf}
% \caption{Same as Figure \ref{fig:obs-vcf} except for two simulations from Embedded generator.
% }
% \label{fig:obs-vem}
% \end{figure}
% \vspace*{-20pt}
% \end{center}

\begin{center}
\begin{figure}[!htbp]
\includegraphics[width=0.65\linewidth]{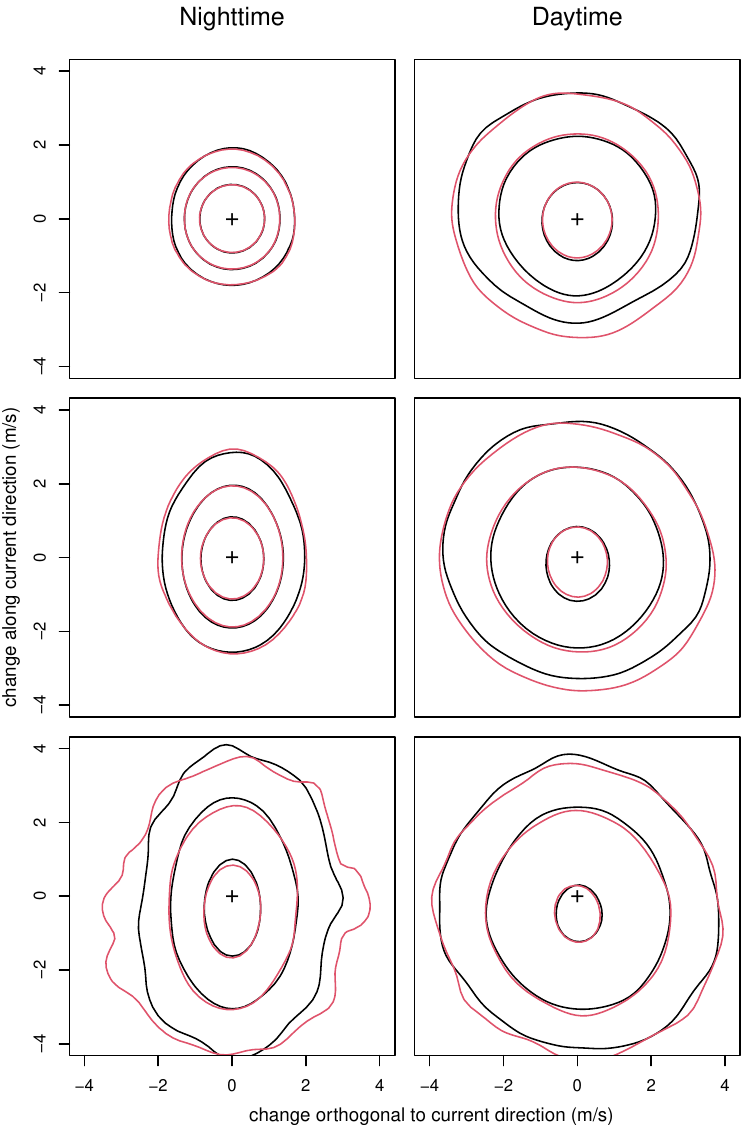}
\caption{Same as Figure \ref{fig:obs-vcf} except for two simulations from Embedded generator.
}
\label{fig:obs-vem}
\end{figure}
\vspace*{-20pt}
\end{center}

%\clearpage

% \begin{figure}[!htbp]
% \includegraphics[width=0.9\linewidth]{figures/App graphical comparison/ew-ns-ch.pdf}
% \caption{Contour plots for nighttime densities of changes in wind vector relative to current wind vector for current wind speed greater than 7.5 m s$^{-1}$.  Left column for east-west winds (current wind direction within $45^\circ$ of east-west axis) and right column for north-south winds (current wind direction within $45^\circ$ of north-south axis). Top row for Features generator, bottom row for Embedded generator.  Colors and levels of contours as in Figure \ref{fig:obs-vcf}.
% }
% \label{fig:ew-ns-ch}
% \end{figure}

%\section{Evaluation Experiments}

\FloatBarrier

\section{Additional quantile-regression diagnostics}
\label{app:qr-diag}

The quantile regressions follow the specification in Sect.~\ref{sec:Stochastic-volatility}.  In brief, we analyze nighttime minutes and omit times with $s_t\le 1$ m s$^{-1}$.  We only use times $t$ for which the full covariate history is observed (i.e., $X_{t-9},\ldots,X_{t+1}$ are all present), so that missingness does not induce differences in the set of regression rows across models.  For the synthetic experiments, we generate $n=5$ independent runs for each generator, and remove minutes corresponding to missing minutes in the observational record before fitting the regressions.  
%Horizontal jitter is used only for visualization of synthetic replicate points and does not affect any computed values of $C_\tau$ or reported reductions.

\begin{figure}[!htbp]
    \centering
    \includegraphics[width=1.0\linewidth]{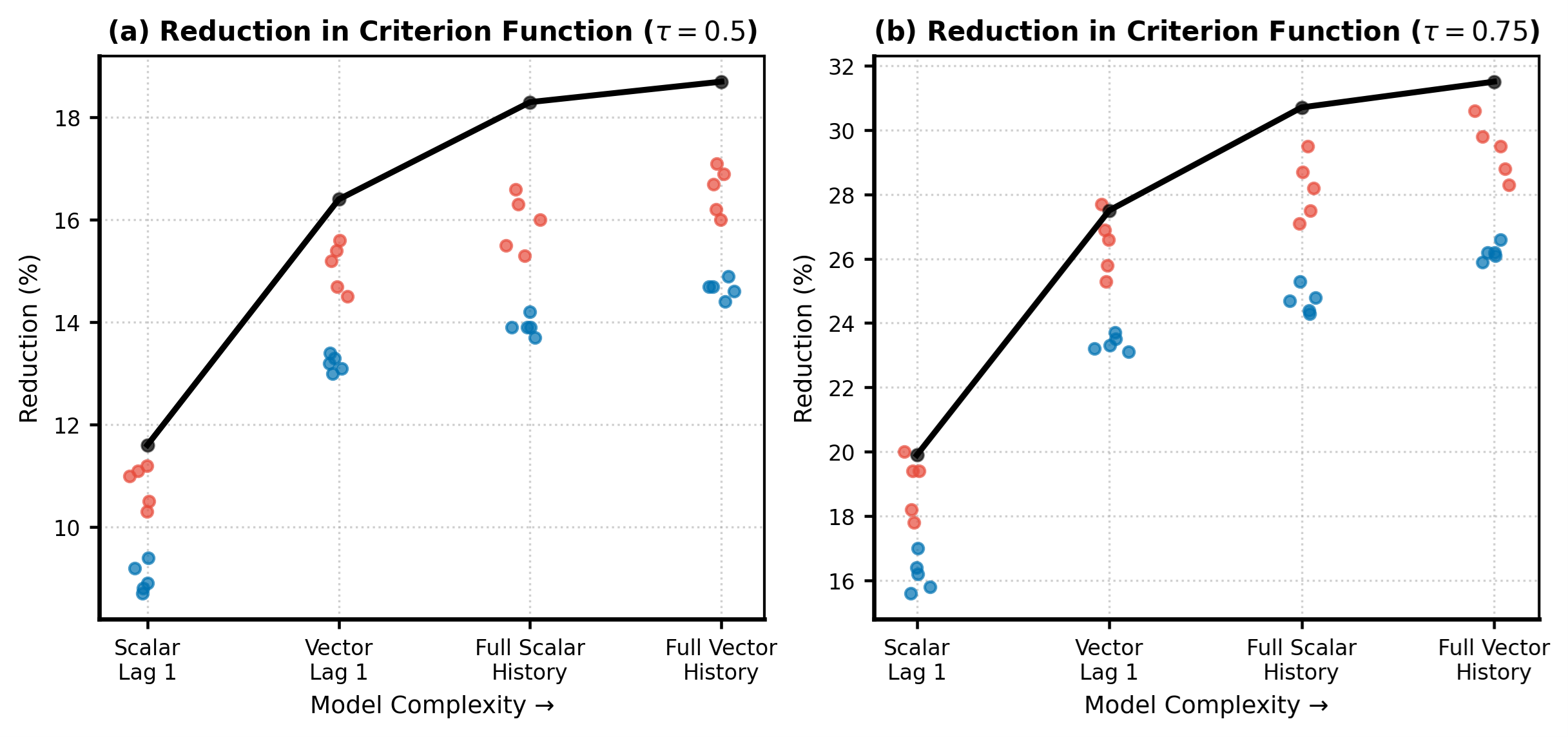}
    \caption{Quantile regression validation for \textbf{(a)} $\tau=0.5$ and \textbf{(b)} $\tau=0.75$. The plots show the relative reduction in the criterion function $C_\tau$ as predictive complexity increases. \textbf{Black:} observed data; \textbf{red:} Weather as Embeddings ($n=5$ runs); \textbf{blue:} Weather as Features ($n=5$ runs). Synthetic replicate points are jittered slightly in the horizontal direction to reduce overlap. The systematic performance gap between the Embeddings and Features models persists across these lower quantiles.}
    \label{fig:quantile_supp}
\end{figure}

The main text gives detailed results on the upper tail ($\tau=0.9$), where heteroskedastic effects are most pronounced.  Here, we provide complementary results for lower quantiles ($\tau=0.5$ and $\tau=0.75$), which show lower values for the reduction in the criterion value $C_\tau$ but qualitatively similar patterns in state- and history-dependent variability.
Figure~\ref{fig:quantile_supp} mirrors the main-paper diagnostic (Figure~\ref{fig:quantile_validation}), but for $\tau=0.5$ and $\tau=0.75$.  As in the main paper, we report the relative reduction in the minimized criterion value $C_\tau$ as predictive complexity increases (from a scalar lag-1 model to a full vector-history model).  
 
% \begin{figure}[!htbp]
%     \centering
%     \includegraphics[width=1.0\linewidth]{figures/Additional qr diagnostics/tao.png}
%     \caption{Quantile regression validation for \textbf{(a)} $\tau=0.5$ and \textbf{(b)} $\tau=0.75$. The plots show the relative reduction in the criterion function $C_\tau$ as predictive complexity increases. \textbf{Black:} observed data; \textbf{red:} Weather as Embeddings ($n=5$ runs); \textbf{blue:} Weather as Features ($n=5$ runs). Synthetic replicate points are jittered slightly in the horizontal direction to reduce overlap. The systematic performance gap between the Embeddings and Features models persists across these lower quantiles.}
%     \label{fig:quantile_supp}
% \end{figure}

Table~\ref{tab:rq_s6_numbers} provides a numerical summary of Figure~\ref{fig:quantile_supp}.  For each model class and quantile, synthetic results are reported as mean (SD) across $n=5$ independent runs; observed values are computed from the training data.

%%%%FIX!
\begin{table}[!htbp]
\caption{Numerical summary for Figure~\ref{fig:quantile_supp}. Entries for synthetic generators are mean (SD) across $n=5$ runs; observed values are from the training data. Values shown are relative reductions in $C_\tau$ (\%).}
\label{tab:rq_s6_numbers}
%\resizebox{0.85\linewidth}{%
\centering
\begin{tabular}{lccc ccc}
\toprule
& \multicolumn{3}{c}{$\tau=0.5$} & \multicolumn{3}{c}{$\tau=0.75$} \\
\cmidrule(lr){2-4}\cmidrule(lr){5-7}
Model & Observed & Embeddings & Features & Observed & Embeddings & Features \\
\midrule
Scalar lag-1         & 11.6 & 10.82 (0.40) &  9.00 (0.29) & 19.9 & 18.96 (0.92) & 16.20 (0.55) \\
Vector lag-1         & 16.4 & 15.08 (0.47) & 13.20 (0.16) & 27.5 & 26.46 (0.94) & 23.36 (0.24) \\
Full scalar history  & 18.3 & 15.94 (0.54) & 13.92 (0.18) & 30.7 & 28.20 (0.95) & 24.70 (0.39) \\
Full vector history  & 18.7 & 16.58 (0.47) & 14.66 (0.18) & 31.5 & 29.40 (0.89) & 26.20 (0.25) \\
\bottomrule
\end{tabular}
\end{table}

\section{Details of Discriminative Evaluation}
\label{app:disc}

The discriminative classifier is implemented using a bidirectional Long Short-Term Memory (Bi-LSTM) network to effectively capture temporal dependencies. The architecture comprises two stacked Bi-LSTM layers, each with a hidden size of 64. Due to the bidirectional configuration, the output dimension of these layers is 128. To aggregate the temporal information into a single feature vector, we apply global average pooling across the time dimension. This pooled output is then passed through a feed-forward network consisting of a fully connected layer mapping 128 inputs to 32 hidden units, followed by a ReLU activation and a dropout layer with a rate of 0.2. A final fully connected layer maps the 32 hidden units to a single scalar output, which is passed through a sigmoid function to obtain the classification probability.

The model is trained using the Binary Cross-Entropy loss function and the Adam optimizer with a learning rate of $1 \times 10^{-3}$. We use a batch size of 64 and train for a maximum of 1000 epochs. Model selection is performed by evaluating the accuracy on the validation set every 5 epochs. We retain the model checkpoint that yields the highest validation discriminative score. All experiments were implemented in PyTorch and executed on a single NVIDIA A100 GPU.

\begin{figure}[!htbp]
    \centering
    \includegraphics[width=0.55\linewidth]{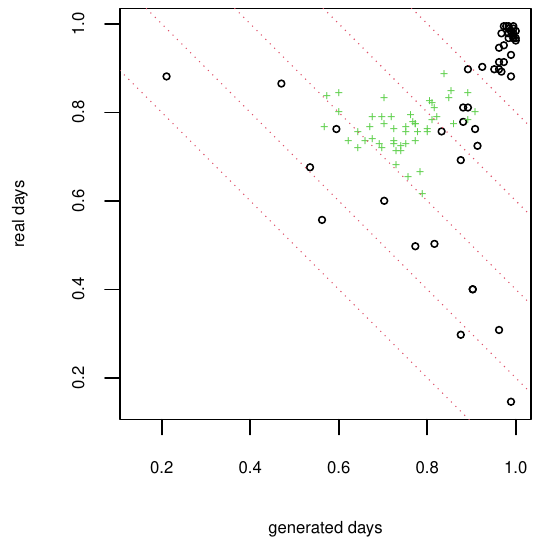}
    \caption{Accuracy of discriminator using generator with embedded weather states and consecutive days.  Green plus signs for discriminator applied to unfiltered data and black circles for discriminator applied to high-pass filtered data. Red dotted lines indicate levels of overall accuracy of discriminator of 0.5, 0.6, 0.7, 0.8 and 0.9.}
    \label{fig:gen-real}
\end{figure}

Figure~\ref{fig:gen-real} gives the probabilities of detecting real and synthetic days using this discriminator to simulations generated using embedded weather states and days simulated consecutively.
Results dramatically differ depending on whether the simulations are first processed through a high-pass filter.
In particular, the discriminator often performs nearly perfectly on the filtered data but occasionally quite poorly.
For the unfiltered data, the discriminator performs much more consistently across synthetic datasets but noticeably worse on average.

% \begin{figure}[!htbp]
%     \centering
%     \includegraphics[width=0.55\linewidth]{figures/Discriminative Score/gen-real.pdf}
%     \caption{Accuracy of discriminator using generator with embedded weather states and consecutive days.  Green plus signs for discriminator applied to unfiltered data and black circles for discriminator applied to high-pass filtered data. Red dotted lines indicate levels of overall accuracy of discriminator of 0.5, 0.6, 0.7, 0.8 and 0.9.}
%     \label{fig:gen-real}
% \end{figure}

\section{Additional Energy Scores Discussion}
\label{app:Add-En}

\begin{figure}[!htbp]
    \centering
    \includegraphics[width=0.8\linewidth]{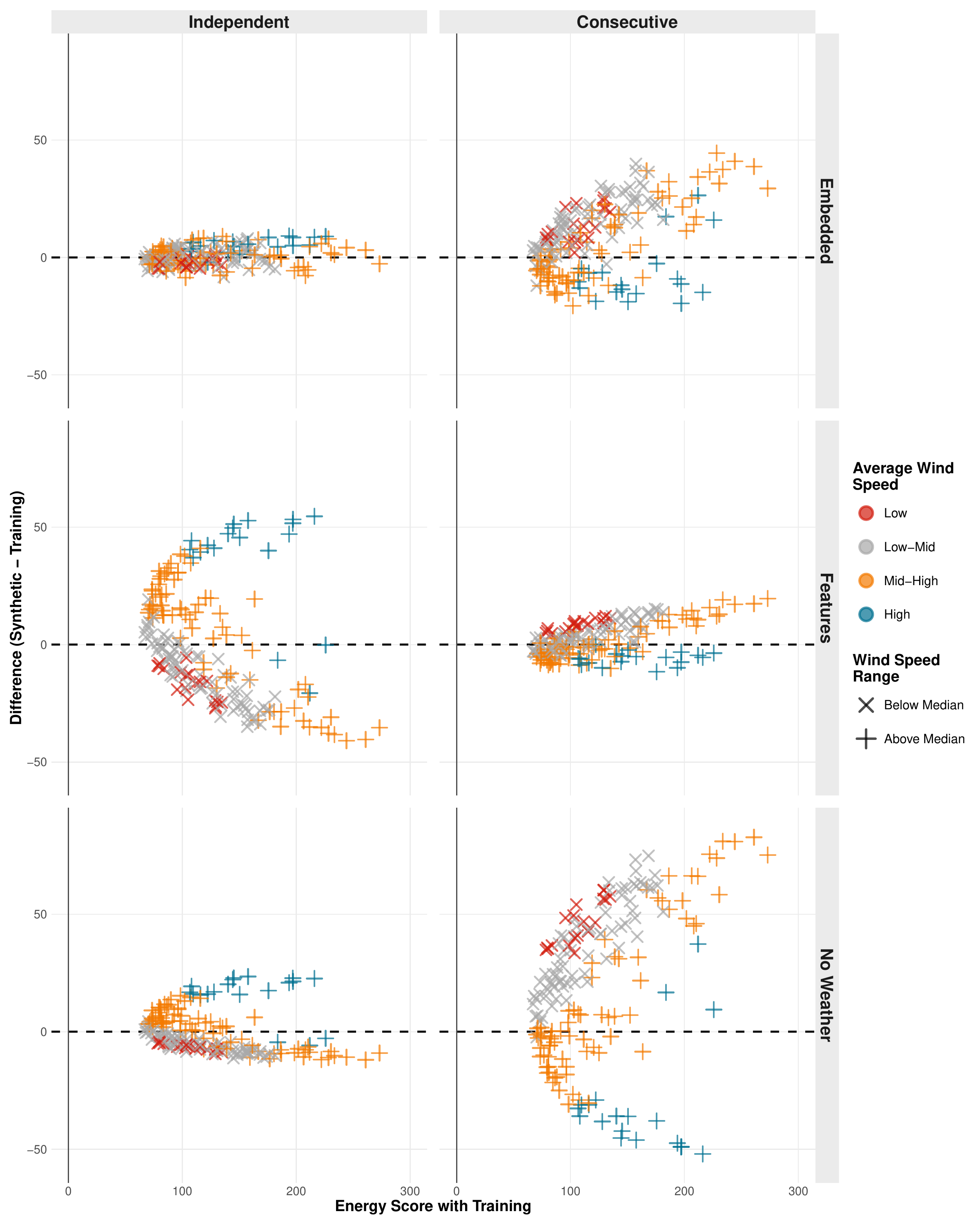}
    \caption{Day-level energy score differences for each generator and sampling type combination. Each point represents one of the 185 test days colored by wind speed over the test day with Low (0-0.1 quantile), Low-Mid (0.1-0.5 quantile), Mid-High (0.5-0.9 quantile), and High (0.9-1 quantile). The vertical axes shows the difference between synthetic and training energy scores versus training energy, with the horizontal dashed line ($y = 0$) indicating perfect agreement. }
    \label{fig:es_day_grid}
\end{figure}

As discussed in Sect.~\ref{sec:energy_scores}, energy scores are used to evaluate the accuracy of the estimated distributions of our generators. Figure~\ref{fig:es_day_grid} shows the difference between the day-level energy score performance for each generator and sampling type combination versus the energy score based on the training data. The bifurcation seen in the Embedded generator with consecutive sampling that led to a ``C'' shaped pattern is present in nearly all generator and sampling type combinations, with only Embedded with independent sampling having nearly no discernible bimodal behavior. Interestingly for the Embedded and No Weather generators, this relationship is far clearer with the consecutive sampling than independent, but with the Features generator, this flips. These patterns also appear for the hourly-level energy scores (results not shown). %\textcolor{blue}{Since we are not showing all 6 hourly plots, I am not sure whether I should mention them at all, but for now I did.}

Figure~\ref{fig:es_day_grid} also shows that for consecutive sampling the largest average wind speeds tend to fall below $y = 0$ and the smallest average wind speeds above $y = 0$, however, for independent sampling the larger tend to be above $y=0$ and the smaller averages below $y =0$. Table~\ref{tab:es_pct_summary} reports the five number summary (minimum, Q1, median, Q3, maximum) of the day-level energy scores expressed as a percentage of training energy scores ($100 \times \frac{\text{Synth} - \text{Train}}{\text{Train}}$) for the six synthetic data types. The Embedded generator with independent sampling had a median of $+0.55\%$ and $[Q_1, Q_3]$ of $[-2.12\%, +3.12\%]$, which is the closest any of the synthetic types got to training performance. These five number summaries continue to drive home the difference in energy score performance for consecutive versus independent sampling. For the Embedded generator, the middle 50\% of percentage differences are $[-2.12\%, +3.12\%]$ for independent and $[-4.39\%, +14.40\%]$ for consecutive sampling. The independent sampling days are more tightly distributed around 0\% and have less of a tendency to produce higher energy scores than the training distribution, which are desirable traits. This difference between independent and consecutive sampling is exhibited in the Embedded and No Weather generators, but for the Features generator, consecutive sampling has the desired traits. We do not have an explanation for these results, but find it noteworthy that there is no simple pattern in which generators perform best in this assessment.

% \begin{figure}[!htbp]
%     \centering
%     \includegraphics[width=0.8\linewidth]{figures/Energy Scores/es_scatter_day_diff_combined.pdf}
%     \caption{Day-level energy score differences for each generator and sampling type combination. Each point represents one of the 185 test days colored by wind speed over the test day with Low (0-0.1 quantile), Low-Mid (0.1-0.5 quantile), Mid-High (0.5-0.9 quantile), and High (0.9-1 quantile). The vertical axes shows the difference between synthetic and training energy scores versus training energy, with the horizontal dashed line ($y = 0$) indicating perfect agreement. }
%     \label{fig:es_day_grid}
% \end{figure}

\begin{table}[!htbp]
\caption{Five-number summary of day-level energy score differences expressed as percentages of training energy scores ($100 \times \frac{\text{Synth} - \text{Train}}{\text{Train}}$) for each generator and sampling type across 185 test days.}
\label{tab:es_pct_summary}
\centering
\begin{tabular}{llrrrrr}
\hline
Generator & Sampling & Min & Q1 & Median & Q3 & Max \\
\hline
\multirow{2}{*}{Embedded}   & Independent &  $-8.33$ &  $-2.12$ &  $0.55$ &  $3.12$ &  $7.04$ \\
                             & Consecutive & $-20.10$ &  $-4.39$ &  $7.32$ & $14.40$ & $25.30$ \\
\hline
\multirow{2}{*}{Features}   & Independent & $-23.00$ & $-14.90$ & $-4.69$ & $16.50$ & $40.80$ \\
                             & Consecutive & $-10.90$ &  $-3.67$ &  $0.91$ &  $5.57$ &  $9.56$ \\
\hline
\multirow{2}{*}{No Weather} & Independent &  $-8.18$ &  $-5.45$ & $-3.12$ &  $4.62$ & $17.90$ \\
                             & Consecutive & $-33.20$ &  $-2.78$ & $22.40$ & $35.70$ & $51.50$ \\
\hline
\end{tabular}
\end{table}

\FloatBarrier

\section{Implementation Detail}
\label{app:imp-detail}

As stated in Sect.~\ref{sec:Methodology}, we consider both independent synthetic and consecutive synthetic generators (conditional on previous day wind vector and weather states).
For independent synthetic, during training for stages 1 and 2, the 474 days of observed wind vectors and weather states in the training dataset are further split: the first 470 days are used for training and the last 4 for validation. For consecutive synthetic, we need to use two consecutive days at a time yielding potentially 20 overlapping two-day periods per year, or 460 in all.  
After taking account of missing days, we are left with 444 available two-day periods in the training data.
The first 440 of these 444 two-day periods are used to train consecutive synthetic generators, leaving 4 two-day periods for validation. The batch size is 16. For training in both stages 1 and 2, we use linear warmup and cosine annealing to schedule the learning rate. 

The learning rate $\eta_t$ at step $t$ is typically defined as follows:

1. During Warmup ($t < T_{warmup}$):$$\eta_t = \eta_{max} \times \frac{t}{T_{warmup}}$$

2. During Cosine Annealing ($t \geq T_{warmup}$):$$\eta_t = \eta_{min} + \frac{1}{2}(\eta_{max} - \eta_{min})\left(1 + \cos\left(\frac{t - T_{warmup}}{T_{total} - T_{warmup}} \pi\right)\right)$$
where 
\begin{itemize}
    \item $\eta_{max}=0.005$: The peak learning rate.
    \item $\eta_{min}=0.0005$: The minimum learning rate.
    \item $\eta_{warmup}=0.05$: The linear warmup rate.
    \item $T_{total}=40000$: The total number of training steps.
    \item $T_{warmup} = \lfloor T_{total} \times \eta_{warmup} \rfloor$: The number of steps allocated for warmup.
\end{itemize}

We use the AdamW optimizer \citep{loshchilov2018}, which performs the standard Adam update first, and then decays the weights with a rate of 0.01  directly.
During MaskGIT training, each sequence of tokens is randomly masked with ratio $\gamma=\cos{\frac{\pi r}{2}}$ and $r$ is uniformly sampled.
\noappendix

\authorcontribution{Much of the work for this paper was done as part of the class STAT690, Data Analysis Project, offered at Rutgers University during Spring of 2025.
The first six authors (listed alphabetically) were the students in this class and the last author was the instructor.
Mingshi Cui and Zern Ke, supervised by Gemma Moran, contributed to the literature review of stochastic weather generators for wind, the modeling and implementation of the derived Time VQ-VAE generative model presented in this paper. Kevin Eng contributed to the weather state implementation and conceptualization.
Justin Greene contributed to the data collection and compilation from ARM, Brownian bridge imputation method, ensuring generators did not memorize training data and energy score analyses.
Abolfazl Sodagartojgi contributed to the stochastic-volatility validation, quantile-regression analysis, figure generation and other numerical summaries in the appendix. 
Zhiqiu Xia conducted the discriminative evaluation by implementing a BiLSTM classifier to distinguish real from synthetic wind sequences.
Michael Stein conducted the exploratory analyses and coordinated the overall effort.
All authors participated in the writing and editing of the paper.}

\codedataavailability{The source code and reproduction materials used to implement the proposed stochastic weather generator and reproduce the numerical experiments are archived in Zenodo \citep{2026windcode}. The development version is available on GitHub at \url{https://github.com/Zernjk/Stochastic-weather-generators-for-high-frequency-wind-vector-time-series}. The wind vector dataset used in this study is archived in Zenodo \citep{2026winddata}.}

\begin{acknowledgements}
Data were provided by the US Department of Energy Advanced Radiation Measurement (ARM) program, which has collected barometric pressure, temperature, relative humidity, wind speed, and wind direction at a rate of 1 Hz at the Southern Great Plains E13 station in Lamont, Oklahoma since 1993. ARM offers data collected from this station, as well as a range of others across globe, for free through the ARM Data Center via Data Discovery. 
\end{acknowledgements}

\smallskip\noindent
\textit{Competing interests.}
The authors declare that they have no conflict of interest.

%\printbibliography
\bibliographystyle{copernicus}
\bibliography{ref}

@inproceedings{
loshchilov2018,
title={Decoupled Weight Decay Regularization},
author={Ilya Loshchilov and Frank Hutter},
booktitle={International Conference on Learning Representations},
year={2019},
url={https://openreview.net/forum?id=Bkg6RiCqY7},
}

@article{yin2026,
title = {Variations in fundamental statistics of wind speeds based on high-frequency meteorological station data in Fukuoka, Japan},
journal = {Journal of Wind Engineering and Industrial Aerodynamics},
volume = {268},
pages = {106286},
year = {2026},
issn = {0167-6105},
doi = {https://doi.org/10.1016/j.jweia.2025.106286},
url = {https://www.sciencedirect.com/science/article/pii/S016761052500282X},
author = {Sankang Yin and Yezhan Li and Naoki Ikegaya}
}

@INPROCEEDINGS{10715230,
  author={Apellaniz, Patricia A. and Parras, Juan and Zazo, Santiago},
  booktitle={2024 32nd European Signal Processing Conference (EUSIPCO)}, 
  title={An Improved Tabular Data Generator with VAE-GMM Integration}, 
  year={2024},
  volume={},
  number={},
  pages={1886-1890},
  doi={10.23919/EUSIPCO63174.2024.10715230}}

@misc{met2021,
  doi = {10.5439/1786358},
  url = {https://www.osti.gov/servlets/purl/1786358/},
  author = {Kyrouac,  Jenni and Shi,  Yan and Tuftedal,  Matt},
  language = {en},
  title = {met.b1},
  publisher = {Atmospheric Radiation Measurement (ARM) Archive,  Oak Ridge National Laboratory (ORNL),  Oak Ridge,  TN (US); ARM Data Center,  Oak Ridge National Laboratory (ORNL),  Oak Ridge,  TN (United States)},
  year = {2021}
}

@article{carta2009,
title = {A review of wind speed probability distributions used in wind energy analysis: Case studies in the Canary Islands},
journal = {Renewable and Sustainable Energy Reviews},
volume = {13},
number = {5},
pages = {933-955},
year = {2009},
issn = {1364-0321},
doi = {https://doi.org/10.1016/j.rser.2008.05.005},
author = {J.A. Carta and P. Ramírez and S. Vel\'azquez}
}

@article{vaswani2017attention,
  title={Attention is all you need},
  author={Vaswani, Ashish and Shazeer, Noam and Parmar, Niki and Uszkoreit, Jakob and Jones, Llion and Gomez, Aidan N and Kaiser, {\L}ukasz and Polosukhin, Illia},
  journal={Neural Information Processing Systems},
  volume={30},
  year={2017}
}

@inproceedings{
lipman2023flow,
title={Flow Matching for Generative Modeling},
author={Yaron Lipman and Ricky T. Q. Chen and Heli Ben-Hamu and Maximilian Nickel and Matthew Le},
booktitle={International Conference on Learning Representations},
year={2023},
url={https://openreview.net/forum?id=PqvMRDCJT9t}
}

@inproceedings{rhudy2024,
author = {Matthew B. Rhudy and Mark Longenberger},
title = {Stochastic Wind Speed Modeling and Prediction Using Historical Wind Data for Aircraft Applications},
booktitle = {AIAA AVIATION FORUM AND ASCEND 2024},
chapter = {},
pages = {},
year = {2024},
doi = {10.2514/6.2024-3849}
}

@inproceedings{masoudian2021,
  title={Stochastic modelling of wind and its implication for wildfire spread predictions},
  author={Masoudian, S and Sharples, J and Jovanoski, Z and Watt, S and Towers, I and Vervoort, RW and Voinov, AA and Evans, JP and Marshall, L},
  booktitle={24th International Congress on Modelling and Simulation, Sydney, NSW, Australia},
  year={2021}
}

@article{skidmore1990,
  title={Stochastic wind simulation for erosion modeling},
  author={Skidmore, EL and Tatarko, J},
  journal={Transactions of the ASAE},
  volume={33},
  number={6},
  pages={1893--1899},
  year={1990},
  publisher={American Society of Agricultural and Biological Engineers}
}

@article{wang2025,
title = {HWGEN: An hourly wind stochastic GENerator},
journal = {International Soil and Water Conservation Research},
year = {2025},
issn = {2095-6339},
doi = {https://doi.org/10.1016/j.iswcr.2025.10.005},
author = {Wang, H. and Liu, J. and Yin, S and Qiao, H. and Zhu, Z. and Hall, J.W.}
}

@article{shah2025,
  title={Generative modeling of microweather wind velocities for urban air mobility},
  author={Shah, Tristan A and Stanley, Michael C and Warner, James E},
  journal={arXiv preprint arXiv:2503.02690},
  year={2025}
}

@article{nikolaev2019,
title = {A regime-switching recurrent neural network model applied to wind time series},
journal = {Applied Soft Computing},
volume = {80},
pages = {723-734},
year = {2019},
issn = {1568-4946},
doi = {https://doi.org/10.1016/j.asoc.2019.04.009},
author = {Nikolay Y. Nikolaev and Evgueni Smirnov and Daniel Stamate and Robert Zimmer}
}

@ARTICLE{yunus2016,
  author={Yunus, Kalid and Thiringer, Torbjörn and Chen, Peiyuan},
  journal={IEEE Transactions on Power Systems}, 
  title={ARIMA-Based Frequency-Decomposed Modeling of Wind Speed Time Series}, 
  year={2016},
  volume={31},
  number={4},
  pages={2546-2556},
  doi={10.1109/TPWRS.2015.2468586}}

@article{brown1984,
  title={Time series models to simulate and forecast wind speed and wind power},
  author={Brown, Barbara G and Katz, Richard W and Murphy, Allan H},
  journal={Journal of Applied Meteorology and Climatology},
  volume={23},
  number={8},
  pages={1184--1195},
  year={1984}
}

@article{wang2024,
    author = {Wang, Kesen and Kim, Minwoo and Castruccio, Stefano and Genton, Marc G},
    title = {Modelling high-resolution spatio-temporal wind with deep echo state networks and stochastic partial differential equations},
    journal = {Journal of the Royal Statistical Society Series C: Applied Statistics},
    pages = {qlaf045},
    year = {2025},
    month = {08},
    issn = {0035-9254},
    doi = {10.1093/jrsssc/qlaf045}
}

@misc{kyrouac2025, 
title={Surface Meteorological Instrumentation (MET), 1993-07-21 to 2025-02-03, Southern Great Plains (SGP), Lamont, OK (Extended and Co-located with C1) (E13)}, 
DOI={10.5439/1786358}, 
journal={Atmospheric Radiation Measurement (ARM) user facility}, 
author={Kyrouac, Jenni and Shi, Yan and Tuftedal, Matt},
year={2025}
}

@article{gneiting2007,
  title={Strictly proper scoring rules, prediction, and estimation},
  author={Gneiting, Tilmann and Raftery, Adrian E},
  journal={Journal of the American Statistical Association},
  volume={102},
  number={477},
  pages={359--378},
  year={2007},
  publisher={Taylor \& Francis}
}

@article{gneiting2008,
  title={Assessing probabilistic forecasts of multivariate quantities, with an application to ensemble predictions of surface winds},
  author={Gneiting, Tilmann and Stanberry, Larissa I and Grimit, Eric P and Held, Leonhard and Johnson, Nicholas A},
  journal={Test},
  volume={17},
  pages={211--235},
  year={2008},
  publisher={Springer}
}

@article{jordan2019,
    author = {Jordan, Alexander and Krüger, Fabian and Lerch, Sebastian},
    title = {Evaluating probabilistic forecasts with scoring rules},
    journal = {Journal of Statistical Software},
    volume = {90},
    issue = {12},
    pages = {1-37},
    year = {2019},
    doi = {10.18637/jss.v090.i12}
}

@article{porte2013,
  title={A numerical study of the effects of wind direction on turbine wakes and power losses in a large wind farm},
  author={Port{\'e}-Agel, Fernando and Wu, Yu-Ting and Chen, Chang-Hung},
  journal={Energies},
  volume={6},
  number={10},
  pages={5297--5313},
  year={2013},
  publisher={MDPI}
}

@misc{Oklahomatornadoes,
  title        = {Oklahoma Tornadoes by County and Month (1950-2024)},
  author       = {{National Weather Service}},
  year         = 2025,
  note         = {\url{https://www.weather.gov/oun/tornadodata-ok-countybymonth} [Accessed: 11/13/2025]}
}

@Article{dallas2024,
AUTHOR = {Dallas, S. and Stock, A. and Hart, E.},
TITLE = {Control-oriented modelling of wind direction variability},
JOURNAL = {Wind Energy Science},
VOLUME = {9},
YEAR = {2024},
NUMBER = {4},
PAGES = {841--867},
DOI = {10.5194/wes-9-841-2024}
}

@article{koenker2017,
  title={Quantile regression: 40 years on},
  author={Koenker, Roger},
  journal={Annual Review of Economics},
  volume={9},
  number={1},
  pages={155--176},
  year={2017},
  publisher={Annual Reviews}
}

@book{koenker2005, 
place={Cambridge}, 
series={Econometric Society Monographs}, 
title={Quantile Regression}, 
publisher={Cambridge University Press}, 
author={Koenker, Roger}, 
year={2005}, 
collection={Econometric Society Monographs}
}

@misc{quantreg2025,
  author = {Koenker, Roger},
  title = {{quantreg}: Quantile Regression},
  year = {2025},
  note = {R package version 6.1, accessed 18 April 2025},
  url = {https://CRAN.R-project.org/package=quantreg}
}

@article{gneiting2006,
  title={Calibrated probabilistic forecasting at the stateline wind energy center: The regime-switching space--time method},
  author={Gneiting, Tilmann and Larson, Kristin and Westrick, Kenneth and Genton, Marc G and Aldrich, Eric},
  journal={Journal of the American Statistical Association},
  volume={101},
  number={475},
  pages={968--979},
  year={2006},
  publisher={Taylor \& Francis}
}

@article{hering2010,
  title={Powering up with space-time wind forecasting},
  author={Hering, Amanda S and Genton, Marc G},
  journal={Journal of the American Statistical Association},
  volume={105},
  number={489},
  pages={92--104},
  year={2010},
  publisher={Taylor \& Francis}
}

@article{zhang2024,
  title={Multi-step ahead forecasting of wind vector for multiple wind turbines based on new deep learning model},
  author={Zhang, Zhendong and Dai, Huichao and Jiang, Dingguo and Yu, Yi and Tian, Rui},
  journal={Energy},
  volume={304},
  pages={131964},
  year={2024},
  publisher={Elsevier}
}

@article{zhu2012,
author = {Zhu, Xinxin and Genton, Marc G},
title = {Short-Term Wind Speed Forecasting for Power System Operations},
journal = {International Statistical Review},
volume = {80},
number = {1},
pages = {2-23},
doi = {https://doi.org/10.1111/j.1751-5823.2011.00168.x},
year = {2012}
}

@article{liu2020,
title = {A combined forecasting model for time series: Application to short-term wind speed forecasting},
journal = {Applied Energy},
volume = {259},
pages = {114137},
year = {2020},
issn = {0306-2619},
doi = {https://doi.org/10.1016/j.apenergy.2019.114137},
author = {Zhenkun Liu and Ping Jiang and Lifang Zhang and Xinsong Niu}
}

@article{jiang2013,
title = {Very short-term wind speed forecasting with Bayesian structural break model},
journal = {Renewable Energy},
volume = {50},
pages = {637-647},
year = {2013},
issn = {0960-1481},
doi = {https://doi.org/10.1016/j.renene.2012.07.041},
author = {Yu Jiang and Zhe Song and Andrew Kusiak}
}

@article{WindSpeedDistributions2021,
  author    = {Shi, Yingyi and Zhao, Wenjing and Guan, Honghong and Kumar, Nirmal},
  title     = {Wind Speed Distributions Used in Wind Energy Assessment: A Review},
  journal   = {Frontiers in Energy Research},
  volume    = {9},
  pages     = {769920},
  year      = {2021},
  doi       = {10.3389/fenrg.2021.769920}
}

@article{hering2015,
  title={A Markov-switching vector autoregressive stochastic wind generator for multiple spatial and temporal scales},
  author={Hering, Amanda S and Kazor, Karen and Kleiber, William},
  journal={Resources},
  volume={4},
  number={1},
  pages={70--92},
  year={2015},
  publisher={MDPI}
}

@article{bessac2016,
AUTHOR = {Bessac, J. and Ailliot, P. and Cattiaux, J. and Monbet, V.},
TITLE = {Comparison of hidden and observed regime-switching autoregressive models for ($u,v$)-components of wind fields in the northeastern Atlantic},
JOURNAL = {Advances in Statistical Climatology, Meteorology and Oceanography},
VOLUME = {2},
YEAR = {2016},
NUMBER = {1},
PAGES = {1--16},
DOI = {10.5194/ascmo-2-1-2016}
}

@inproceedings{lee2023vector,
  title={Vector Quantized Time Series Generation with a Bidirectional Prior Model},
  author={Lee, Daesoo and Malacarne, Sara and Aune, Erlend},
  booktitle={International Conference on Artificial Intelligence and Statistics},
  pages={7665--7693},
  year={2023},
  organization={PMLR}
}

@article{yoon2019time,
  title={Time-series generative adversarial networks},
  author={Yoon, Jinsung and Jarrett, Daniel and Van der Schaar, Mihaela},
  journal={Neural Information Processing Systems},
  year={2019}
}

@article{van2017neural,
  title={Neural discrete representation learning},
  author={Van Den Oord, Aaron and Vinyals, Oriol and others},
  journal={Neural Information Processing Systems},
  volume={30},
  year={2017}
}

@inproceedings{rombach2022high,
  title={High-resolution image synthesis with latent diffusion models},
  author={Rombach, Robin and Blattmann, Andreas and Lorenz, Dominik and Esser, Patrick and Ommer, Bj{\"o}rn},
  booktitle={Proceedings of the IEEE/CVF Conference on Computer Vision and Pattern Recognition},
  pages={10684--10695},
  year={2022}
}

@inproceedings{chang2022maskgit,
  author    = {Huiwen Chang and Han Zhang and Lu Jiang and Ce Liu and William T. Freeman},
  title     = {{MaskGit}: Masked Generative Image Transformer},
  booktitle = {Proceedings of the IEEE/CVF Conference on Computer Vision and Pattern Recognition (CVPR)},
  year      = {2022},
  pages     = {11315--11325},
}

@article{desai2021timevae,
  title={Timevae: A variational auto-encoder for multivariate time series generation},
  author={Desai, Abhyuday and Freeman, Cynthia and Wang, Zuhui and Beaver, Ian},
  journal={arXiv preprint arXiv:2111.08095},
  year={2021}
}

@inproceedings{
liu2025sundial,
title={Sundial: A Family of Highly Capable Time Series Foundation Models},
author={Yong Liu and Guo Qin and Zhiyuan Shi and Zhi Chen and Caiyin Yang and Xiangdong Huang and Jianmin Wang and Mingsheng Long},
booktitle={International Conference on Machine Learning},
year={2025},
url={https://openreview.net/forum?id=LO7ciRpjI5}
}

@misc{2026windcode,
  author       = {Cui, Mingshi and
                  Eng, Kevin and
                  Greene, Justin T. and
                  Ke, Zern and
                  Sodagartojgi, Abolfazl and
                  Xia, Zhiqiu and
                  Moran, Gemma E. and
                  Stein, Michael L.},
  title        = {Zernjk/Stochastic-weather-generators-for-high-
                   frequency-wind-vector-time-series: Version 1.0.0
                   for Copernicus manuscript submission
                  },
  month        = may,
  year         = 2026,
  publisher    = {Zenodo},
  version      = {v1.0.0},
  doi          = {10.5281/zenodo.20421182},
  url          = {https://doi.org/10.5281/zenodo.20421182},
}

@misc{2026winddata,
  author       = {Cui, Mingshi and
                  Eng, Kevin and
                  Greene, Justin T. and
                  Ke, Zern and
                  Sodagartojgi, Abolfazl and
                  Xia, Zhiqiu and
                  Moran, Gemma E. and
                  Stein, Michael L.},
  title        = {Wind Vector Data for Stochastic Wind Vector
                   Generation
                  },
  month        = may,
  year         = 2026,
  publisher    = {Zenodo},
  doi          = {10.5281/zenodo.20421238},
  url          = {https://doi.org/10.5281/zenodo.20421238},
}

\end{document}